\documentclass[longauth]{aa} % for the long lists of affiliations 
\pdfoutput=1
\usepackage{graphicx}
\usepackage{txfonts}
\usepackage[utf8]{inputenc}
\usepackage{array}
\usepackage{float}
\usepackage{multirow}
\usepackage[]{natbib}
\usepackage[colorlinks=true,linkcolor=blue,citecolor=blue]{hyperref}
\usepackage{color}

\begin{document} 

    \title{First light of the VLT planet finder SPHERE}
    \subtitle{II. The physical properties and the architecture of the young systems PZ~Tel and HD~1160 revisited\thanks{Based on data collected at the European Southern Observatory, Chile, during the commissioning of the SPHERE instrument and ESO programs 085.C-0277, 087.C-0109, 087.C-0535, and 060.A-9026.}}

   \author{A.-L. Maire\inst{1}, M. Bonnefoy\inst{2,3}, C. Ginski\inst{4}, A. Vigan\inst{5,6}, S. Messina\inst{7}, D. Mesa\inst{1}, R. Galicher\inst{8}, R. Gratton\inst{1}, S. Desidera\inst{1}, T. G. Kopytova\inst{9,10}, M. Millward\inst{11}, C. Thalmann\inst{12}, R.~U. Claudi\inst{1}, D. Ehrenreich\inst{13}, A. Zurlo\inst{5,1,14,15}, G. Chauvin\inst{2,3}, J. Antichi\inst{16}, A. Baruffolo\inst{1}, A. Bazzon\inst{12}, J.-L. Beuzit\inst{2,3}, P. Blanchard\inst{5}, A. Boccaletti\inst{8}, J. de Boer\inst{6}, M. Carle\inst{5}, E. Cascone\inst{17}, A. Costille\inst{5}, V. De Caprio\inst{17}, A. Delboulb\'e\inst{2,3}, K. Dohlen\inst{5}, C. Dominik\inst{18}, M. Feldt\inst{9}, T. Fusco\inst{19,5}, J.~H. Girard\inst{6,2,3}, E. Giro\inst{1}, D. Gisler\inst{20}, L. Gluck\inst{2,3}, C. Gry\inst{5}, T. Henning\inst{9}, N. Hubin\inst{21}, E. Hugot\inst{5}, M. Jaquet\inst{5}, M. Kasper\inst{21,2,3}, A.-M. Lagrange\inst{2,3}, M. Langlois\inst{22,5}, D. Le Mignant\inst{5}, M. Llored\inst{5}, F. Madec\inst{5}, P. Martinez\inst{23}, D. Mawet\inst{24}, J. Milli\inst{6,2,3}, O. M\"oller-Nilsson\inst{9}, D. Mouillet\inst{2,3}, T. Moulin\inst{2,3}, C. Moutou\inst{5,25}, A. Orign\'e\inst{5}, A. Pavlov\inst{9}, C. Petit\inst{19}, J. Pragt\inst{26}, P. Puget\inst{2,3}, J. Ramos\inst{9}, S. Rochat\inst{2,3}, R. Roelfsema\inst{26}, B. Salasnich\inst{1}, J.-F. Sauvage\inst{19,5}, H. M. Schmid\inst{12}, M. Turatto\inst{1}, S. Udry\inst{13}, F. Vakili\inst{23}, Z. Wahhaj\inst{6,5}, L. Weber\inst{13}, and F. Wildi\inst{13}}

   \institute{INAF -- Osservatorio Astronomico di Padova, Vicolo dell'Osservatorio 5, 35122 Padova, Italy\\
              \email{annelise.maire@oapd.inaf.it}  
              \and
                          Universit\'e Grenoble Alpes, IPAG, 38000 Grenoble, France\\
                          \email{mickael.bonnefoy@obs.ujf-grenoble.fr}
              \and
              CNRS, IPAG, F-38000 Grenoble, France            
                          \and
                          Leiden Observatory, Leiden University, P.O. Box 9513, 2300 RA Leiden, The Netherlands\\
                          \email{ginski@strw.leidenuniv.nl}
                          \and
                          Aix Marseille Universit\'e, CNRS, LAM (Laboratoire d'Astrophysique de Marseille) UMR 7326, 13388, Marseille, France\\
                          \email{arthur.vigan@lam.fr} 
                          \and
                          European Southern Observatory, Alonso de Cordova 3107, Casilla 19001 Vitacura, Santiago 19, Chile
                          \and
                          INAF Catania Astrophysical Observatory, via S. Sofia 78, 95123 Catania, Italy\\
                          \email{sergio.messina@oact.inaf.it} 
              \and
              LESIA, CNRS, Observatoire de Paris, Universit\'e Paris Diderot, UPMC, 5 place J. Janssen, 92190 Meudon, France
              \and
              Max-Planck-Institut f\"ur Astronomie, K\"onigstuhl 17, 69117 Heidelberg, Germany
              \and
              International Max Planck Research School for Astronomy and Space Physics, Heidelberg, Germany
              \and
              York Creek Observatory, Georgetown, Tasmania, Australia
              \and
              Institute for Astronomy, ETH Zurich, Wolfgang-Pauli-Strasse 27, 8093 Zurich, Switzerland              
              \and
              Geneva Observatory, University of Geneva, Chemin des Maillettes 51, 1290 Versoix, Switzerland
              \and
              N\'ucleo de Astronom\'ia, Facultad de Ingenier\'ia, Universidad Diego Portales, Av. Ejercito 441, Santiago, Chile
              \and
              Millennium Nucleus ``Protoplanetary Disk'', Departamento de Astronom\'ia, Universidad de Chile, Casilla 36-D, Santiago, Chile
              \and
              INAF -- Osservatorio Astrofisico di Arcetri, largo E. Fermi 5, 50125 Firenze, Italy 
              \and
              INAF -- Osservatorio Astronomico di Capodimonte, Salita Moiariello 16, 80131 Napoli, Italy
              \and
              Anton Pannekoek Astronomical Institute, University of Amsterdam, PO Box 94249, 1090 GE Amsterdam, The Netherlands
              \and
              ONERA, The French Aerospace Lab BP72, 29 avenue de la Division Leclerc, 92322 Ch\^atillon Cedex, France
              \and
              Kiepenheuer-Institut f\"ur Sonnenphysik, Sch\"oneckstr. 6, D-79104 Freiburg, Germany
              \and
              European Southern Observatory, Karl Schwarzschild St, 2, 85748 Garching, Germany
              \and
              Centre de Recherche Astrophysique de Lyon, CNRS/ENS-L/Universit\'e Lyon 1, 9 av. Ch. Andr\'e, 69561 Saint-Genis-Laval, France
              \and
              Laboratoire Lagrange, UMR 7293, Observatoire de la C\^ote d'Azur (OCA), Universit\'e de Nice-Sophia Antipolis (UNS), CNRS, Bd. de l'Observatoire, 06304 Nice, France
              \and
              Department of Astronomy, California Institute of Technology, 1200 E. California Blvd, MC 249-17, Pasadena, CA 91125 USA
              \and
              CNRS, Canada-France-Hawaii Telescope Corporation, 65-1238 Mamalahoa Hwy., Kamuela, HI-96743, USA
              \and
              NOVA Optical-Infrared Instrumentation Group at ASTRON, Oude Hoogeveensedijk 4, 7991 PD Dwingeloo, The Netherlands
              }

   \date{Received 25 May 2015; accepted 13 October 2015}

% \abstract{}{}{}{}{} 
% 5 {} token are mandatory
 
  \abstract
   {The young systems PZ~Tel and HD~1160, hosting known low-mass companions, were observed during the commissioning of the new planet finder of the Very Large Telescope (VLT) SPHERE with several imaging and spectroscopic modes.}
   {We aim to refine the physical properties and architecture of both systems.}
   {We use SPHERE commissioning data and dedicated Rapid Eye Mount (REM) observations, as well as literature and unpublished data from VLT/SINFONI, VLT/NaCo, Gemini/NICI, and Keck/NIRC2.}
   {We derive new photometry and confirm the short-term ($P$\,=\,0.92~d) photometric variability of the star PZ~Tel~A with values of 0.06 and 0.14~mag at optical and near-infrared wavelengths, respectively. We note from the comparison to literature data spanning 38~yr that the star also exhibits a long-term variability trend with a brightening of $\sim$0.25~mag. The 0.63--3.8~$\muup$m spectral energy distribution of PZ~Tel~B (separation $\sim$25~AU) allows us to revise its physical characteristics: spectral type M7$\pm$1, $T_{\rm{eff}}$\,=\,2700\,$\pm$\,100~K, log($g$)\,$<$\,4.5~dex, luminosity log($L/L_{\odot}$)\,=\,$-$2.51\,$\pm$\,0.10~dex, and mass 38--72~$M_{\rm{J}}$ from ``hot-start'' evolutionary models combining the ranges of the temperature and luminosity estimates. The 1--3.8~$\muup$m SED of HD~1160~B ($\sim$85~au) suggests a massive brown dwarf or a low-mass star with spectral type M6.0$^{+1.0}_{-0.5}$, $T_{\rm{eff}}$\,=\,3000\,$\pm$\,100~K, subsolar metallicity [M/H]\,$=$\,$-$0.5--0.0~dex, luminosity log($L/L_{\odot}$)\,=\,$-$2.81\,$\pm$\,0.10~dex, and mass 39--168~$M_{\rm{J}}$. The physical properties derived for HD~1160~C ($\sim$560~au) from $K_sL^{\prime}$-band photometry are consistent with the discovery study. The orbital study of PZ~Tel~B confirms its deceleration and the high eccentricity of its orbit ($e$\,$>$\,0.66). For eccentricities below 0.9, the inclination, longitude of the ascending node, and time of periastron passage are well constrained. In particular, both star and companion inclinations are compatible with a system seen edge-on. Based on ``hot-start'' evolutionary models, we reject other brown dwarf candidates outside 0.25$''$ for both systems, and giant planet companions outside 0.5$''$ {that are more massive than 3~$M_{\rm{J}}$} for the PZ~Tel system. {We also show that $K1-K2$ color can be used along with $YJH$ low-resolution spectra to identify young L-type companions, provided high photometric accuracy ($\leq$0.05~mag) is achieved.}}
   {SPHERE opens new horizons in the study of young brown dwarfs and giant exoplanets using direct imaging thanks to high-contrast imaging capabilities at optical (0.5--0.9~$\muup$m) and near-infrared (0.95--2.3~$\muup$m) wavelengths, as well as high signal-to-noise spectroscopy in the near-infrared domain (0.95--2.3~$\muup$m) from low resolutions ($R$\,$\sim$\,30--50) to medium resolutions ($R$\,$\sim$\,350).}


\authorrunning{A.-L. Maire et al.}
\titlerunning{Physical properties and architecture of the PZ~Tel and HD~1160 systems revisited with VLT/SPHERE.}

   \maketitle
%
%________________________________________________________________

\section{Introduction}

Direct imaging of young ($\lesssim$300~Myr) stars has revealed a population of giant planet and brown dwarf companions at wide ($>$5~au) separations \citep[e.g.,][]{Chauvin2005a, Chauvin2005b, Itoh2005, Luhman2006, Luhman2007, Marois2008c, Thalmann2009, Lagrange2010b, Biller2010, Mugrauer2010, Lafreniere2010, Marois2010b, Carson2013, Kuzuhara2013, Rameau2013b, Bailey2014}. Several formation mechanisms have been proposed to account for the diversity of these objects: core accretion \citep{Pollack1996}, gravitational instability \citep{Boss1997}, and binary-like formation. Direct imaging surveys attempted to place first constraints on the occurrence of giant planets ($\sim$10--20\%) and/or brown dwarfs ($\sim$1--3\%) and their formation mechanisms for separations beyond $\sim$10--20~au \citep[e.g.,][]{Lafreniere2007b, Chauvin2010, Ehrenreich2010, Vigan2012b, Rameau2013a, Wahhaj2013a, Nielsen2013, Biller2013, Brandt2014, Chauvin2015}. A new generation of high-contrast imaging instruments designed to search for and characterize giant exoplanets at separations as close as the snow line ($\sim$5~au), such as the Spectro-Polarimetric High-contrast Exoplanet REsearch \citep[SPHERE,][]{Beuzit2008} and the Gemini Planet Imager \citep[GPI,][]{Macintosh2014}, have recently started operations. 

We present in this paper SPHERE first light observations of the young systems PZ~Tel \citep{Biller2010, Mugrauer2010} and HD~1160 \citep{Nielsen2012}. This includes near-infrared images and integral field spectroscopy at low resolution ($R$\,$\sim$\,30) for the brown dwarf companions PZ~Tel~B and HD~1160~B, as well as near-infrared long-slit spectroscopy at low and medium resolutions ($R$\,$\sim$\,50 and 350) for PZ~Tel~B.

PZ~Tel is a G9IV star \citep{Torres2006} member of the young \citep[21\,$\pm$\,4~Myr,][]{Binks2014} stellar association $\beta$~Pictoris \citep{Zuckerman2001, Torres2008} and located at a distance $d$\,=\,51.5\,$\pm$\,2.6~pc \citep{VanLeeuwen2007}. The host star was first known to be single \citep{Balona1987, Innis1988} and to show activity-related to starspots with a rotational period of 0.94486~d \citep{Coates1980} and lightcurve amplitude variations up to $\Delta V$\,=\,0.22~mag \citep{Innis1990}. A late-M brown dwarf companion to PZ~Tel was discovered independently by two teams \citep{Biller2010, Mugrauer2010}. Using Gemini/NICI photometry in the $J$, $H$, and $K_s$ bands, \citet{Biller2010} estimate a spectral type of M5--M9, an effective temperature of $T_{\rm{eff}}$\,=\,2700\,$\pm$\,84~K, and a surface gravity of log($g$)\,=\,4.20\,$\pm$\,0.11~dex. \citet{Mugrauer2010} derive photometry in the same bands from VLT/NaCo data and find a spectral type and an effective temperature consistent with the results of \citet{Biller2010}: M6--M8 and 2500--2700~K, respectively. Astrometric follow-up of the companion shows significant orbital motion (separation increase of $\sim$35~mas/yr) and suggests that its orbit is highly eccentric \citep[$>$0.6,][]{Biller2010, Mugrauer2012, Ginski2014}. \citet{Mugrauer2012} detect a deceleration of the companion using NaCo data taken in 2007 and 2009, suggesting that the companion gets closer to its apoastron. \citet{Ginski2014} note a similar trend using 2012 NaCo data, although at low significance ($\sim$2$\sigma$), and advocate for further astrometric monitoring. Near-infrared (NIR) spectroscopic observations with VLT/SINFONI in the $H$ + $K$ bands at medium resolution ($R$\,=\,1500) of the host star and the companion indicate spectral types of G6.5 and M6--L0, respectively \citep{Schmidt2014}. \citet{Schmidt2014} also derive for the companion $T_{\rm{eff}}$\,=\,2500$^{+138}_{-115}$~K, log($g$)\,=\,3.5$^{+0.51}_{-0.30}$~dex, a metallicity enhancement [M/H]\,=\,0.30$^{+0.00}_{-0.30}$~dex, a radius $R$\,=\,2.42$^{+0.28}_{-0.34}$~$R_{\rm{J}}$ (Jupiter radii), and a range for the mass $M$\,=\,3.2--24.4~$M_{\rm{J}}$ (Jupiter masses) with a most likely value of 21~$M_{\rm{J}}$. \citet{Schmidt2014} stress the need for spectra of PZ~Tel~B with better quality and wider coverage to further constrain its physical properties. Using HARPS, \citet{Lagrange2013} measure for PZ~Tel A a projected rotational velocity $v\,sin(i)$\,=\,80~km\,s$^{-1}$, but the short time baseline and the small amplitude of the radial velocity variations {prevent a trend} due to PZ~Tel B. \citet{Rebull2008} find an excess at 70~$\muup$m using \textit{Spitzer}/MIPS but not at 24~$\muup$m. \citet{Rebull2008} attribute the 70-$\muup$m excess to a low-mass ($\sim$0.3~lunar masses) and cold ($T_{\rm{eff}}$\,$\sim$\,41~K) debris disk spanning a range of separations 35--165~au. However, this hypothesis is rejected by \citet{RiviereMarichalar2014}, who find no infrared excesses with \textit{Herschel}/PACS at 70, 100, and 160~$\muup$m.

\begin{table*}[t]
\caption{Log of the SPHERE observations.}
\label{tab:obs}
\begin{center}
\begin{tabular}{l c c c c c c c c}
\hline\hline
Object & UT date & Seeing ($''$) & Mode & Bands & DIT\,(s)\,$\times$\,NDIT & DIT\,(s)\,$\times$\,NDIT & $N_{\mathrm{exp}}$ & $\Delta$PA ($^{\circ}$) \\
 & & & & & IRDIS or ZIMPOL & IFS & & \\
\hline
PZ~Tel & 2014/07/15 & 0.88--1.11 & IRDIFS & $H2H3$+$YJ$ & 20$\times$6 & 60$\times$2 & 16 & 7.6 \\
PZ~Tel & 2014/08/06 & 0.5--0.7 & IRDIS-LRS & $YJHK_s$ & 20$\times$10 & -- & 6 & -- \\
PZ~Tel & 2014/08/06 & 0.6--0.7 & IRDIS-MRS & $YJH$ & 30$\times$5 & -- & 6 & -- \\
PZ~Tel & 2014/08/08 & 0.78--1.05 & IRDIFS\_EXT & $K1K2$+$YJH$ & 12$\times$5 & 30$\times$2 & 16 & 9.0 \\
PZ~Tel & 2014/08/16 & 1.12--1.58 & ZIMPOL & $R^{\prime}I^{\prime}$ & 50$\times$12 & -- & 3 & 14.1 \\
PZ~Tel & 2014/10/11 & 1.12--1.97 & IRDIFS & $H2H3$+$YJ$ & 8$\times$1\tablefootmark{a} & 32$\times$20\tablefootmark{a} & 1 & 3.7\tablefootmark{b} \\
HD~1160 & 2014/08/13 & 0.54--0.84 & IRDIFS\_EXT & $K1K2$+$YJH$ & 4$\times$20 & 8$\times$15 & 16 & 18.4 \\
\hline
\end{tabular}
\end{center}
\tablefoot{The seeing is the value measured by the differential image motion monitor (DIMM) at 0.5~$\muup$m. DIT {(detector integration time)} refers to the single exposure time, NDIT {(Number of Detector InTegrations)} to the number of frames in a single data cube, $N_{\rm{exp}}$ to the number of data cubes, and $\Delta$PA to the amplitude of the parallactic rotation.\\ 
\tablefoottext{a}{This sequence was a test and we only used the coronagraphic images with the satellite spots for measuring the astrometry of the companion.} \\
\tablefoottext{b}{For the IFS data.}} \\
\end{table*}

HD~1160 is a moderately young (50$^{+50}_{-40}$~Myr) A0V star located at a distance $d$\,=\,108.5\,$\pm$\,5.0~pc \citep{VanLeeuwen2007}, which is not classified as a member of any of the known young stellar associations \citep{Nielsen2012}. This star hosts two low-mass companions at projected separations $\sim$85 and $\sim$560~au \citep{Nielsen2012} discovered as part of the Gemini NICI Planet-Finding Campaign \citep{Liu2010}. \citet{Nielsen2012} show that both companions are gravitationally bound using archival data from VLT/NaCo and VLT/ISAAC spanning almost a decade (2002--2011) and obtained for purposes of photometric calibration. They also present Gemini/NICI photometry in the $J$, $H$, and $K_s$ bands and Keck/NIRC2 photometry in the $L^{\prime}$ and $M_s$ bands for both companions and NIR IRTF/SpeX spectra for the farthest companion. Using the NICI photometry, these authors find that both companions have colors at odds with field M dwarfs but consistent with colors of giant M stars, although the companions cannot be giant stars based on their absolute magnitude. \citet{Nielsen2012} are not able to explain this discrepancy, but note that an error in the calibration of the NICI photometry could account for it. For HD~1160~C, the spectrum indicates a low-mass star of spectral type M3.5$\pm$0.5 and mass $M$\,=\,0.22$^{+0.03}_{-0.04}$~solar masses ($M_{\odot}$). For HD~1160~B, the photometry suggests a brown dwarf of spectral type L0$\pm$2 and mass $M$\,=\,33$^{+12}_{-9}$~$M_{\rm{J}}$. \citet{Bonnefoy2014a} argue that HD~1160~B may have formed according to the gravitational instability scenario \citep{Boss1997} in a massive disk ($\sim$20\% of the host star mass).

We describe the observations and the data reduction in Sects.~\ref{sec:observations} and \ref{sec:datareduction}. Then we present new analyses of the physical properties of PZ~Tel~A (Sect.~\ref{sec:photometrypztela}), the spectral energy distribution (SED) of the companions for both systems (Sect.~\ref{sec:sed}), the orbit of PZ~Tel~B (Sect.~\ref{sec:orbitpztelb}), as well as further constraints on putative additional companions in each system (Sect.~\ref{sec:detlims}).

\section{Observations}
\label{sec:observations}

The SPHERE planet-finder instrument installed at the VLT \citep{Beuzit2008} is a highly specialized instrument dedicated to high-contrast imaging and the spectroscopy of young giant exoplanet. It was built by a large consortium of European institutes and is based on the SAXO system \citep[Sphere Adaptive Optics for eXoplanet Observation,][]{Fusco2006, Fusco2014, Petit2014, Sauvage2014}, which includes a 41$\times$41-actuator wavefront control, pupil stabilization, differential tip tilt control and stress polished toric mirrors \citep{Hugot2012} for beam transportation to the coronagraphs and science subsystems. Several coronagraphs for stellar diffraction suppression are provided, including apodized pupil Lyot coronagraphs \citep{Soummer2005} and achromatic four-quadrant phase masks \citep{Boccaletti2008c}. The instrument is equipped with three science channels: an infrared dual-band imager and spectrograph \citep[IRDIS,][]{Dohlen2008a}, an integral field spectrometer \citep[IFS,][]{Claudi2008}, and a rapid-switching imaging polarimeter \citep[ZIMPOL,][]{Thalmann2008}. We refer to Beuzit et al., in prep. for detailed descriptions of the subsystems and observing modes.

We observed PZ~Tel and HD~1160 during four nights of the SPHERE commissioning runs (Table~\ref{tab:obs}). The observations of PZ~Tel include all the three science subsystems: IRDIS in both dual-band imaging mode \citep{Vigan2010} and long-slit spectroscopy mode \citep{Vigan2008}, IFS, and ZIMPOL in dual-band imaging mode. HD~1160 was observed with IRDIS in dual-band imaging mode and IFS in parallel. 

\subsection{Simultaneous IRDIS and IFS observations}

The observations with IRDIS in dual-band imaging mode and IFS are simultaneous (IRDIFS and IRDIFS$\_$EXT modes in Table~\ref{tab:obs}). In the IRDIFS mode, {IRDIS is operated in the filter pair $H23$ (Table~\ref{tab:photometry}}) and IFS in the bands $YJ$ (0.95--1.35~$\muup$m, $R$\,$\sim$\,54). In the IRDIFS\_EXT mode, {IRDIS is operated in the filter pair $K12$ (Table~\ref{tab:photometry}}) and IFS in the bands $YJH$ (0.95--1.65~$\muup$m, $R$\,$\sim$\,33).

PZ~Tel was observed twice in the IRDIFS mode and once in the IRDIFS\_EXT mode. The IRDIFS sequences were obtained in poor observing conditions (July 2014) or for purposes of technical tests (October 2014). The IRDIFS\_EXT sequence was acquired in better observing conditions. HD~1160 was observed in the IRDIFS\_EXT mode in good observing conditions.

The sequences are obtained using the following strategy:
\begin{itemize}
\item The star is centered on the coronagraphic mask. The coronagraph used for all the sequences is an apodized pupil Lyot coronagraph \citep{Soummer2005} with inner working angle IWA\,$\sim$\,0.09$''$, except for the July 2014 IRDIFS sequence for which IWA\,$\sim$\,0.07$''$.
\item A coronagraphic image with four crosswise faint replicas of the star is acquired for estimating the star location, based on the original concept proposed by \citet{Marois2006c} and \citet{Sivaramakrishnan2006}. These replicas are produced by applying a 2-D periodic modulation on the deformable mirror of SPHERE. This calibration is critical for accurate registration of the frames before the application of the angular differential imaging processing \citep[ADI,][]{Marois2006a} and for precise relative astrometry of the detected companions.
\item An unsaturated image of the star (point-spread function or PSF hereafter) is recorded for purposes of photometric calibration. This image is obtained by shifting the star out of the coronagraphic mask using a tip-tilt mirror. A neutral density filter located in the common path and infrastructure of SPHERE (so common to IRDIS and IFS) is inserted into the light beam (average transmission $\sim$1/100)\footnote{\label{note:ndf} The transmission curves of the neutral density filters can be found in the User Manual available at \url{www.eso.org/sci/facilities/paranal/instruments/sphere/doc.html}.}.
\item The science coronagraphic images are acquired in pupil-stabilized mode to take advantage of the ADI technique \citep{Marois2006a}. For IRDIS, a dither pattern of 4$\times$4 positions with shifts of one pixel is applied to handle the detector bad pixels. For IFS, no dither pattern is used.
\item A second set of a PSF and coronagraphic images with four satellite spots is recorded. This second set of calibrations allows us to evaluate the photometric error due to the stellar flux variations and the stability of the star centering.
\end{itemize}
After each sequence, six sky backgrounds are acquired, with the same exposure times used for the coronagraphic images and the PSF. All the other calibration data used in the data reduction (Sect.~\ref{sec:datareduction}) are obtained the following day.

For calibrating the distortion, plate scale, and orientation of the IRDIS images, a field in the outer regions of the 47 Tuc globular cluster was observed during each of the observing runs with the same instrument setup (filter and coronagraph), except for the $K12$ filter pair. For the latter configuration, {we used the calibration} in the $H23$ filter pair. The 47~Tuc field is selected because it has a bright star for adaptive optics guiding and accurate \textit{Hubble Space Telescope} (HST) astrometry (A. Bellini \& J. Anderson, private comm.; data collected as part of the program GO-10775: P.I. Sarajedini). No IFS data of astrometric calibrators were obtained during the commissioning runs.

\begin{figure*}
\centering
\includegraphics[trim = 0mm 4.6mm 0mm 0mm, clip, height=0.28\textheight]{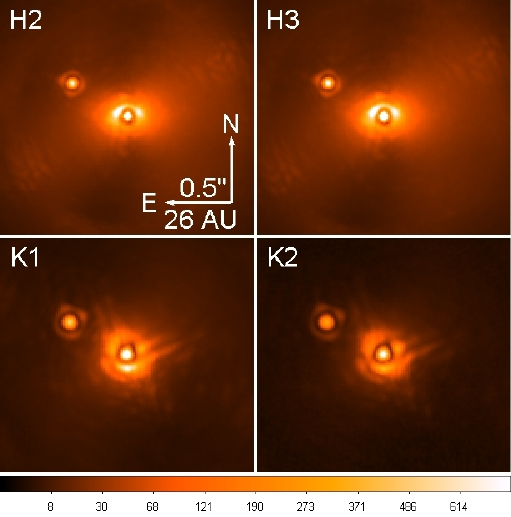}
\includegraphics[trim = 80mm 0mm 10mm 8mm, clip, height=0.28\textheight]{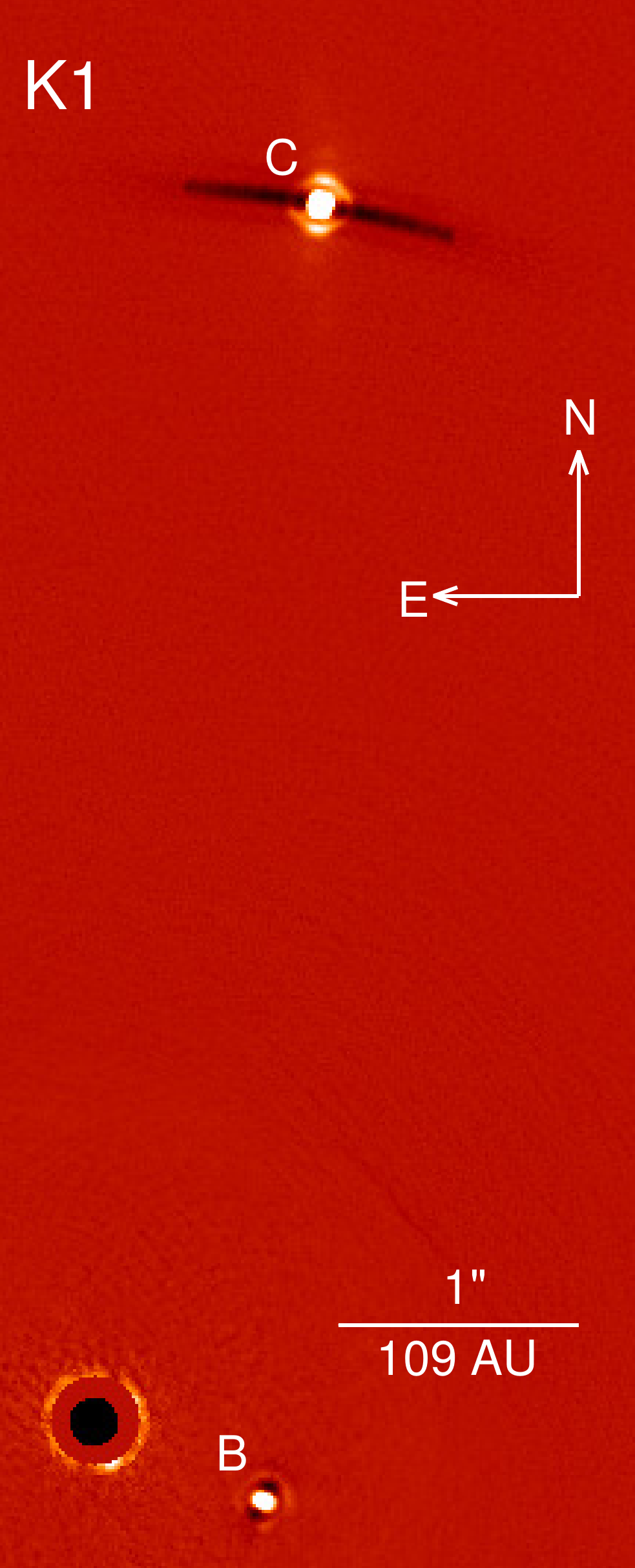}
\includegraphics[trim = 80mm 0mm 10mm 8mm, clip, height=0.28\textheight]{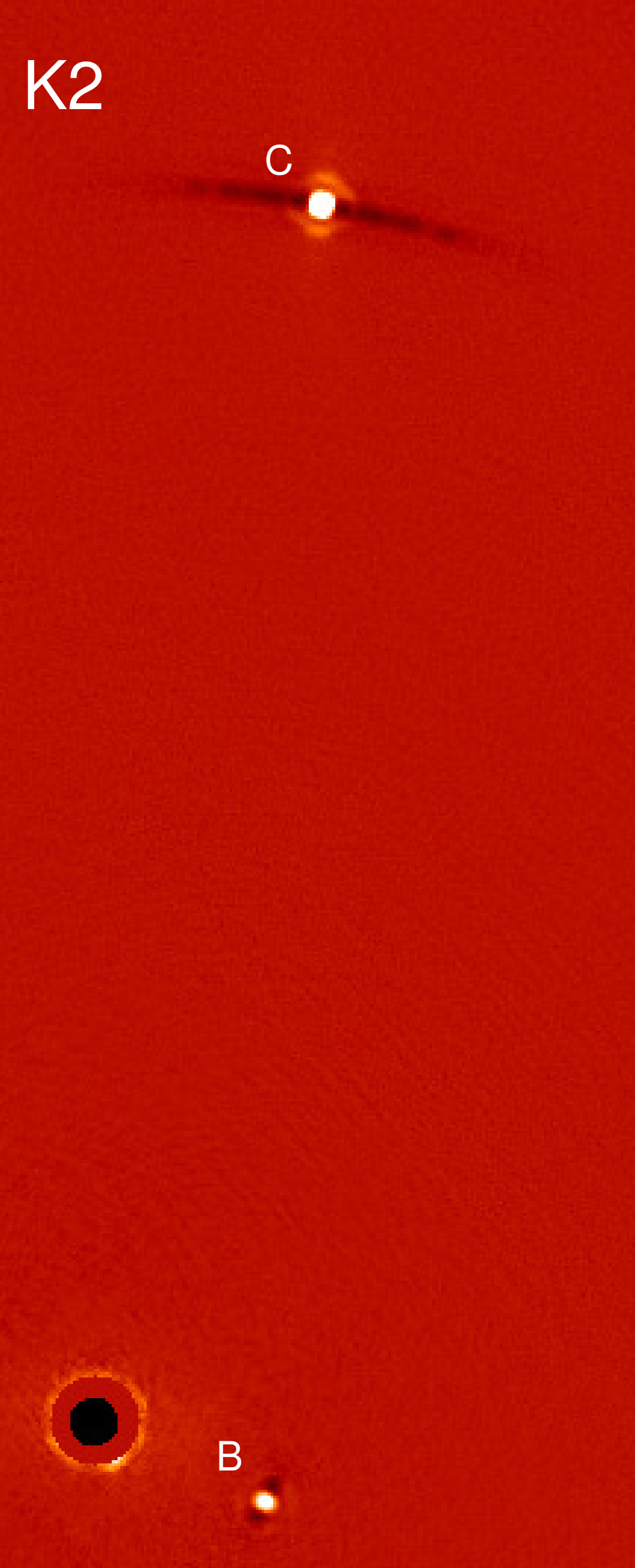} 
\caption{IRDIS images of PZ~Tel~B in the $H23$ and $K12$ filter pairs (\textit{left}) and HD~1160~BC in the $K12$ filter pair (\textit{right}). The intensity scale is the same for a given target and is square root.
} 
\label{fig:irdisimages}
\end{figure*}

\subsection{IRDIS long-slit spectroscopy}
PZ~Tel~B was observed with the long-slit spectroscopy (LSS) mode of IRDIS with two setups (Table~\ref{tab:obs}). The LSS mode includes a low-resolution mode (LRS), which covers the $YJHK_{s}$ bands at resolutions of 35--50 in one shot, and a medium-resolution mode (MRS), which covers the $YJH$ bands at a resolution of $\sim$350 \citep{Vigan2012a}. The LSS observations are always performed in field-stabilized mode to keep the object within the slit during the complete integration. {A sequence consists of images with the star behind the coronagraph}, an off-axis reference PSF, in which the star is shifted out of the coronagraph within the slit, and finally a series of sky backgrounds. The sky backgrounds are particularly important for the LRS mode that covers the $K_{s}$ band, for which the thermal background of the instrument and sky are much higher. The sequences do not differ between the LRS and MRS modes except for the change of the dispersive element. For the observations reported in this paper, we used the configuration Wide slit and Large mask (WL), where the width of the slit is 120~mas and the radius of the coronagraphic mask is 0.2$''$. No spectrophotometric calibrator was observed after the sequences. Other calibrations (darks, flat fields) were acquired during the morning following the observations.

\subsection{ZIMPOL imaging}
PZ~Tel~B was observed simultaneously in the $R^{\prime}$ and $I^{\prime}$ broad-band filters with the two camera channels of ZIMPOL (Table~\ref{tab:obs}). {To implement ADI \citep{Marois2006a}, the sequence of science observations was obtained in pupil-stabilized mode. A classical Lyot coronagraph \citep{Lyot1932} of diameter 155~mas and the satellite spots were also used.} The sequence was started about 30~min after the meridian passage of PZ~Tel. For photometric calibration, the unsaturated PSF of the primary star was acquired once after the science sequence. For this purpose, a tip-tilt mirror was used to move the star away from the coronagraph. No neutral density filter was inserted into the light beam. To avoid saturation, the exposure time for the PSF was set to 2~s. A total of twenty subintegrations in one data cube were taken in this case. The companion is well detected in the raw (without any preprocessing) individual $I^{\prime}$-band science images and in the preprocessed $I^{\prime}$-band PSF (after cosmetic correction, frame registering, and median combination). For data reduction purposes, special domeflats with the coronagraph in place were taken right after the observations. In addition, bias frames were taken during the following day. 

\subsection{SINFONI integral field spectroscopy}
\label{subsec:SINFOobs}
For comparison purposes, we compare the SPHERE IFS and LSS spectra of PZ~Tel~B to archival SINFONI data. SINFONI \citep{2003SPIE.4841.1548E, 2004Msngr.117...17B} was operated with the adaptive optics system of the instrument locked on the star. The system was observed on September 5, 9, and 13, 2011 with the $J$-band and $H+K$ band grating and the preoptics leading to a 25$\times$50~mas sampling of the 0.8$''\times$\,0.8$''$ field of view (ESO program 087.C-0535, {P.I. Tecza}). Additional observations were obtained in the $H+K$ band on September 8, 2011. {We also re-analyzed} the observations made on August 22, 2011 (ESO program 087.C-0109, P.I. Mugrauer) with the $H+K$ band grating and the preoptics leading to 25$\times$50~mas spaxels and presented by \cite{Schmidt2014}. {With respect to the spectra presented in \cite{Schmidt2014}, the combined SINFONI observations enabled us to extend the wavelength coverage to the $J$ band and to improve the signal-to-noise ratio (S/N) in the $H$ band from 5 to 10 (no S/N improvement in the $K$ band, S/N\,=\,11)}.

\section{Data reduction and analysis}
\label{sec:datareduction}

\subsection{IRDIS imaging}
\label{sec:irdisreduction}

All the data reported in Table~\ref{tab:obs} were preprocessed in the same way: subtraction of the sky background, correction of the flat field, processing of the bad pixels, and registration of the frames using the coronagraphic images with the satellite spots.
Then the data were processed using derotation and median-stacking of the temporal
frames in each filter separately in the case of the PZ~Tel data sets (Fig.~\ref{fig:irdisimages},
\textit{left}) and differential imaging for the HD~1160 data set. {For the
PZ Tel data sets, we checked that the derived photometric values agree with
the photometry extracted using the TLOCI differential imaging algorithm (see
below).}

\begin{table*}[t]
\caption{Photometric measurements of PZ~Tel~B and HD~1160~BC relative to the host star.}
\label{tab:photometry}
\begin{center}
\begin{tabular}{l c c c c c}
\hline\hline
Filter & $\lambda$ ($\muup$m) & $\Delta \lambda$\tablefootmark{a} ($\muup$m) & PZ~Tel~B & HD~1160~B & HD~1160~C \\
\hline
\multicolumn{6}{c}{SPHERE} \\
\hline
$R^{\prime}$ & 0.6263 & 0.1486 &9.76$^{+0.22}_{-0.30}$ & -- & -- \\
$I^{\prime}$ & 0.7897 & 0.1525 & 7.529\,$\pm$\,0.108 & -- & -- \\
$H2$ & 1.5888 & 0.0531 & 5.34\,$\pm$\,0.18 & -- & -- \\
$H3$ & 1.6671 & 0.0556 & 5.28\,$\pm$\,0.18 & -- & -- \\
$K1$ & 2.1025 & 0.1020 & 5.29\,$\pm$\,0.08 & 7.03\,$\pm$\,0.05 & 5.35\,$\pm$\,0.06 \\
$K2$ & 2.2550 & 0.1090 & 5.02\,$\pm$\,0.09 & 6.77\,$\pm$\,0.04 & 5.16\,$\pm$\,0.04 \\
\hline
\multicolumn{6}{c}{Other instruments} \\
\hline
$L^{\prime}$ & 3.8000 & 0.6200 & 5.15\,$\pm$\,0.15\tablefootmark{b} & 6.54\,$\pm$\,0.10\tablefootmark{c} & 4.69\,$\pm$\,0.05\tablefootmark{c} \\
\hline
\end{tabular}
\end{center}
\tablefoot{\tablefoottext{a}{Full width at half maximum.} \tablefoottext{b}{NaCo; {Beust et al., in prep.}} \tablefoottext{c}{NaCo; this work.}}
\end{table*}

\begin{table*}[t]
\caption{Astrometric measurements of PZ~Tel~B and HD~1160~BC relative to the host star.}
\label{tab:astrometry}
\begin{center}
\begin{tabular}{l c c c c c}
\hline\hline
Object & Instrument & Epoch & Band & Separation (mas) & Parallactic angle ($^{\circ}$) \\
\hline
PZ~Tel~B & IFS & 2014.53 & $YJ$ & 478.22\,$\pm$\,0.70 & 59.71\,$\pm$\,0.19 \\
PZ~Tel~B & IFS & 2014.60 & $YJH$ & 479.53\,$\pm$\,0.69 & 59.62\,$\pm$\,0.14 \\
PZ~Tel~B & IFS & 2014.78 & $YJ$ & 482.60\,$\pm$\,0.93 & 59.44\,$\pm$\,0.15 \\
PZ~Tel~B & IRDIS & 2014.53 & $H2$ & 478.48\,$\pm$\,2.10 & 59.58\,$\pm$\,0.48 \\
PZ~Tel~B & IRDIS & 2014.53 & $H3$ & 476.44\,$\pm$\,2.09 & 60.06\,$\pm$\,0.49 \\
PZ~Tel~B & IRDIS & 2014.60 & $K1$ & 479.69\,$\pm$\,0.34 & 59.71\,$\pm$\,0.47 \\
PZ~Tel~B & IRDIS & 2014.60 & $K2$ & 479.61\,$\pm$\,0.34 & 60.17\,$\pm$\,0.47 \\
PZ~Tel~B & IRDIS & 2014.78 & $H2$ & 483.87\,$\pm$\,0.34 & 59.49\,$\pm$\,0.16 \\
PZ~Tel~B & IRDIS & 2014.78 & $H3$ & 483.87\,$\pm$\,0.29 & 59.51\,$\pm$\,0.16 \\
HD~1160~B & IFS & 2014.62 & $YJH$ & 780.87\,$\pm$\,1.06 & 244.25\,$\pm$\,0.13 \\
HD~1160~B & IRDIS & 2014.62 & $K1$ & 780.97\,$\pm$\,0.47 & 243.89\,$\pm$\,0.21 \\
HD~1160~C & IRDIS & 2014.62 & $K1$ & 5149.75\,$\pm$\,2.69 & 349.17\,$\pm$\,0.10 \\
\hline
\end{tabular}
\end{center}
\tablefoot{The IFS and IRDIS measurements were obtained from the coronagraphic images with the satellite spots (Sect.~\ref{sec:observations}).} \\\end{table*}

In the second case, two independent algorithms were considered for calibrating the stellar residuals. The main algorithm was an upgrade of the Template Locally Optimized Combination of Images algorithm \citep[TLOCI,][]{Marois2014}. A second algorithm, which was used for checks for the photometry of the companions and provided consistent results within the error bars with respect to the TLOCI pipeline, consisted in derotating and median-combining the frames followed by a spatial filtering in boxes of size 5~$\lambda/D$\,$\times$\,5~$\lambda/D$. For the TLOCI algorithm, we considered the dual-band data set in each filter separately and performed ADI \citep{Marois2006a} in order to avoid ambiguities in the photometric calibration \citep{Maire2014}. The data were also binned temporally to reduce the number of frames, hence the computing time. The TLOCI pipeline selected the 80 most correlated frames for which the total self-subtraction, estimated using the measured PSF, was at maximum 70\% \citep{Marois2014}. Then it found the best linear combination for subtracting the speckles in annuli of width $\sim$1~full width at half maximum (FWHM). Finally, we derotated the frames to align north up and median-combined them (Fig.~\ref{fig:irdisimages}, \textit{right}).

The relative photometry of HD~1160~BC was derived using the method of the ``negative synthetic companions'' \citep{Marois2010a, Bonnefoy2011}. We subtracted a synthetic companion in the preprocessed data at the measured location of the companions based on the median of the observed PSF (Sect.~\ref{sec:observations}). We then processed the data assuming the TLOCI coefficients computed on the data without the synthetic companions to account for the ADI biases. The subpixel position and the flux of the modeled images were adjusted to optimize the subtraction of the model to the real image within a disk of radius 1.5~FWHM centered on the real image. For the photometry, the 1$\sigma$ error bar of the fitting was the excursion that increased the residuals in the 1.5-FWHM area by a factor of $\sqrt{2}$. This factor was empirically determined in \citet{Galicher2011a}. The error bars include the variations in the PSF, the variations in the stellar flux during the sequence (estimated from the intensity fluctuations of the stellar residuals), and the fitting accuracy of the model companion images to their measured images. For the relative photometry of PZ~Tel~B, the same sources of error are included in the error bars, except for the fitting error.

The astrometry of the companions was measured in the coronagraphic images with the satellite spots. These images were corrected for the distortion of the telescope and the instrument (Sect.~\ref{sec:astromcalib}). The astrometric calibration is described in Sect.~\ref{sec:astromcalib}.

We report the measured relative photometry and astrometry in Tables~\ref{tab:photometry} and \ref{tab:astrometry}. For the photometry of PZ~Tel~B in the $H23$ filter pair, we only considered the July 2014 data set, because the October 2014 observation was a short technical test so the data quality was poorer. For the updated orbital analysis of PZ~Tel~B (Sect.~\ref{sec:orbitpztelb}), we only considered the IRDIS $H2$ measurement obtained in July 2014.

The S/N map was estimated using the TLOCI pipeline. Each pixel value was divided by the standard deviation of the flux inside the annulus of same angular separation with a 1 FWHM width. The algorithm throughput was {accounted for by injecting} synthetic companions in the preprocessed data at regular separations between 0.15$''$ and 6$''$ and performing the TLOCI analysis. The process was repeated several times using {different position angles} in order to average the effects of random speckle residuals in the estimation of the algorithm throughput. The S/N maps were used to derive the detection curves (Sect.~\ref{sec:detlims}).

\subsection{IRDIS LSS}
\label{subsec:LSSanalysis}

The LRS and MRS data of PZ~Tel were reduced in a similar way. {The reduction and analysis procedures are very similar to those presented in \citet{Hinkley2015a}. A standard set of calibration data} (dark, flat field, bad pixel maps) was produced using the preliminary release (v0.14.0--2) of the Data Reduction and Handling software \citep[DRH,][]{Pavlov2008}, the official pipeline of the SPHERE instrument. Then the science frames were dark-subtracted and divided by the flat field. Bad pixels were replaced with the median of neighboring good pixels. The remaining bad pixels were filtered using a sigma-clipping algorithm. Each of the science frames were combined using a median, producing a 2-D spectrum from which the companion signal was extracted.

For the subtraction of the stellar halo and speckles at the position of the companion, we compared two approaches that gave similar results. The first method was based on spectral differential imaging \citep{Racine1999, Sparks2002} adapted for LSS data, as presented in \citet{Vigan2008} and demonstrated on-sky in \citet{Vigan2012a}. The second technique was much simpler and {consisted of subtracting} the symmetric part of the halo at the position of the companion with respect to the star. Given that the level of the halo and speckles was a factor $\sim$18 fainter than the peak signal of the companion, this approach gave good results and introduced a negligible bias after the subtraction.

The 1-D spectrum of the companion and the off-axis primary were extracted in the following way: for each spectral channel, the signal was summed in an aperture centered on the object position with a width of $\epsilon~\lambda/D$. When $\epsilon$ was varied from 0.5 to 1.0, there were only very small variations in the output spectrum. The local noise was estimated by summing the residual speckle noise in an aperture at the same separation as the companion, but on the other side of the primary. We checked that the spectra extracted with the two subtraction schemes and the different aperture sizes lay within these error bars.

The flux spectrum of the companion was converted into a contrast spectrum by dividing by the flux spectrum of the primary extracted in an aperture of the same width. To properly measure the contrast, we took the transmission of the neutral density filter used for acquiring the off-axis PSF data into account. The neutral density filter used for the LRS and MRS observations had an average transmission of 1/3160 and 1/100, respectively (see Note~\ref{note:ndf}).

The comparison of the LSS spectra of PZ Tel B to those of late-M objects (Sect.~\ref{subsubsec:empstud}) and those obtained with the SINFONI instrument (Sect.~\ref{subsec: SINFONIred}) revealed a residual error in wavelength dispersion relation of the SPHERE spectra. For the MRS spectrum, we fitted a new dispersion relation matching ten telluric features with those found into a synthetic spectrum of the Earth atmosphere transmission curve \citep{2012A&A...543A..92N, 2013A&A...560A..91J}. For the LRS spectrum, we applied a linear correction factor to the wavelengths derived by the DRH pipeline. The factors were found by comparing the PZ Tel B LRS spectrum to the SINFONI spectrum and to the spectrum of the M8 standard VB10 \citep{1961AJ.....66..528V, 2004AJ....127.2856B}.

\subsection{IFS spectroscopy}
\subsubsection{SPHERE IFS}
\label{sec:ifsdataanalysis}

The raw data were preprocessed using the SPHERE DRH software \citep{Pavlov2008} up to the extraction of the calibrated data cubes. {The design of the SPHERE IFS differs in two main points from the IFS of the high-contrast exoplanet imagers P1640 and GPI. First, the cross talk between adjacent pixels is minimized \citep{Antichi2009}. Then, the spectra are aligned with the detector columns \citep[see Fig. 2 in][]{Mesa2015}. These differences allow for a simpler reduction procedure for the SPHERE IFS data, which does not rely on determining the instrument PSF \citep{Zimmerman2011, Perrin2014}.}

The raw data were first corrected for the dark and the detector flat. Then the positions of the spectra were defined from an image in which the whole integral field unit (IFU) was uniformly illuminated with a white calibration lamp. Each pixel of the detector was assigned a first-guess wavelength. In a second step, the wavelength was further refined based on the illumination of the IFU with three (resp. four) monochromatic lasers with known wavelength for the $YJ$ (resp. $YJH$) data. Finally, the data cubes (science, PSF, and satellite spots) were extracted and corrected for the variations in the response of the IFU lenslets. Each data cube had 39 monochromatic images. Right before the extraction of the spectral data cubes, an additional step exploiting custom IDL routines \citep{Mesa2015} was performed for improving the correction of the bad pixels and spectral cross-talk.

\begin{figure}
\centering
\includegraphics[width=0.24\textwidth]{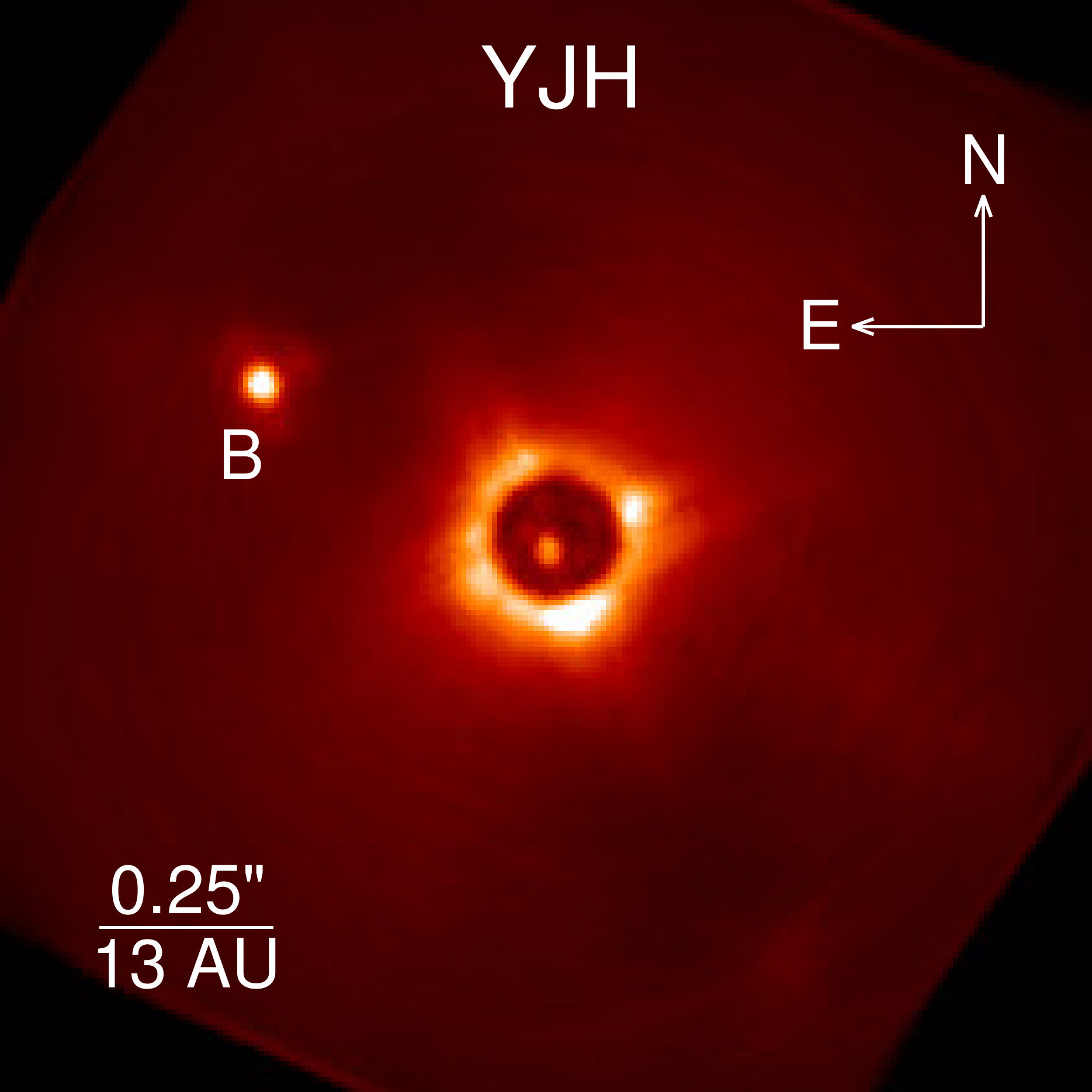} 
\includegraphics[width=0.24\textwidth]{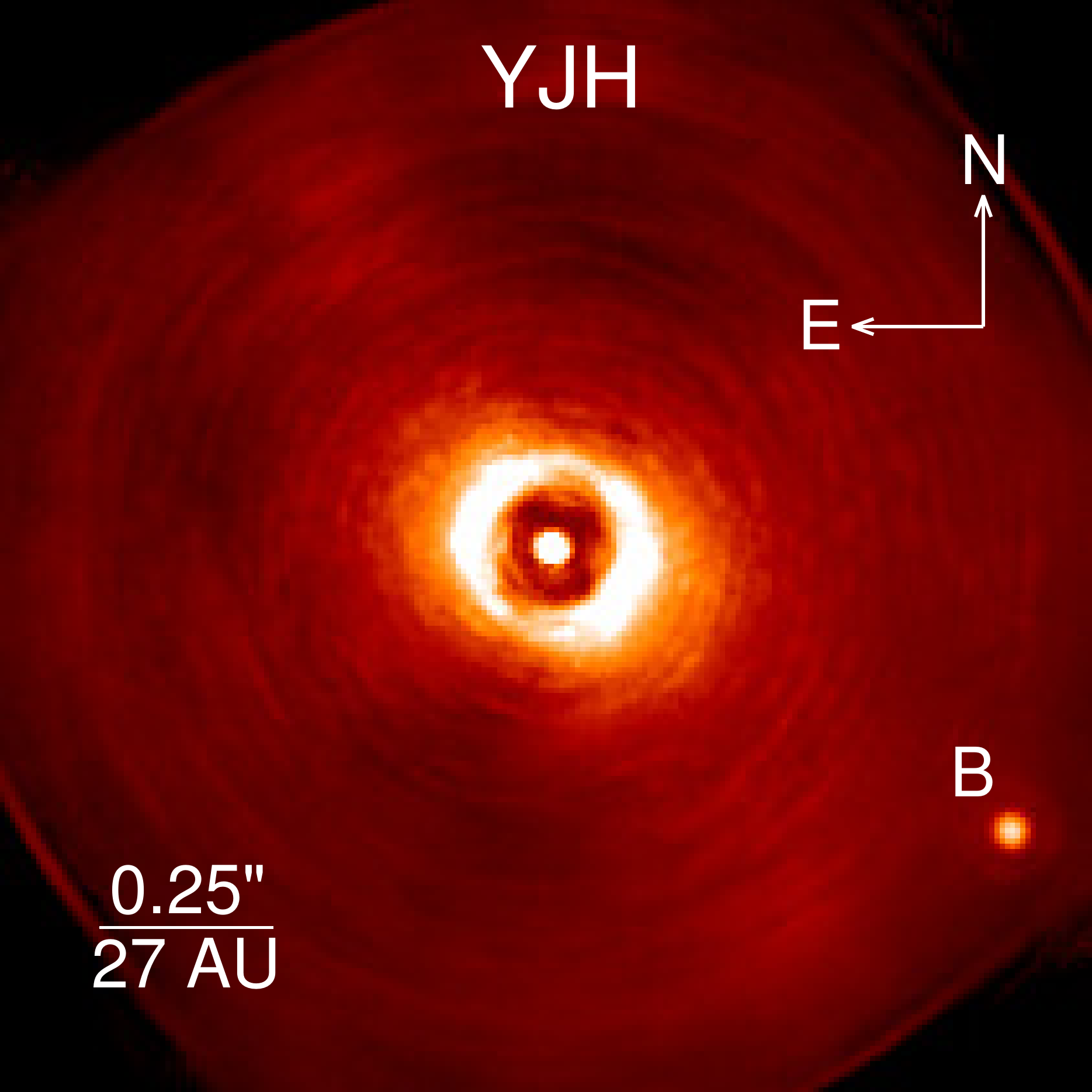} 
\caption{{IFS median-collapsed images of PZ~Tel~B (\textit{left}) and HD~1160~B (\textit{right}) after derotation and stack of the temporal frames. The square root intensity scales have different ranges.}} 
\label{Fig:ifsimages}
\end{figure}

We used the coronagraphic images with the satellite spots (Sect.~\ref{sec:observations}) of the data sets listed in Table~\ref{tab:obs} to obtain the astrometry of PZ~Tel~B and HD~1160~B (Table~\ref{tab:astrometry}).

For extracting the spectrum of the companions, we considered each spectral channel separately and performed derotation and median-stacking of the frames (Fig.~\ref{Fig:ifsimages}). {For the purpose of searching for additional companions in the system}, we also considered each spectral channel separately but subtracted the stellar residuals with a principal component analysis approach \citep[PCA,][]{Soummer2012, Amara2012}, as illustrated in \citet{Maire2014}.

The 1-D detection limits (Sect.~\ref{sec:detlims}) were derived as the standard deviation of the residual flux in annuli of 1~$\lambda/D$ width at each angular separation. We estimated the PCA subtraction of off-axis point sources using synthetic companions injected into the preprocessed data at several separations and position angles. We optimized the number of PCA modes to maximize the detection performance.

\begin{figure*}
\centering
\includegraphics[scale=0.6]{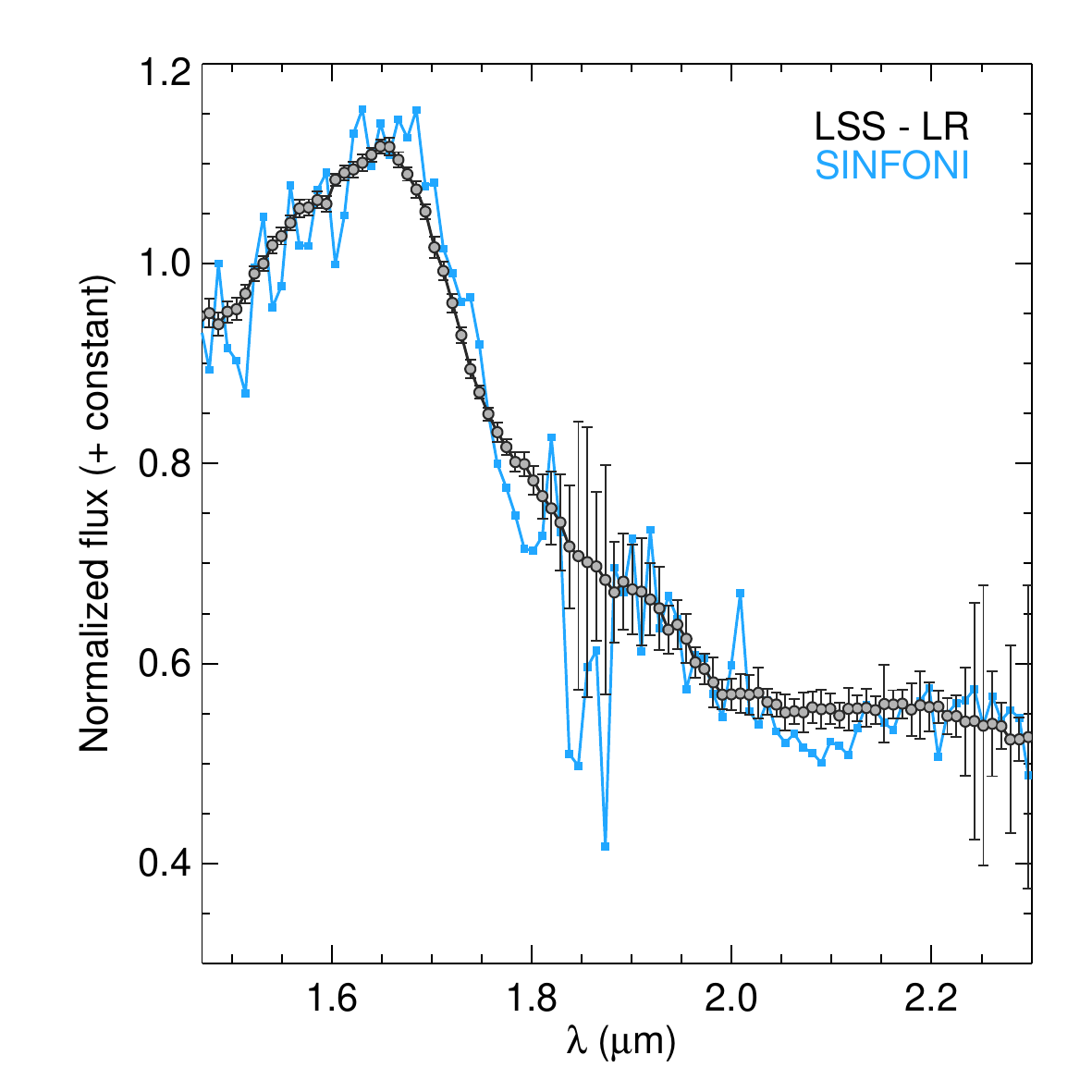} 
\includegraphics[scale=0.6]{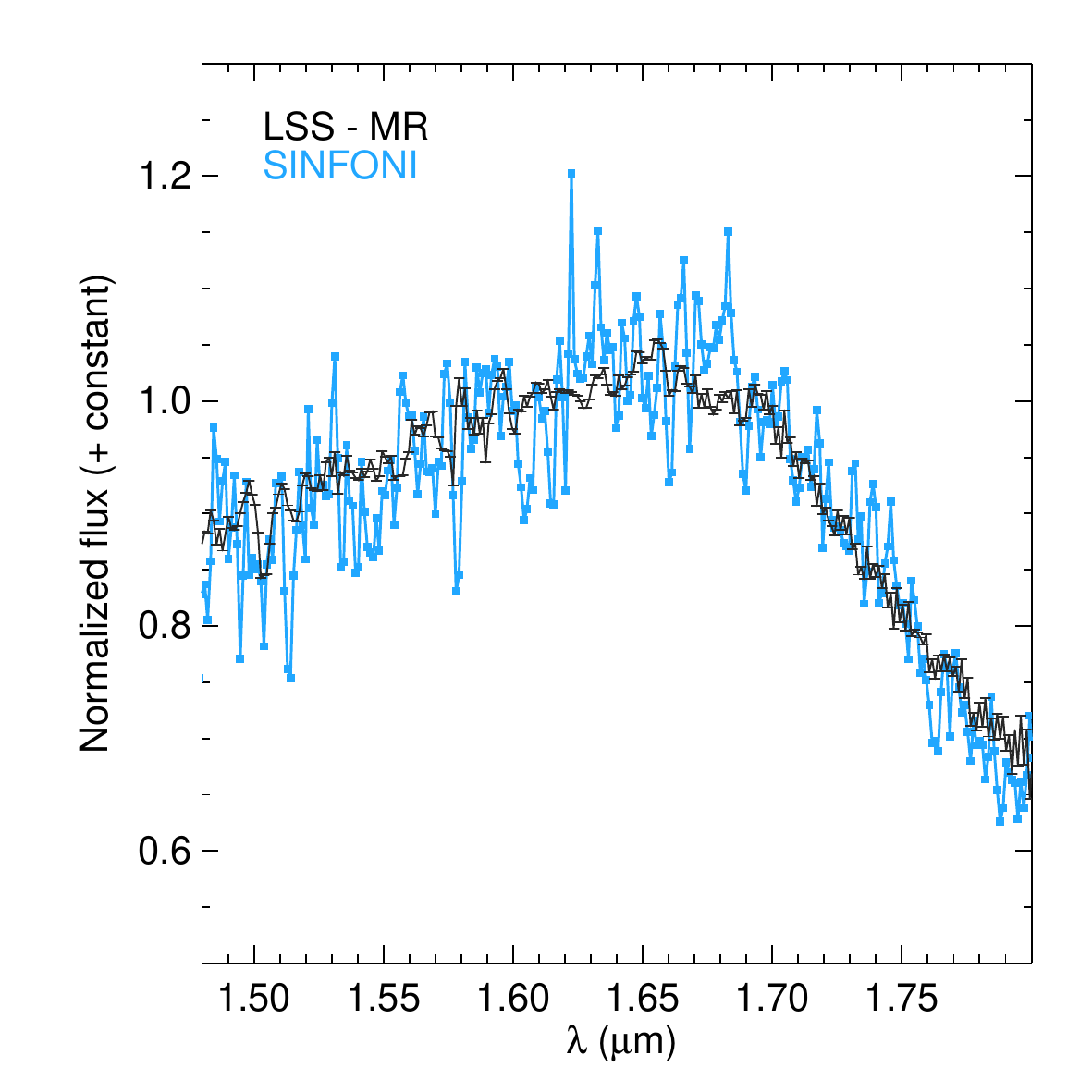} 
\caption{Comparison of the SINFONI $H$+$K$ band spectrum and the SPHERE long-slit spectra of PZ~Tel~B at comparable resolutions.} 
\label{Fig:FigcompSINF}
\end{figure*}

\begin{figure*}
\centering
\includegraphics[scale=0.6]{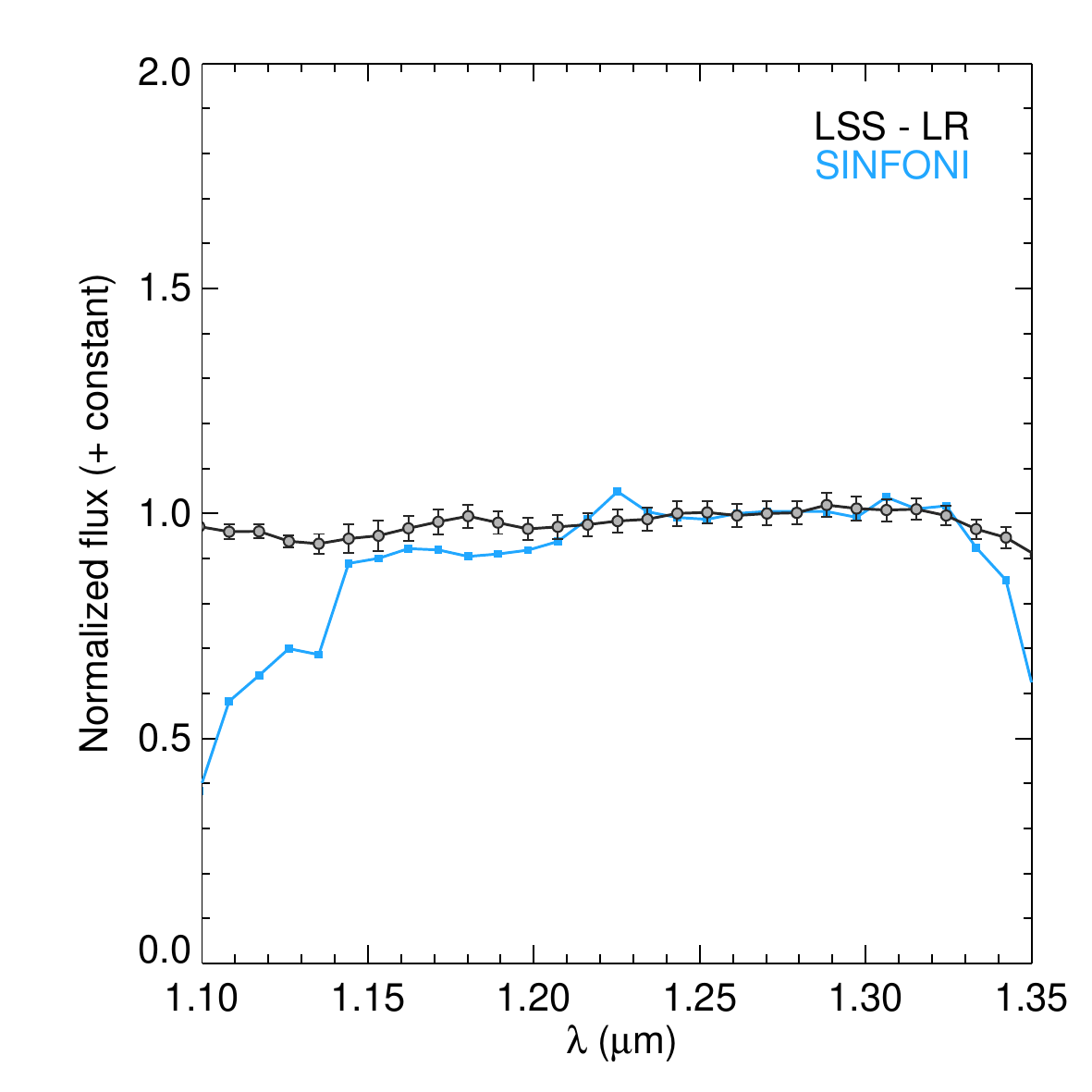} 
\includegraphics[scale=0.6]{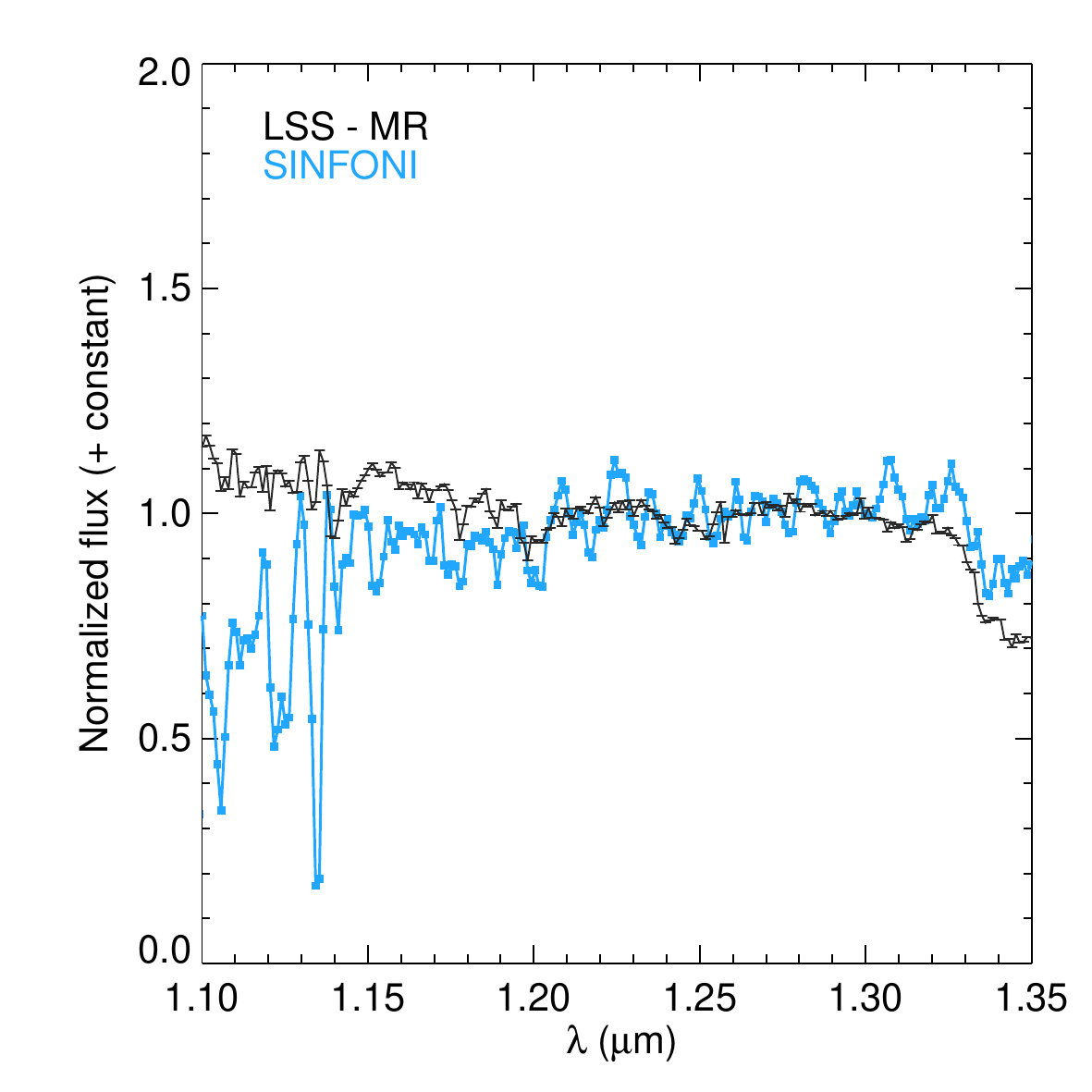} 
\caption{Comparison of the SINFONI and the SPHERE long-slit spectra of PZ~Tel~B in the $J$ band.} 
\label{Fig:FigcompSINFjband}
\end{figure*}

\subsubsection{SINFONI spectra of PZ~Tel B}
\label{subsec: SINFONIred}

We reduced the data corresponding to the SINFONI observations of PZ Tel B (Sect.~\ref{subsec:SINFOobs}) with the ESO data reduction pipeline \citep{2007astro.ph..1297M} version 2.5.2. The pipeline built the bad pixel mask and flat field associated to each observation date and setup. {It corrected each raw frame for the instrument distortion}, derived a map of wavelength associated to each detector pixel, and found the {position of the spectra (slitlets)} on the detector. After these steps, we obtained the final data cube from the raw 2D science frames. {For the August 2011 data}, the sky emission was evaluated and removed using exposures of an empty field obtained during the observation sequence. No sky exposure was obtained for the remaining observations. Therefore, we instead subtracted a dark frame {from the raw science images} in order to remove the bias and the detector ramp effect, and applied the algorithms of \cite{2007MNRAS.375.1099D} implemented into the pipeline to remove the sky contribution (thermal background and emission lines).

PZ Tel A was located inside the field of view of the instrument for the September 2011 observations. This was not the case for the August 2011 observations \citep{Schmidt2014}, for which the PSF core of PZ Tel A was located a few pixels outside the field of view. We removed the stellar halo using a radial profile for both cases. {For the August 2011 data}, we chose the approximate star position that minimized the halo residuals. The companion was always masked (circular mask of radius 7~pixels) while computing the radial profile.

The flux of PZ Tel B was measured in the halo-removed data cubes using aperture photometry (circular aperture of radius 5 or 6 pixels). The aperture radius was determined by eye with respect to the encircled energy of a given observation and the level of residuals from the halo subtraction close to the companion. All extracted spectra {were corrected for telluric absorptions} using those of B-type standard stars observed after PZ Tel ({hereafter Method 1}). {For the case of the data acquired on Sept. 11, 2014}, we also used the spectrum of PZ Tel A extracted with circular apertures of similar size as an alternative correction ({hereafter Method 2}). This enabled the estimation of a companion-to-star contrast ratio at each wavelength that can be compared directly to the SPHERE spectra and that was less sensitive to differential flux losses due to the limited size of the apertures. We converted the contrast obtained following {Method 2} to a flux-calibrated spectrum following the method described in Sect.~\ref{subsubsec:confluxes}.

Both methods give results that agree within error bars estimated from the standard deviation of the flux at each wavelength obtained from the different epochs. The $HK$-band spectrum obtained with {Method 1} benefits from the increased exposure time from all combined data sets. It looks identical to the spectrum obtained by \cite{Schmidt2014} except for the 1.45--1.6 $\muup$m range, where it has less pronounced H$_{2}$O absorptions. It nevertheless reproduces all the features of the low-resolution LSS spectra of the source obtained with SPHERE well (Fig.~\ref{Fig:FigcompSINF}).

The $J$-band SINFONI spectrum and low-resolution LSS spectrum of PZ Tel B have an identical slope from 1.2 to 1.3~$\muup$m (Fig.~\ref{Fig:FigcompSINFjband}). The LSS spectrum is, however, not affected by strong telluric residuals shortward of 1.2~$\muup$m, which probably arise from an improper subtraction of the stellar halo in the SINFONI data at these wavelengths. The medium-resolution LSS spectrum has a bluer slope than the SINFONI spectrum. The origin of this slope is unclear. The preprocessed data were checked for saturation and contamination by the star signal. We are still analyzing the data, as well as other LSS data sets from the commissioning runs, to determine the science performance and limitations of this observing mode. The results will be presented in a forthcoming paper. The SPHERE spectra supersede the SINFONI spectrum inside this wavelength range because of their higher S/N at a comparable spectral resolution.

\begin{figure*}
\centering
\includegraphics[trim = 0mm 1mm 0mm 0mm,clip,scale=0.3]{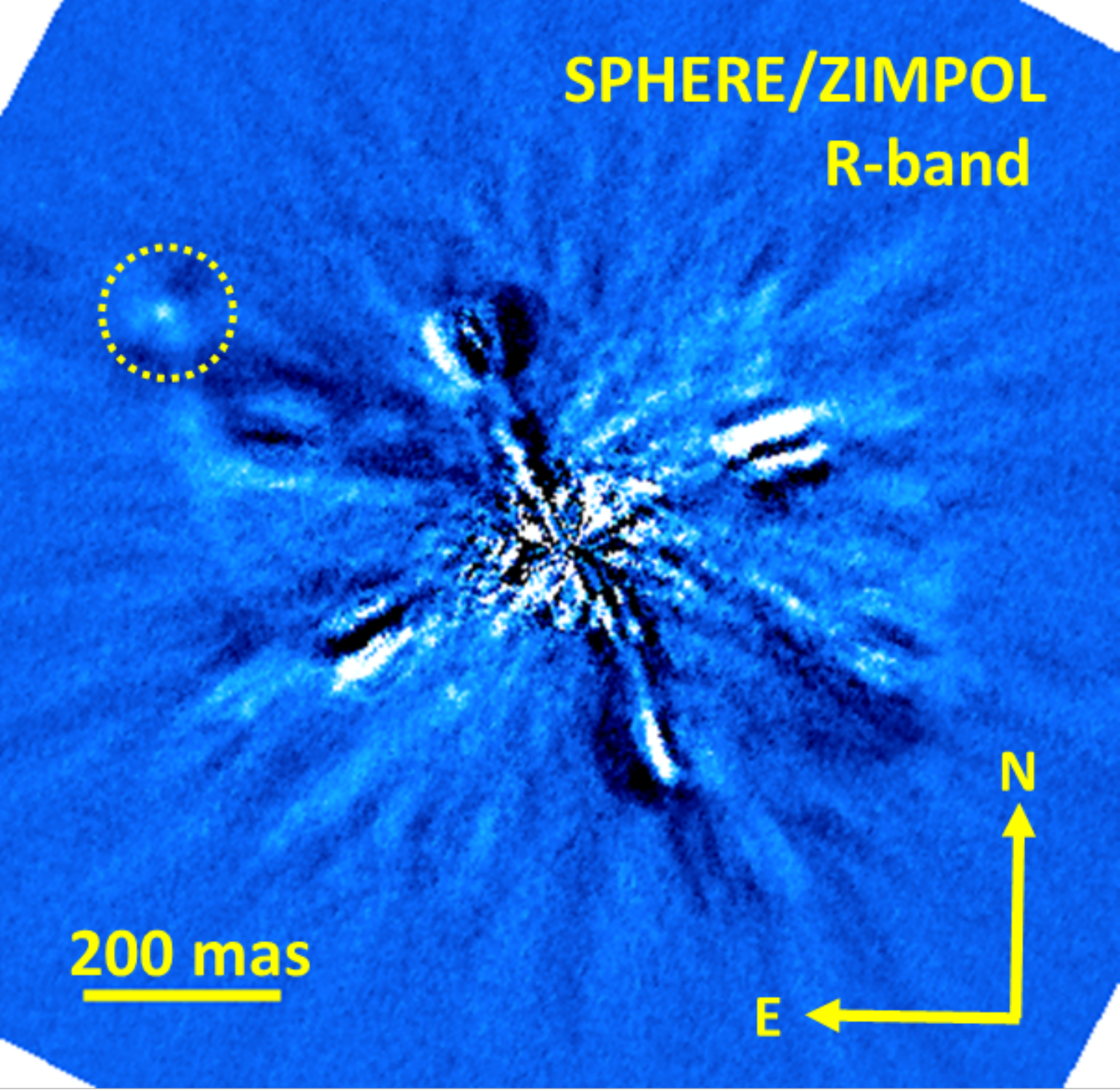}
\includegraphics[trim = 0mm 1mm 0mm 0mm,clip,scale=0.3]{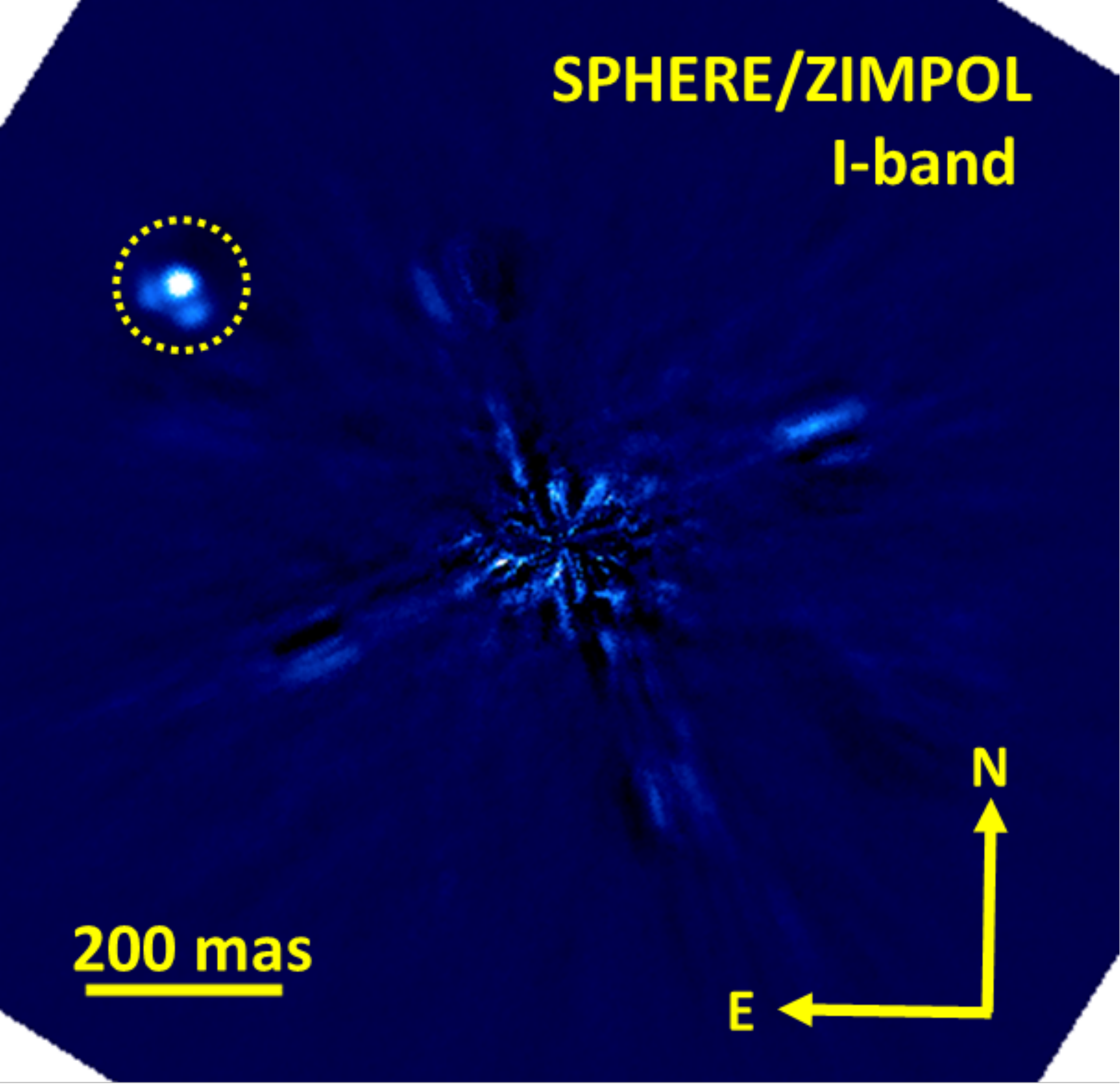}
\caption[]{ADI-reduced $I^{\prime}$-band (\textit{left}) and $R^{\prime}$-band coronagraphic ZIMPOL images of PZ~Tel~B obtained simultaneously in the imaging mode (see text). The companion is marked in both images with a dashed circle. The artifact seen around the companion signal in both images is due to an intermittent instrumental effect currently under study, possibly related to the telescope.} 
\label{zimpol-images}
\end{figure*}

\subsection{ZIMPOL imaging}

The raw data were preprocessed using the SPHERE DRH software \citep{Pavlov2008}. This included bias subtraction and flat-fielding with coronagraphic flats taken after the science sequences, as well as ZIMPOL specific tasks, such as the cropping of the image and the interpolation of pixels in the image $y$ direction in order to produce a square image. Using the timestamps in the image header, for each individual exposure the parallactic angle was computed and the images combined using a simple ADI approach \citep{Marois2006a}. A reference image of the stellar signature was generated by median combination of all preprocessed images. Owing to the field rotation during the pupil-stabilized observations, the large majority of the flux of the companion was rejected in this process. This reference image of the stellar signature was then subtracted from all individual frames, which were subsequently derotated to align north up and combined. {The true north was estimated using the IRDIS astrometry of the companion measured during the same run.} Since the coordinate system in ZIMPOL camera 1 is flipped in the $y$ direction with respect to the sky, all $R^{\prime}$-band images were first flipped in the $y$ direction before derotation and image combination. The resulting images are shown in Fig.~\ref{zimpol-images}.

We used relative aperture photometry to measure the flux of PZ~Tel~B in both bands. For all measurements, the Aperture Photometry Tool \citep{Laher2012} was utilized. The aperture size was set to a radius of five~pixels for all measurements. This was chosen to exclude the bright diffraction-related features in the companion PSF. The companion flux was measured in the ADI-reduced images, while the star flux was measured in the reference images and then rescaled to the longer observation time of the science images. Care was taken to choose regions for background subtraction that were not contaminated by the stellar halo to avoid oversubtraction. Since ADI imaging suffers from self-subtraction of the companion flux, we corrected for this effect as well. For this purpose, we used the unsaturated images of PZ~Tel~A to generate fake companion PSF, which we injected in the preprocessed data. We then measured the amount of self-subtraction of these fake planets. Care was taken to inject the fake companions at the same angular separation as the real companion PZ~Tel~B and various position angles, to ensure similar self-subtraction effects during ADI processing. The resulting photometric measurements are summarized in Table~\ref{tab:photometry}. The uncertainties include the statistical uncertainties from the aperture photometry, the uncertainties from the calibration of the ADI self-subtraction, and the temporal variations of the stellar flux during the sequence (measured using one of the simultaneous satellite spots). The self-subtraction uncertainties were estimated according to the range of correction factors calculated from several fake companions. The photometric error due to the PSF variations could not be estimated since only one PSF was recorded. The astrometry of the companion was not considered for this paper, because of the availability of IRDIS and IFS data close in time. The ZIMPOL astrometry will be discussed in a forthcoming paper.

\begin{table*}[t]
\caption{Mean plate scale and true north orientation for the SPHERE commissioning runs measured on the 47~Tuc field-stabilized observations in the IRDIS images in the $H23$ filter pair obtained with the APLC coronagraph (Sect.~\ref{sec:observations}).}
\label{tab:astrometriccalib}
\begin{center}
\begin{tabular}{l c c c c}
\hline\hline
& \multicolumn{2}{c}{$H2$} & \multicolumn{2}{c}{$H3$} \\
Com. date & Plate scale (mas/pix) & True north ($^{\circ}$) & Plate scale (mas/pix) & True north ($^{\circ}$) \\
\hline
July 2014 & 12.252\,$\pm$\,0.006 & $-$1.636\,$\pm$\,0.013 & 12.247\,$\pm$\,0.006 & $-$1.653\,$\pm$\,0.005\\
August 2014 & 12.263\,$\pm$\,0.006 & $-$1.636\,$\pm$\,0.013 & 12.258\,$\pm$\,0.006 & $-$1.653\,$\pm$\,0.005\\
October 2014 & 12.259\,$\pm$\,0.006 & $-$1.788\,$\pm$\,0.008 & 12.254\,$\pm$\,0.005 & $-$1.795\,$\pm$\,0.008\\
\hline
\end{tabular}
\end{center}
\tablefoot{To align the IRDIS science images so that north is oriented up and east left, an offset accounting for the zeropoint of the derotator in pupil-stabilized mode of 135.87$\pm$0.03$^{\circ}$ has to be added to the parallactic angles indicated in the fits header. This offset is measured on data of 47~Tuc acquired in July 2014 in both stabilization modes and assumed to be constant between the runs. For the astrometric calibration of the IFS science images, an additional offset accounting for the relative orientation between the IFS and IRDIS fields of $-$100.46\,$\pm$\,0.13$^{\circ}$ also has
to be added. This value is measured on simultaneous data of distortion grids of both instruments.}
\end{table*}

\subsection{NaCo $L^{\prime}$-band images of HD~1160}
\label{subsec: NaCoim}

The Keck/NIRC2 $L^{\prime}$-band images reported in \cite{Nielsen2012} appear to have poorer quality than the NaCo \citep[Nasmyth Adaptive Optics System and Near-Infrared Imager and Spectrograph,][]{Rousset2003, Lenzen2003} $L^{\prime}$-band images obtained at multiple epochs and available in the ESO public archive (ESO program 60.A-9026). We made use of these NaCo images to check the consistency of the Keck/NIRC2 $L^{\prime}$-band photometry. We reduced five epochs of observations (Dec. 23, 2005; Sept. 17, 2010; Jul. 11, 2011; Sept. 2, 2011; Nov. 8, 2011) with the ESO \texttt{Eclipse} software \citep{1997Msngr..87...19D}. HD~1160~B was resolved at each epoch, but the strehl and residual background close to the star appeared to have better quality in the images obtained in November 2011. We chose to perform aperture photometry (circular apertures of radius 160~mas) on the star and its companion on these images and found $\Delta L^{\prime}_{B/A}$\,=\,6.54\,$\pm$\,0.10~mag and $\Delta L^{\prime}_{C/A}$\,=\,4.69\,$\pm$\,0.05~mag for HD~1160~B and C, respectively. The values were consistent with those derived from the four other epochs within the error bars. The NaCo photometry also agrees within the error bars (3$\sigma$) with the Keck/NIRC2 photometry reported in \cite{Nielsen2012}. Nevertheless, we preferred to use the NaCo photometry below, since it enabled a direct comparison to the photometry of PZ~Tel~B. The NaCo $L^{\prime}$ photometry brings the $K_s-L^{\prime}$ band color of HD~1160~B ({$K_s-L^{\prime}$\,=\,0.53\,$\pm$\,0.12~mag}) into better agreement with those of mid- to late-M field dwarfs. Using the $L$-band magnitude of HD~1160~A reported in \cite{1996A&AS..119..547V}, we found $L^{\prime}_B$\,=\,13.60\,$\pm$\,0.10~mag and $L^{\prime}_C$\,=\,11.76\,$\pm$\,0.05~mag.

We note that, independently of our new photometry, the $L^{\prime}-M_s$ color of HD~1160~B \citep{Nielsen2012} appears to be much redder with respect to M or early-L dwarf companions \citep[Fig. 5,][]{Bonnefoy2014a}. Therefore, we decided not to use the Keck/NIRC2 photometry of the system below.

\subsection{IRDIS and IFS astrometric calibration}
\label{sec:astromcalib}

The IRDIS astrometric measurements reported in Table~\ref{tab:astrometry} were derived using the plate scale and the true north orientation measured for the 47~Tuc data (Table~\ref{tab:astrometriccalib}). We recall that no relevant observations of astrometric calibrators were obtained for IFS (Sect.~\ref{sec:observations}). For the IFS data we estimated a plate scale of 7.46\,$\pm$\,0.01~mas/pix and a relative orientation to IRDIS data of $-$100.46\,$\pm$\,0.13$^{\circ}$ from simultaneous data of distortion grids of both instruments. Using this calibration and correcting for the instrument distortion (see below), we estimated the astrometry of the companions in IFS data listed in Table~\ref{tab:astrometry}.

The 47~Tuc IRDIS data were reduced and analyzed using the SPHERE DRH software \citep{Pavlov2008} and custom IDL routines. The star positions were measured using the centroid IDL routine \texttt{cntrd}\footnote{\url{http://idlastro.gsfc.nasa.gov/ftp/pro/idlphot/}.} derived from the DAOphot software \citep{Stetson1987}. The measured positions were then compared to HST positions corrected for the differential proper motions of the individual stars between the HST observations (March 13, 2006) and the SPHERE observations (A. Bellini \& J. Anderson, private comm.; see \citet{Bellini2014} for the description of the methods used for deriving the catalog positions and the individual stellar proper motions). The typical accuracy of the catalog positions is $\sim$0.3~mas and takes the time baseline between the HST and SPHERE observations into account. Typically, more than 50 stars were cross-identified and used for the analysis. The distortion measured on-sky is dominated by an anamorphism (0.60\,$\pm$\,0.02\%) between the horizontal and the vertical directions of the detector. The on-sky distortion is similar to the distortion measured in laboratory, suggesting that the distortion from the telescope is negligible with respect to the distortion of SPHERE. The anamorphism is produced by cylindrical mirrors in the common path and infrastructure of the instrument, hence common to IRDIS and IFS, except that the anamorphism is rotated for the IFS data. The plate scale and the true north orientation given in Table~\ref{tab:astrometriccalib} were corrected for the anamorphism.

\begin{table*}
\caption{Comparison stars used for the REM photometry.}
   \begin{center}
   \begin{tabular}{@{} llccc @{}}
      \hline
      \hline
      & Name                     & RA (J2000.0)  & DEC (J2000.0) & $V$\\
            &                  &  (hh mm ss)  &  ($^\circ$ $^\prime$ $^{\prime\prime}$) & (mag)\\
      \hline
C1 & TYC\,8381-2548-1 &  18 52 55.39 & $-$50 09 13.26 & 10.487$\pm$0.016  \\

C2 & CD-50\,12193  & 18 53 28.02 & $-$50 13 08.26  & 10.173$\pm$0.015  \\
C3 & 2MASS\,J18531926-5014042 &  18 53 19.26   & $-$50 14 04.20  & 12.233$\pm$0.060  \\
\hline
   \end{tabular}
   \end{center}
   \label{comparison}
\end{table*}

\section{Updated photometry of PZ~Tel A}
\label{sec:photometrypztela}

Young, late-type stars are known to show photometric variability on several timescales, hourly due to flares, daily due to rotational modulations of active regions on the stellar surface, and on longer timescales due to, for instance, reconfiguration of active regions and activity cycles. Long-term variations of these stars were studied less intensively than short-term variations due to the long time baseline needed for the analysis.

The characterization of substellar objects like those presented in this paper are typically obtained differentially with respect to the host stars. Ideally, the magnitude of the host star should be derived simultaneously to the high-angular resolution observations, but this is difficult to achieve in practice. {Photometric monitoring in the same observing season as the high-contrast observations insures that the latter are properly calibrated.} From the analysis of literature data (Sect.~\ref{sec:literaturepztela}), PZ~Tel~A is shown to have photometric variations of up to 0.2~mag over several decades. We present below new photometric observations of the star taken in the same season as the SPHERE data and the study of its photometric variability on various timescales.

\subsection{REM observations}
\label{subsec:REMobs}

We observed PZ~Tel~A with the 60-cm Rapid Eye Mount \citep[REM,][]{Chincarini2003} telescope (La Silla, ESO, Chile) with both the NIR REMIR and the optical ROS2 cameras. The REMIR camera hosts a 512~pix\,$\times$\,512~pix CCD and has a field of view of 10$^{\prime}\times10^{\prime}$ and a plate scale of 1.2$^{\prime\prime}$/pix. The ROS2 camera hosts a 2048~pix\,$\times$\,2048~pix CCD and has a field of view of 8$^{\prime}\times$8$^{\prime}$ and a plate scale of 0.58$^{\prime\prime}$/pix. We could observe the star {from October 29, 2014 until December 5, 2014} for a total of 30 nights. After discarding frames collected in very poor seeing conditions and frames missing one or more comparison stars (due to inaccurate telescope pointing), we were left with a total of 173, 131, and 133 frames in the $J$, $H$, and $K$ filters, respectively, and 205, 166, and 219 frames in the $g$, $r$, and $i$ filters. The exposure time was fixed to 1~s in the ROS2 camera, whereas five consecutive dither frames of 1~s each were collected with the REMIR camera on each telescope pointing. All frames were first bias-subtracted and then flat-fielded. Aperture photometry was used to extract the magnitudes of PZ~Tel~A and other stars in the field to be used as comparison stars. {All reduction steps were performed} using the tasks within IRAF\footnote{IRAF is distributed by the National Optical Astronomy Observatory, which is operated by the Association of the Universities for Research in Astronomy, inc. (AURA) under cooperative agreement with the National Science Foundation.}. After removing a few outliers from the magnitude time series, using a 3$\sigma$ filtering, we averaged consecutive data collected within one hour, and finally we were left with 24, 18, and 19 averaged magnitudes in the $J$, $H$, and $K$ filters, respectively, and 38, 26, and 34 averaged magnitudes in the $g$, $r$, and $i$ filters, respectively.

% Requires the booktabs if the memoir class is not being used
\begin{table}
        \caption{Average differential magnitudes from 2MASS catalog and REM observations.}
   \centering
   %\topcaption{Table captions are better up top} % requires the topcapt package
   \begin{tabular}{@{} lrrc @{}} % Column formatting, @{} suppresses leading/trailing space

      \hline
      \hline
                         & 2MASS                       & REM                         & Filter\\
                                                 & (mag)                       & (mag)                         &  \\
          \hline  
       C1$-$C2          &  1.00$\pm$0.04          & 1.01$\pm$0.03          & $J$\\
      C1$-$C3           &$-$0.24$\pm$0.04      & $-$0.26$\pm$0.04          & $J$\\    
      C2$-$C3           &$-$1.24$\pm$0.03      & $-$1.26$\pm$0.02          & $J$\\     
      \hline
      C1$-$C2  & 1.12$\pm$0.06         &  1.17$\pm$0.03           & $H$\\      C1$-$C3& $-$0.01$\pm$0.05     &$-$0.02$\pm$0.04          & $H$\\
      C2$-$C3  & $-$1.13$\pm$0.04     & $-$1.17$\pm$0.04          & $H$\\            \hline
      C1$-$C2  & 1.28$\pm$0.03 &  1.25$\pm$0.02          & $K$\\
      C1$-$C3& 0.13$\pm$0.04      &0.11$\pm$0.03          & $K$\\
      C2$-$C3 & $-$1.14$\pm$0.03 & $-$1.12$\pm$0.02          & $K$\\
            \hline
   \end{tabular}
   \label{comp-diff}
\end{table}

The mean standard deviations associated to the average magnitudes were $\sigma_{\rm J}$\,=\,0.017, $\sigma_{\rm H}$\,=\,0.018, $\sigma_{\rm K}$\,=\,0.013~mag, and $\sigma_{\rm g}$\,=\,0.035, $\sigma_{\rm r}$\,=\,0.021, $\sigma_{\rm i}$\,=\,0.018~mag. 
Analyzing the All Sky Automated Survey \citep[ASAS,][]{Pojmanski1997} photometric timeseries of the brighter stars in the field of PZ~Tel~A,
 we identified three stars (Table\,\ref{comparison}) whose light curves
were very stable, and they were therefore suitable as comparison stars. Their differential magnitudes also during our observing run were found to 
be constant within our photometric precision in both optical and the NIR photometric bands (Table\,\ref{comp-diff}).

We compared the average differential values of the comparison stars (C1$-$C2, C1$-$C3, and C2$-$C3) with respect to the differential values derived from the Two Micron All Sky Survey \citep[2MASS,][]{2003tmc..book.....C} {listed in Table~\ref{2mass}. Table~\ref{comp-diff} lists the differential values. We note that they are all consistent with each other within the uncertainties, verifying that these stars do not vary also in the NIR}. Only the comparison C3 exhibits slightly larger magnitude variations with respect to the 2MASS values. More precisely, it remained constant during the observation run, but probably has a long-term (on the order of years) small amplitude variation.

\begin{figure*}
\begin{centering}
\includegraphics[scale=.4,angle=90,trim= 5mm 0mm 10mm 14mm,clip]{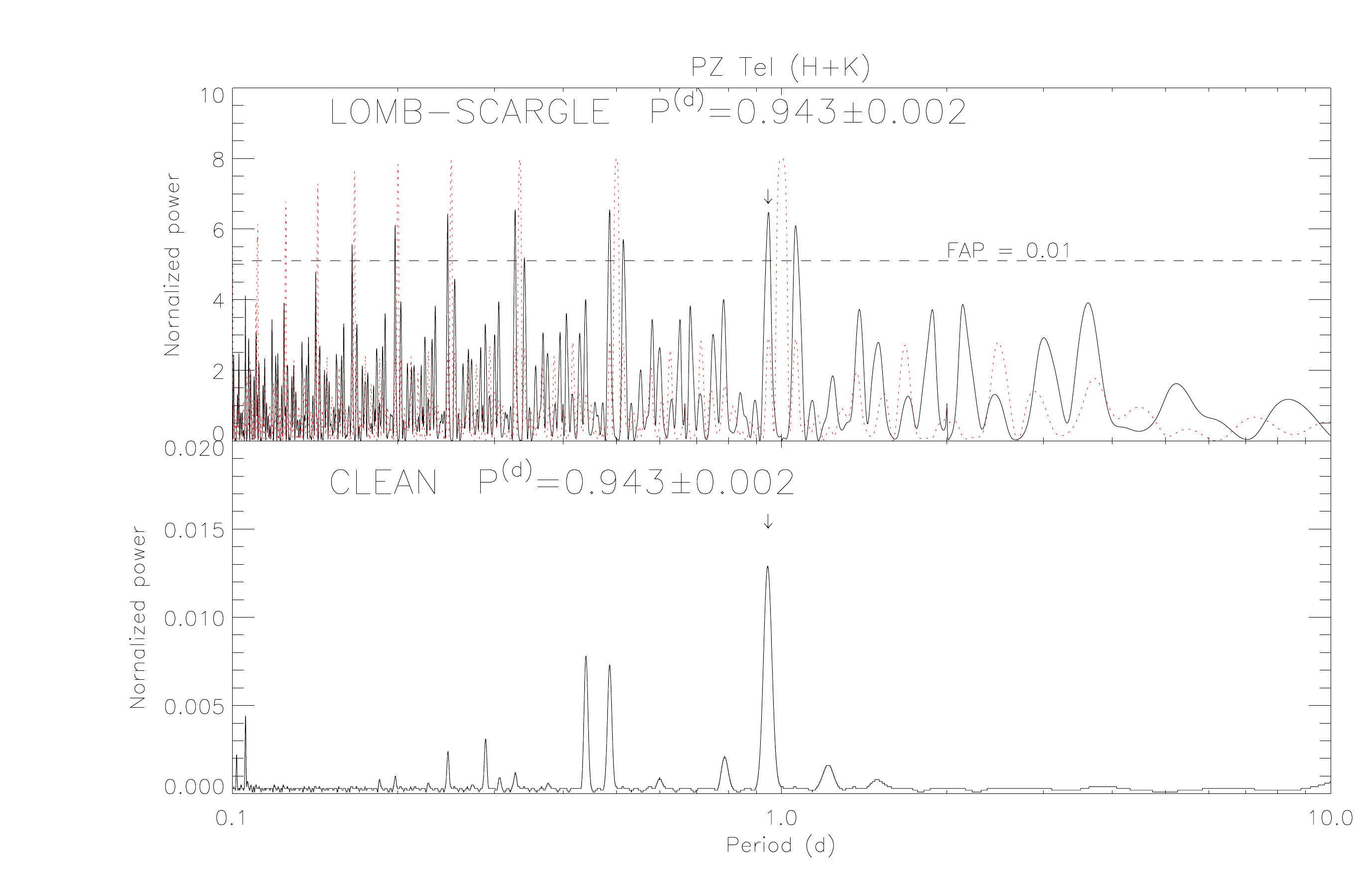}\\
\includegraphics[scale=.4,angle=90,trim= 5mm 0mm 5mm 14mm,clip]{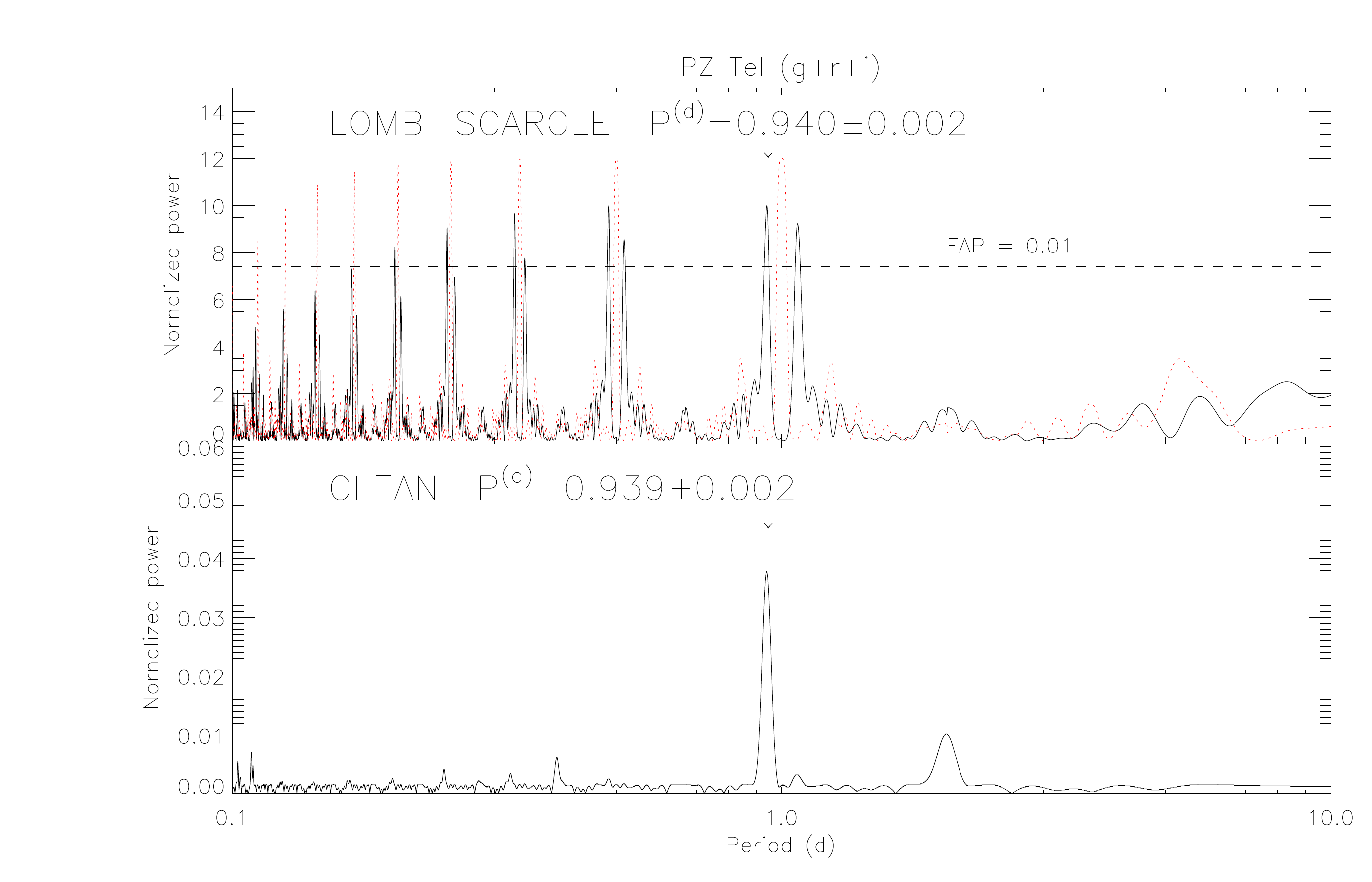}\\
\caption{LS and CLEAN periodograms of the $H$+$K$ timeseries (\textit{top panels}) and of the $g$+$r$+$i$ timeseries (\textit{bottom panels}) (see text).}
\label{fig:periodogram}
\end{centering}
\end{figure*}

Once we checked that the comparison stars were not variable, we could use the C1 and C2 2MASS magnitudes (Table~\ref{2mass}) to transform the PZ~Tel~A differential magnitudes into absolute values. We finally obtained the average magnitudes: $J$\,=\,6.84\,$\pm$\,0.04~mag, $H$\,=\,6.430\,$\pm$\,0.045~mag, and 
$K$\,=\,6.27\,$\pm$\,0.03~mag in the NIR, $g$\,=\,8.51\,$\pm$\,0.04~mag, $r$\,=\,8.11\,$\pm$\,0.04~mag, and $i$\,=\,7.70\,$\pm$\,0.04~mag in the optical.\\

\subsection{York Creek Observatory observations}

PZ~Tel~A was observed from November 18, 2014 to December 2, 2014, for a total of six nights at the York Creek Observatory (41$^{\circ}$\,06$^{\prime}$\,06$^{\prime\prime}$S; 146$^{\circ}$\,50$^{\prime}$\,33$^{\prime\prime}$E, Georgetown, Tasmania) using a f/10 25-cm Takahashi Mewlon reflector, equipped with a QSI 683ws-8 camera, and $B$, $V$, and $R$ standard Johnson-Cousins filters. The telescope has a field of view of 24.5$^{\prime}\times$18.5$^{\prime}$. The plate scale is 0.44$^{\prime\prime}$/pix. A total of 48 frames in each filter were collected using an integration time of 30~s.
Data reduction was performed as described in the previous section for the REM data. The photometric accuracies we could achieve are $\sigma_{\rm B}$ = 0.009, $\sigma_{\rm V}$ = 0.007, and $\sigma_{\rm R}$ = 0.006~mag. {We derived new photometry for PZ Tel~A using the differential $B$, $V$, and $R$ light curves of the star, as well as literature $B$ and $V$ magnitudes of the  comparison stars C1 and C2 (see Sect.~\ref{subsec:REMobs}). The average magnitudes are $B$\,=\,9.05\,$\pm$\,0.05~mag and $V$\,=\,8.30\,$\pm$\,0.05~mag.}

\begin{table}[t]
   \centering
   \caption{2MASS magnitudes of comparison stars adopted to calibrate the REM magnitudes.} % requires the topcapt package
   \begin{tabular}{@{} lrrr @{}} % Column formatting, @{} suppresses leading/trailing space
   \hline
   \hline
                        &       \multicolumn{1}{c}{$J$ (mag)}   &       \multicolumn{1}{c}{$H$ (mag)}  &       \multicolumn{1}{c}{$K$ (mag)}\\
                        \hline
      C1                & 8.493$\pm$0.035  &  7.904$\pm$0.044 & 7.801$\pm$0.029   \\
      C2                & 7.495$\pm$0.023  & 6.781$\pm$0.036  & 6.525$\pm$0.020  \\
      C3                & 8.734$\pm$0.023  &  7.915$\pm$0.020 & 7.667$\pm$0.023  \\
      
      \hline
    \end{tabular}
   \label{2mass}
\end{table}

\subsection{Rotation period search}

We used the Lomb-Scargle \citep{Scargle1982} and CLEAN \citep{Roberts1987} periodogram
analyses on our own observations to search for the rotation period of PZ~Tel~A.
As expected, we found that the star exhibits variability in all bands. A correlation study showed that the
magnitude variations among all $J$+$H$+$K$ bands are correlated (r$_{\rm JH}$ = 0.70, r$_{\rm JK}$ = 0.42, r$_{\rm HK}$ = 0.84~mag) with 
significance level $>$99\% (with the $H$ and $K$ magnitudes exhibiting
a higher degree of correlation) and  also among all $g$+$r$+$i$ bands (r$_{\rm gr}$ = 0.87, r$_{\rm gi}$ = 0.53, r$_{\rm ri}$ = 0.46~mag) 
with 
significance level of $>$95\% (with the $r$ and $i$ magnitudes exhibitinga smaller degree of correlation). To improve the S/N of our time series, we built a new light curve by averaging the $H$ and $K$ light curves
(which is justified by the high correlation between the light curves) and by averaging the $g$, $r$, and $i$ light curves.

Both LS and CLEAN periodograms showed that the major power peak in the averaged NIR light curve is at $P$\,=\,0.943\,$\pm$\,0.002\,d 
and with false alarm probability FAP\,$<$\,1\%.
The FAP, which is the probability that a power peak of that height simplyarises from Gaussian noise in the data, was estimated
using a Monte-Carlo method, i.e., by generating 1000 artificial light
curves obtained from the real one, keeping the date but scrambling
the magnitude values. 
The results are summarized in Fig.~\ref{fig:periodogram}, where in the top panels we plot the LS periodograms and the CLEAN periodograms in the
lower panels. 
In the LS periodogram we also note a number of secondary power peaks
at a high significance level. These peaks are beats of
the rotation period according to the
relation $B$\,=\,$(1/P)$\,$\pm$\,$n$, where $B$ is the beat period and $n$ an integer.
{They arise from a one-day sampling interval imposed} by the rotation of the Earth and
the fixed longitude of the observation site.
It can be noted that in the CLEAN periodograms,
the power peak arising from the light rotational
modulation dominates, whereas all secondary peaks are effectively removed. The horizontal dashed line indicates the power level associated to a FAP = 1\%,
whereas the red dotted line shows the spectral window arising from the data time sampling.

\begin{figure}
\begin{centering}
\includegraphics[trim = 3mm 70mm 15mm 20mm, clip, width=.4\textwidth]{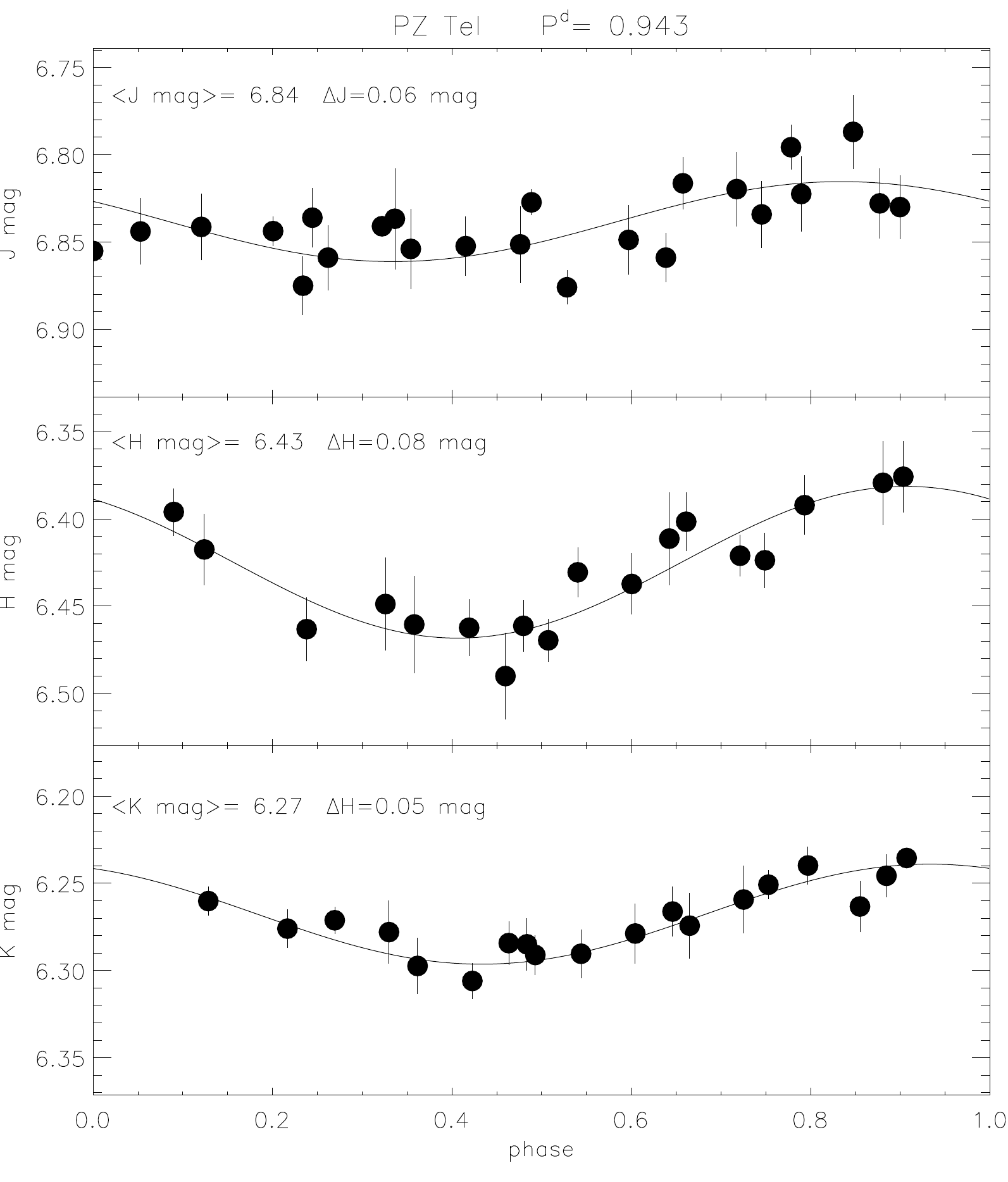}\\
\caption{$J$, $H$, and $K$ light curves of PZ~Tel~A collected with the REM telescope and phased with the $P$\,=\,0.943~d rotation period. Sinusoidal fits are overplotted (solid lines). Average magnitude and peak-to-peak amplitudes are given in the labels.}
\label{series_JHK}
\end{centering}
\end{figure}

In Fig.~\ref{series_JHK}, we plot the NIR light curves of PZ~Tel~A that are phased
with the 0.943-d rotation period. Solid lines are sinusoidal fits with the rotation period. {The amplitude of the light curves}
(measured from the amplitude of the sinusoid) are $\Delta J$ = 0.06, $\Delta H$ = 0.08, and $\Delta K$ = 0.05~mag.

\begin{figure}[t]
\begin{centering}
\includegraphics[trim = 2mm 70mm 15mm 20mm, clip, width=.4\textwidth]{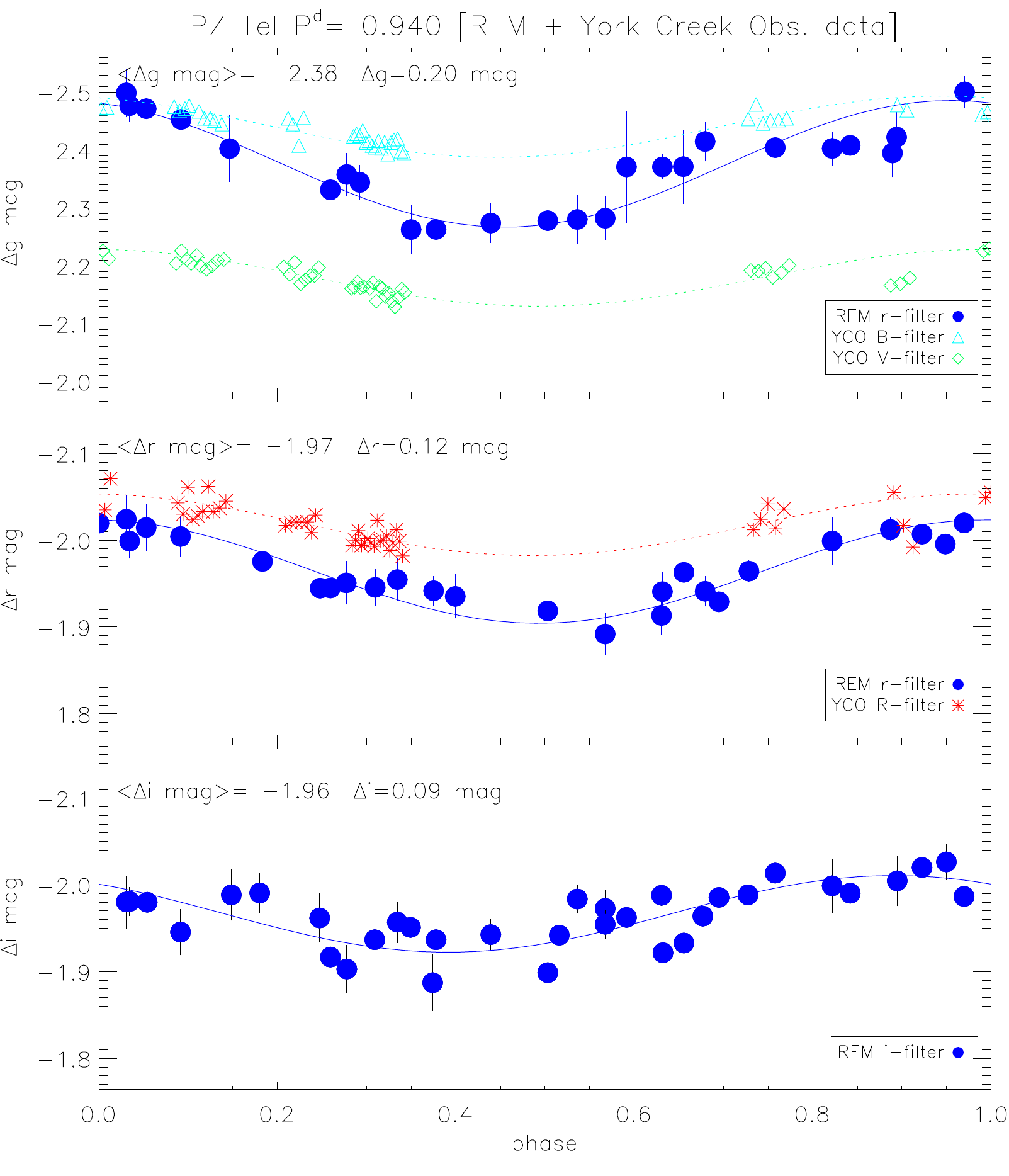}\\
\caption{$g$, $r$, and $i$ light curves of PZ~Tel~A collected with the REM telescope (blue bullets) and phased with the $P$\,=\,0.940~d rotation period. Sinusoidal fits are overplotted (solid lines). Average magnitudes and peak-to-peak amplitudes are given in the labels. Triangles, diamonds, and asterisks represent the $B$, $V$, and $R$ data, respectively, collected at the York Creek Observatory.}
\label{series_gri}
\end{centering}
\end{figure}

The same periodogram analysis in the composed optical light curve
showed that the major power peak is at $P$\,=\,0.940\,$\pm$\,0.002~d and with FAP\,$<$\,1\%.
In Fig.~\ref{series_gri}, we plot the optical $g$, $r$, and $i$ light curves of PZ~Tel~A that are phased
with the 0.940-d rotation period. Solid lines are sinusoidal fits with the rotation period. {The amplitudes of the light curves} are $\Delta g$ = 0.20, $\Delta r$ = 0.12, and $\Delta i$ = 0.09~mag.\\
$B$, $V$, and $R$ data also indicate that PZ~Tel~A is variable (Fig.~\ref{series_gri}). However, owing to the short data set, we
did not search for the rotation period. They
appear to be well in phase with the $g$, $r$, and $i$ light curves, although the light curve minimum was uncovered by the observations.
From the sinusoidal fits, we infer the light curve amplitudes: $\Delta B$ = 0.11, $\Delta V$ = 0.10, and $\Delta R$ = 0.07~mag.

\subsection{Literature information}
\label{sec:literaturepztela}

The rotation period of PZ Tel was first discovered by \citet{Coates1980, Coates1982} who found $P$ = 0.942\,d.  More precise determinations were 
subsequently obtained by \citet{LloydEvans1987} ($P$ = 0.9447\,d), by \citet{Innis1990} ($P$ = 0.94486\,d), and \citet{Cutispoto1998} ($P$ = 0.9447\,d).
Most recently, a period $P$ = 0.9457\,d was derived by \citet{Kiraga2012} from the analysis of about ten years of ASAS photometry. The fast rotation is consistent with the very high
values of the projected rotational velocity reported in the literature by various authors.
\citet{Randich1993} measure $v\sin{i}$ = 70~km\,s$^{-1}$, \citet{Soderblom1998} $v\sin{i}$ = 58~km\,s$^{-1}$, \citet{Barnes2000} $v\sin{i}$ = 68~km\,s$^{-1}$, \citet{Cutispoto2002} $v\sin{i}$ = 70~km\,s$^{-1}$, \citet{delaReza2004} $v\sin{i}$ = 67~km\,s$^{-1}$, \citet{Torres2006} $v\sin{i}$ = 69~km\,s$^{-1}$, and \citet{Scholz2007} $v\sin{i}$ = 77.50~km\,s$^{-1}$. 
Assuming an unspotted magnitude $V$ = 8.25~mag, derived from our own photometry, a distance $d$ = 47$\pm$7\,pc from Hipparcos, a bolometric correction BC$_{\rm V}$ = $-$0.26~mag, and $T_{\rm eff}$ = 5210\,K from \citet{Pecaut2013}, we derive a luminosity of $L$ = 1.16$\pm$0.10\,$L_\odot$ and a radius $R$ = 1.32$\pm$0.14\,$R_\odot$. Combining stellar radius and rotation period, we infer an inclination for the stellar equator (with respect to the direction perpendicular to the line of sight) from the standard formula $\sin{i}$\,=\,$(P\,\times\,v\sin{i})/(R\,\times\,k)$, where $k$ is a constant $k$\,=\,50.578, $R$ {in solar units}, $P$ in days, and $v\sin{i}$ in km\,s$^{-1}$. We find 75$^{\circ}$\,$<$\,$i$\,$<$\,90$^{\circ}$; i.e., we see PZ~Tel~A almost from its equator.

This configuration generally allows a large rotational modulation of the spot visibility, producing light variations with large amplitudes. However, when we compare
the stellar average light curve amplitude ($\Delta V$ = 0.10~mag), we see that {it lies close to the} lower boundary of the light curve amplitude distributions of the $\beta$~Pic members
\citep[see][]{Messina2010}. We have only one season (1982.47) when the star exhibited a light curve amplitude $\Delta V$ = 0.22~mag. A reasonable scenario that we can draw from the available photometry is that PZ~Tel~A has a dominant fraction of spots or spot groups, either uniformly distributed in longitude or located at very high latitudes, which is decreasing in time, making the star brighter. PZ~Tel~A also possesses a smaller fraction of spots unevenly distributed in longitude that accounts for the relatively small light curve amplitude generally observed.

\begin{figure*}
\begin{centering}
\includegraphics[width=.38\textwidth,angle=90,trim= 10mm 0mm 10mm 0mm]{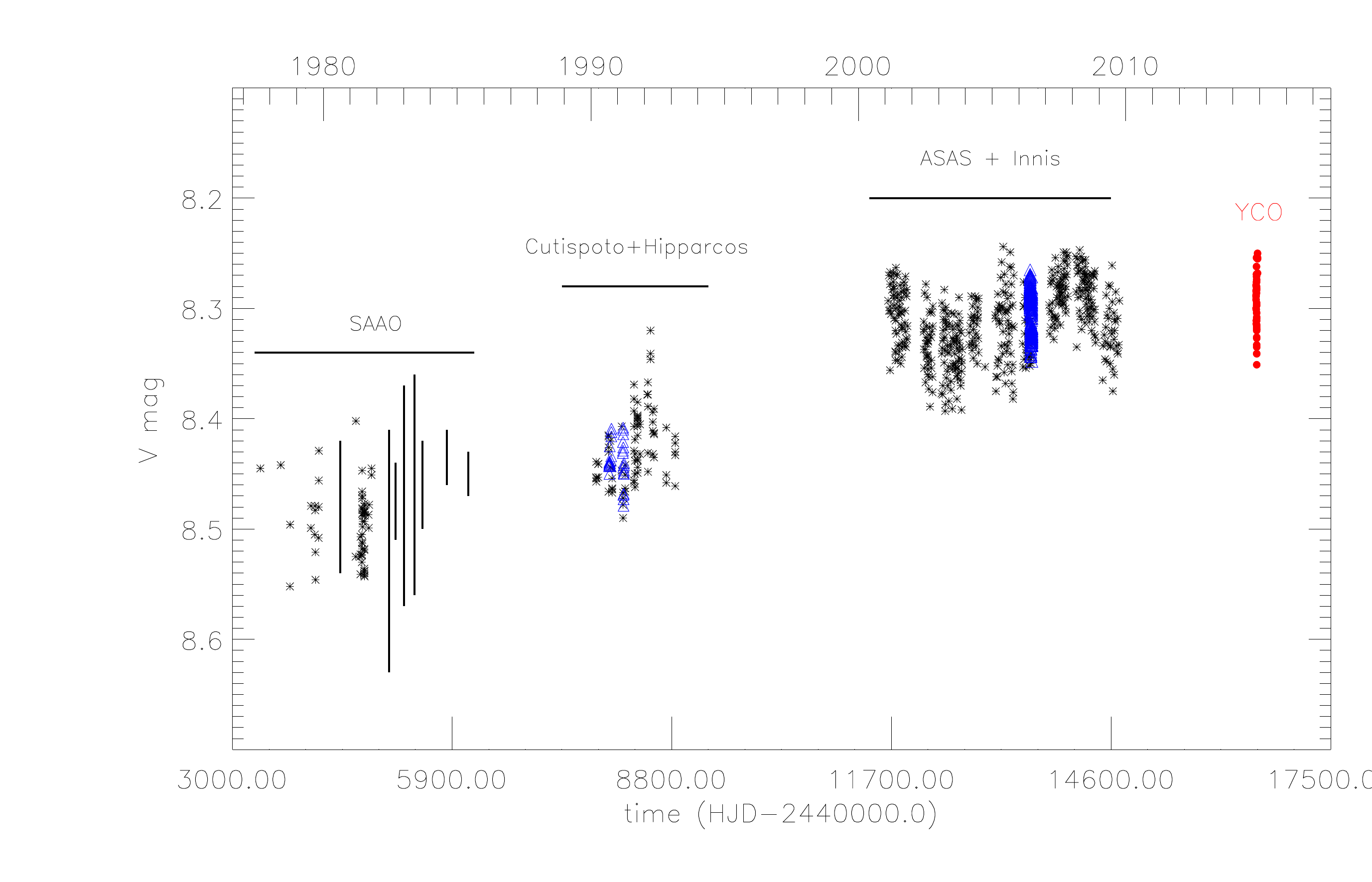}\\
\vspace{0cm}
\caption{Historical 38-yr long time series of $V$-band magnitudes of PZ~Tel~A from 1977 to present. Vertical bars denote epochs for which only brightest and faintest magnitudes are available in the literature.}
\label{long-term}
\end{centering}
\end{figure*}

In Fig.\,\ref{long-term}, we plot the complete series of $V$-band magnitudes, spanning a time interval of almost 38~yr. Data were retrieved from a number of sources: \citet{LloydEvans1987}, \citet{Coates1980}, and \citet{Innis1983} who collected data at SAAO; \citet{Cutispoto1997} and \citet{Cutispoto1998} who collected data at ESO; the Hipparcos Epoch photometry data \citep{ESA1997};  the ASAS public archive \citep{Pojmanski1997}; and \citet{Innis2007}. We note that the star exhibits long-term variability, with a brightening of the average magnitude of $\sim$0.25~mag during the last 38~yr, reaching the brightest value $V$~=~8.25~mag in the most recent years. The light curve amplitude has remained about constant, whereas the component of spots uniformly distributed in longitude has gradually decreased, as if PZ~Tel~A was approaching some sort of starspot activity minimum.

\begin{figure*}
\centering
\caption{{Spectral energy distribution of PZ Tel B. The photometric data from SPHERE, NaCo, and NICI are indicated in light green, blue, and purple, respectively, along with the transmission curves of the associated filters.}}
\includegraphics[width=.9\textwidth, trim = 1mm 2mm 5mm 3mm, clip]{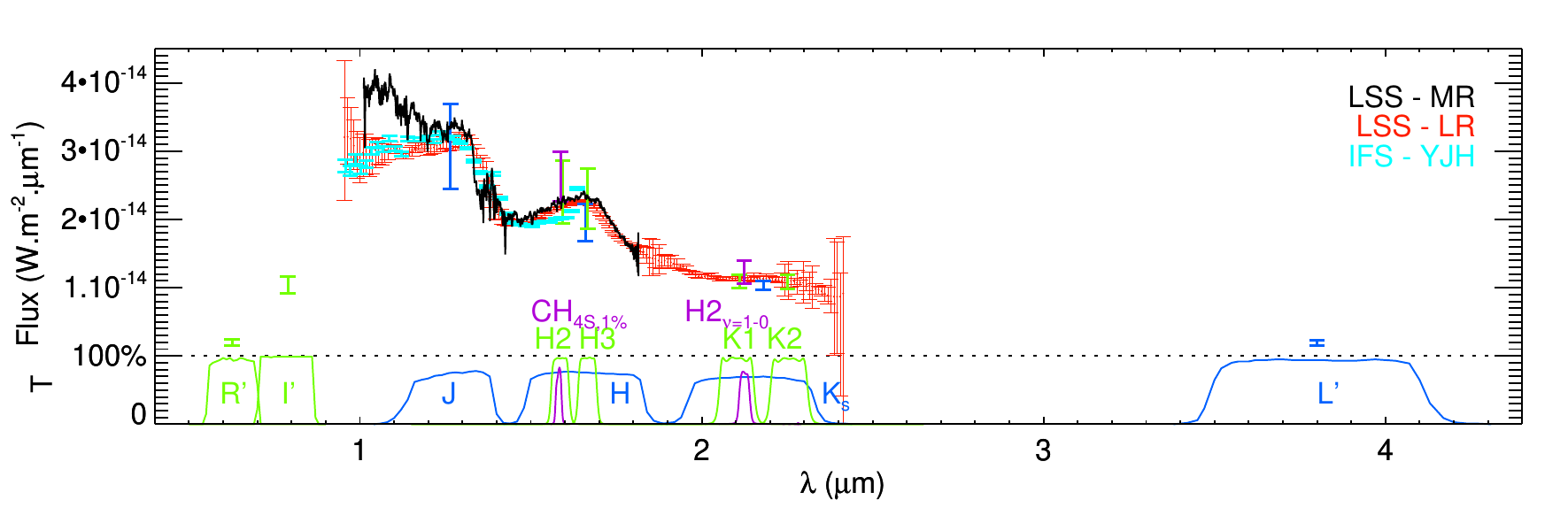}
\label{Fig:SEDPZTELB}
\end{figure*}

\section{Physical properties of PZ~Tel~B and HD~1160~BC}
\label{sec:sed}

      \subsection{Conversion to fluxes}
      \label{subsubsec:confluxes}

The contrast factors extracted for the companions from the IFS and IRDIS data in Sect.~\ref{sec:datareduction} were converted to fluxes in order to allow for a characterization of their SED. We first fit a synthetic spectrum from the GAIA-COND library \citep{2005ESASP.576..565B} onto the photometry of PZ~Tel~A and HD~1160~A: 

\begin{itemize}
\item {{For the fitting of the photometry of PZ~Tel~A}, only the $J$, $H$, and $K$ band photometry derived in Sect.~\ref{subsec:REMobs} was used. It appears close to the 2MASS values \citep{2003tmc..book.....C}. This indicates that it is not deeply affected by the photometric variability of the star, contrary to the optical photometry reported in the same section. The photometry was converted to fluxes using the REM instrument passbands and a flux-calibrated model spectrum of Vega \citep{1994A&A...281..817C}. We investigated the effect of the atmospheric extinction using the sky simulator\footnote{\url{http://www.eso.org/observing/etc/bin/gen/form?INS.MODE=swspectr+INS.NAME=SKYCALC}.} SKYCALC \citep{2012A&A...543A..92N, 2013A&A...560A..91J} and found it to be negligible. We considered models with $T_{\rm{eff}}$ in the range 5000--5600~K, log($g$) between 3.5 and 5.5~dex, and [Fe/H]\,=\,0~dex, for example, with atmospheric parameters bracketing the different values reported in \cite{1999A&A...352..555A, 2006ApJ...638.1004A, 2008ApJ...689.1127M,  2010A&A...515A.111S, 2011MNRAS.411..435B}. A best fit was reached for $T_{\rm{eff}}$\,=\,5000~K and log($g$)\,=\,4.0~dex.}

\item {The HD~1160~A 0.4--22 $\muup$m SED was built from \cite{2001KFNT...17..409K}, \cite{2003tmc..book.....C},\cite{2013yCat.2328....0C}. A model with $T_{\rm{eff}}$\,=\,9200~K and log($g$)\,=\,4.0~dex provided the best fit to the apparent fluxes.}  
\end{itemize}

These flux-calibrated stellar spectra were used to retrieve the companion spectra. We also used them to determine the fluxes of the components of HD~1160 in the IRDIS passbands (Table~\ref{Tab:magcompHD1160}) from the definition of the passbands in Vigan et al. (2015, in press).

\begin{table}[t]
\caption{Available photometry for HD~1160~B and C.}
\label{Tab:magcompHD1160}
\centering
\begin{tabular}{ccccc}
\hline \hline 
Filter &$\lambda$ &$\Delta \lambda$\tablefootmark{a} &   \multicolumn{2}{c}{$F_{\lambda}$ ($10^{-15}$ W\,m$^{-2}$\,$\muup$m$^{-1}$)} \\
  &($\muup$m) &($\muup$m) & B & C \\
\hline
$J$                & 1.250 & 0.180 & $1.361\pm0.131$ & $13.861\pm0.520$ \\
$H$                & 1.650 & 0.290 & $1.582\pm0.121$ & $11.047\pm0.309$ \\
$K1$                & 2.110 & 0.105 & $1.152\pm0.054$ & $5.411\pm0.307$ \\
$K_s$               & 2.200 & 0.330 & $0.938\pm0.044$ & $5.605\pm0.318$ \\
$K2$                & 2.251 & 0.112 & $1.127\pm0.042$ & $4.964\pm0.186$ \\
$L^{\prime}$              & 3.800 & 0.620 & $0.196\pm0.019$ & $1.078\pm0.051$ \\
\hline
\end{tabular}
\tablefoot{\tablefoottext{a}{Full width at half maximum.}} \\
\end{table}

\begin{table}[t]
\caption{Available photometry for PZ Tel B.}
\label{Tab:magcomp}
\centering
\begin{tabular}{ccccc}
\hline \hline 
Filter &$\lambda$ &$\Delta \lambda$\tablefootmark{a} &  Magnitude & $F_{\lambda}$ \\
  &($\muup$m) &($\muup$m) & (mag) &($10^{-14}$ W\,m$^{-2}$\,$\muup$m$^{-1}$) \\
\hline
$R^{\prime}$               & 0.626 & 0.149  &   $17.84^{+0.22}_{-0.31}$ & $0.18^{+0.06}_{-0.03}$ \\
$I^{\prime}$               & 0.790 & 0.153   & $15.16\pm0.12$ & $1.04\pm0.12$  \\
$J$                & 1.265 & 0.250 & $12.47\pm0.20$  & $3.07\pm0.62$\\
CH$_{4}$ 1\%  & 1.587 & 0.015    & $11.68\pm0.14$  & $2.63\pm0.36$\\
$H2$               & 1.593 & 0.053     & $11.78\pm0.19$ & $2.41\pm0.46$ \\
$H$                & 1.660 & 0.330   & $11.93\pm0.14$ & $1.95\pm0.27$ \\
$H3$               & 1.667 & 0.056   & $11.65\pm0.19$ & $2.30\pm0.44$ \\
$K1$               & 2.110 & 0.105   & $11.56\pm0.09$ & $1.09\pm0.10$ \\
H$_{2}$ 1--0 & 2.124 & 0.026 & $11.39\pm0.14$ & $1.23\pm0.17$ \\
$K_s$                & 2.180 & 0.350 & $11.53\pm0.07$ & $1.03\pm0.07$ \\
$K2$               & 2.251 & 0.112 & $11.29\pm0.10$ & $1.08\pm0.10$ \\
$L^{\prime}$               & 3.800 & 0.620 & $11.05\pm0.18$  & $0.19\pm0.04$ \\
\hline
\end{tabular}
\tablefoot{\tablefoottext{a}{Full width at half maximum.}} \\
\end{table}

The remaining photometric data points of PZ~Tel~B were converted using a dedicated procedure. We used the model spectrum of the host star, a model spectrum of Vega, and the atmospheric transmission (SKYCALC) to estimate a photometric shift between the measured ZIMPOL and ROS2 and the IRDIS and REMIR photometry. We applied the same procedure to re-estimate the magnitude of the star and the companion in the NaCo $J$, $H$, $K$ band filters and the NICI CH$_{4}$ 1\% and H$_{2}$ 1--0 filters \citep{Biller2010}. The companion magnitudes were then converted to fluxes using the flux-calibrated spectrum of Vega, the model of the atmospheric transmission, and the corresponding filter passbands. The results are summarized in Table~\ref{Tab:magcomp}. 

The SED of PZ~Tel~B is shown in Fig.~\ref{Fig:SEDPZTELB}. The IRDIS and NICI narrow-band photometry is in good agrement with the flux level of the LSS spectra. The figure highlights the excess of flux of the medium-resolution long-slit spectrum at the shortest wavelengths (Sect.~\ref{subsec: SINFONIred}). The IFS spectrum of the companion reproduces the features of the LRS spectrum quite well, although its slope seems to be flatter in the $H$ band. We decided not to use this spectrum since it is superceded by the LRS spectrum.

\begin{figure}[t]
\centering
\includegraphics[width=.4\textwidth]{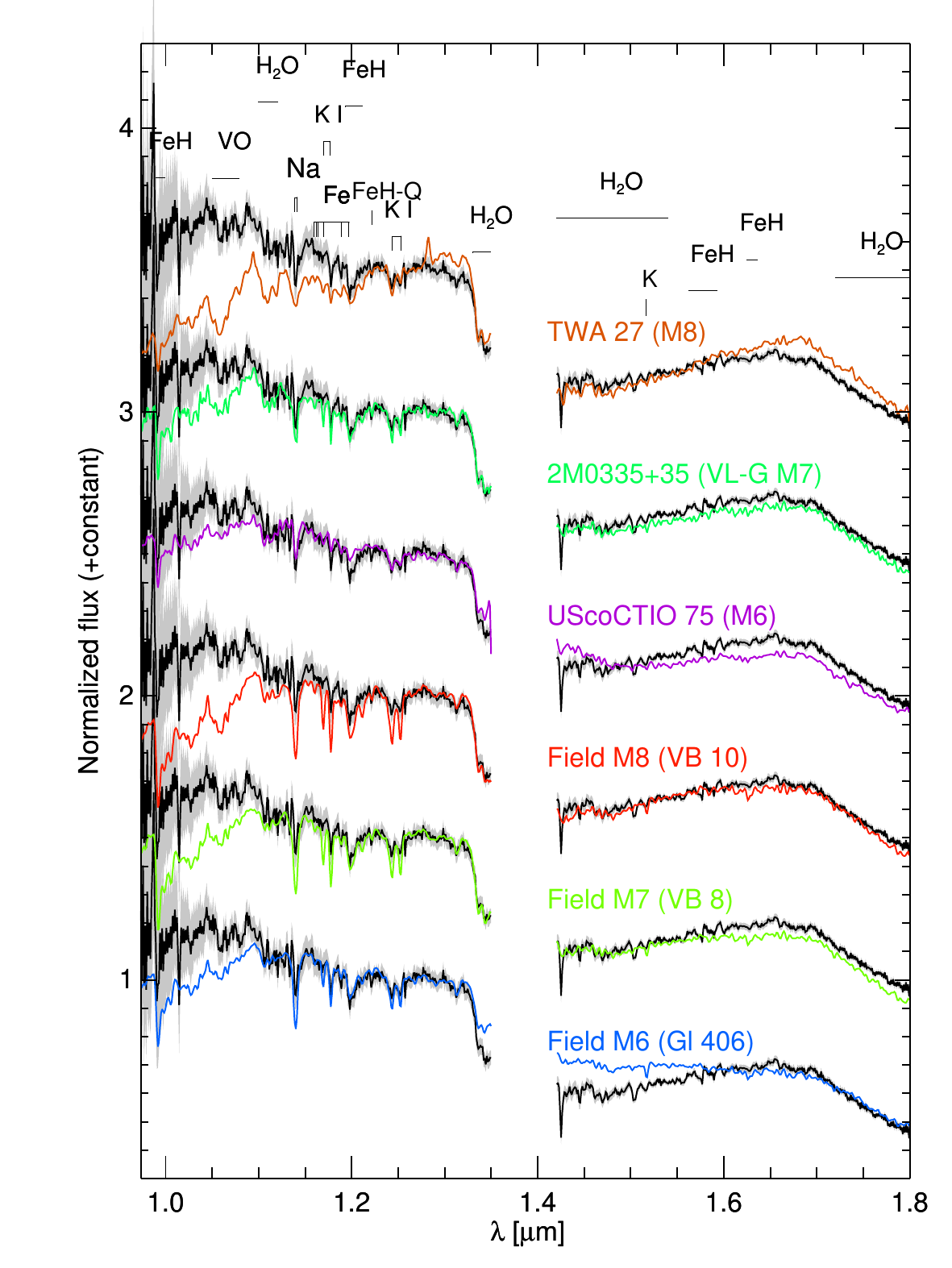}
\caption{{Comparison of the MRS spectrum of PZ~Tel~B to those of field and young M6--M8 dwarfs.}}
\label{Fig:MRScomp}
\end{figure}

\begin{figure}[t]
\centering
\includegraphics[width=.4\textwidth]{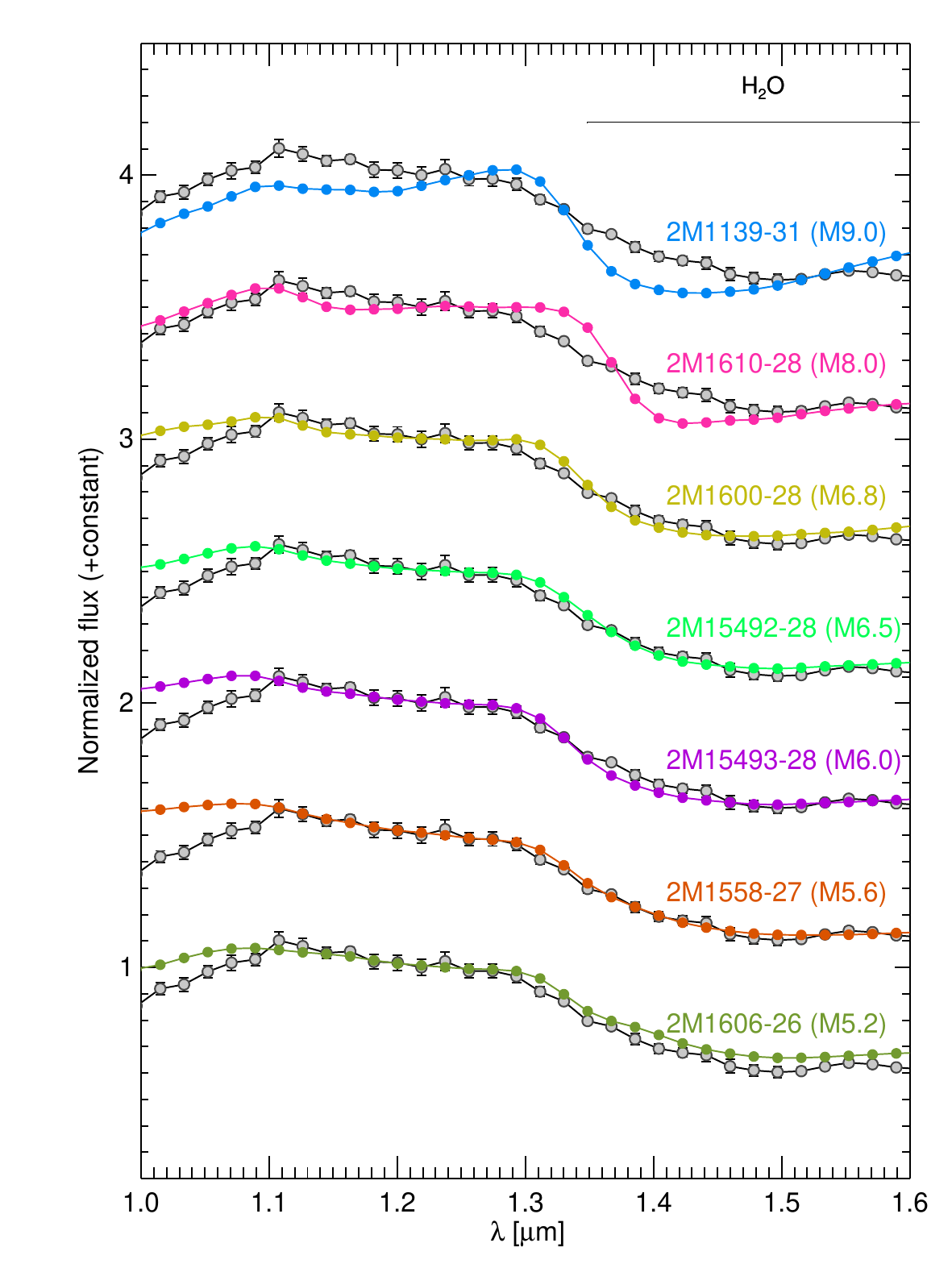}
\caption{Comparison of HD~1160~B IFS spectrum to the spectra of Upper Scorpius members \citep{2014MNRAS.442.1586D} and to \object{2MASS J11395113-3159214} (TWA~26).}
\label{Fig:empHD1160B}
\end{figure}

\subsection{Empirical study}
\label{subsubsec:empstud}

The normalized LRS spectrum of PZ~Tel~B was compared to the {low-resolution spectra of M, L, and T standard dwarfs} of the SpeXPrism library \citep{2014ASInC..11....7B}. The evolution of the $\chi^{2}$ with the spectral type indicates that the companion spectral type is in the M6--M8 range. We find a best match for the field M8 standard \object{VB10} \citep{1961AJ.....66..528V}.
In {the bottom part of Fig.~\ref{Fig:MRScomp}} we compare the MRS spectrum of the companion to those of M6--M8 field dwarf spectra from the SpeXPrism library. {Then, we add in the top part of Fig.~\ref{Fig:MRScomp}} a comparison to the spectra of the young M8 dwarf \object{TWA 27} \citep[TWA member - $\sim$8~Myr, ][]{2002ApJ...575..484G, 2013ApJ...772...79A}, of the low-gravity dwarf \object{2MASS J03350208+2342356} \citep[classified as M7 in the NIR,][]{2000AJ....120.1085G, 2013ApJ...772...79A}, and of the M6 Upper Scorpius (3--11~Myr) member UScoCTIO 75 \citep[\object{2MASS J16003023-2334457}]{2000AJ....120..479A, 2013ApJ...772...79A}. The companion 1.2--1.8~$\muup$m pseudo-continuum is represented best by the pseudo-continuum of the M7 dwarfs. The gravity-sensitive 1.138~$\muup$m Na I doublet, as well as the K I lines (1.169, 1.177, 1.243, 1.253, and 1.516~$\muup$m) are weaker than the lines of field dwarfs and match those of young objects better. We therefore confirm the M7$\pm$1 spectral type and the low surface gravity of the companion.

The IFS spectrum of HD~1160~B (Fig.~\ref{Fig:empHD1160B}) has a pseudo-continuum slope from 1 to 1.6~$\muup$m that is reproduced best by the SpeX \citep{2003PASP..115..362R} spectra of M5.6--M6.8 Upper Sco members classified in the NIR \citep{2014MNRAS.442.1586D, 2007ApJ...669L..97L}. Spectra of later or earlier-type objects than the interval quoted above do not reproduce the depth of the 1.33--1.5~$\muup$m water band, even if the companion spectrum has an absorption band shortward of 1.1~$\muup$m that can only be reproduced by M8 dwarfs. We chose to assign a spectral type of M$6.0^{+1.0}_{-0.5}$ based on this comparison.

\begin{table*}[t]
\centering
\caption{\label{Tab:atmomodchar} Characteristics of the atmospheric model grids adjusted on the full SED of the substellar companions.}
\begin{tabular}{llllllll}
\hline \hline
Model name & $T_{\rm{eff}}$ & $\Delta T_{\rm{eff}}$ & log($g$) & $\Delta$log($g$) & [M/H] & $[\alpha]$  \\
        & (K) & (K) & (dex)  & (dex)  & (dex)  & (dex)  \\
\hline
BT-Settl13 & 400--3500 & 100  & 3.5--5.5 & 0.5 & 0.0 & 0.0 \\  
BT-Settl13 & 800--3000 & 100  & 4.5--5.5 & 0.5 & +0.5 & 0.0 \\  
BT-Settl13 & 1100--3000 & 100  & 3.5--6.0 & 0.5 & $-$0.5 & 0.2 \\  
\hline
BT-Settl14 & 500--3000 & 50 & 3.5--5.5 & 0.5 & 0.0 & 0.0 \\
BT-Settl14 & 500--2800 & 100 & 4.0--5.5 & 0.5 & 0.0 & 0.3 \\ 
\hline
DRIFT-PHOENIX & 1500--3100 & 100        & 3.0--6.0 & 0.5        & $-$0.5,0,+0.5 & 0.0 \\
\hline
\end{tabular}
\end{table*}

\begin{figure}[t]
\centering
\includegraphics[width=.4\textwidth]{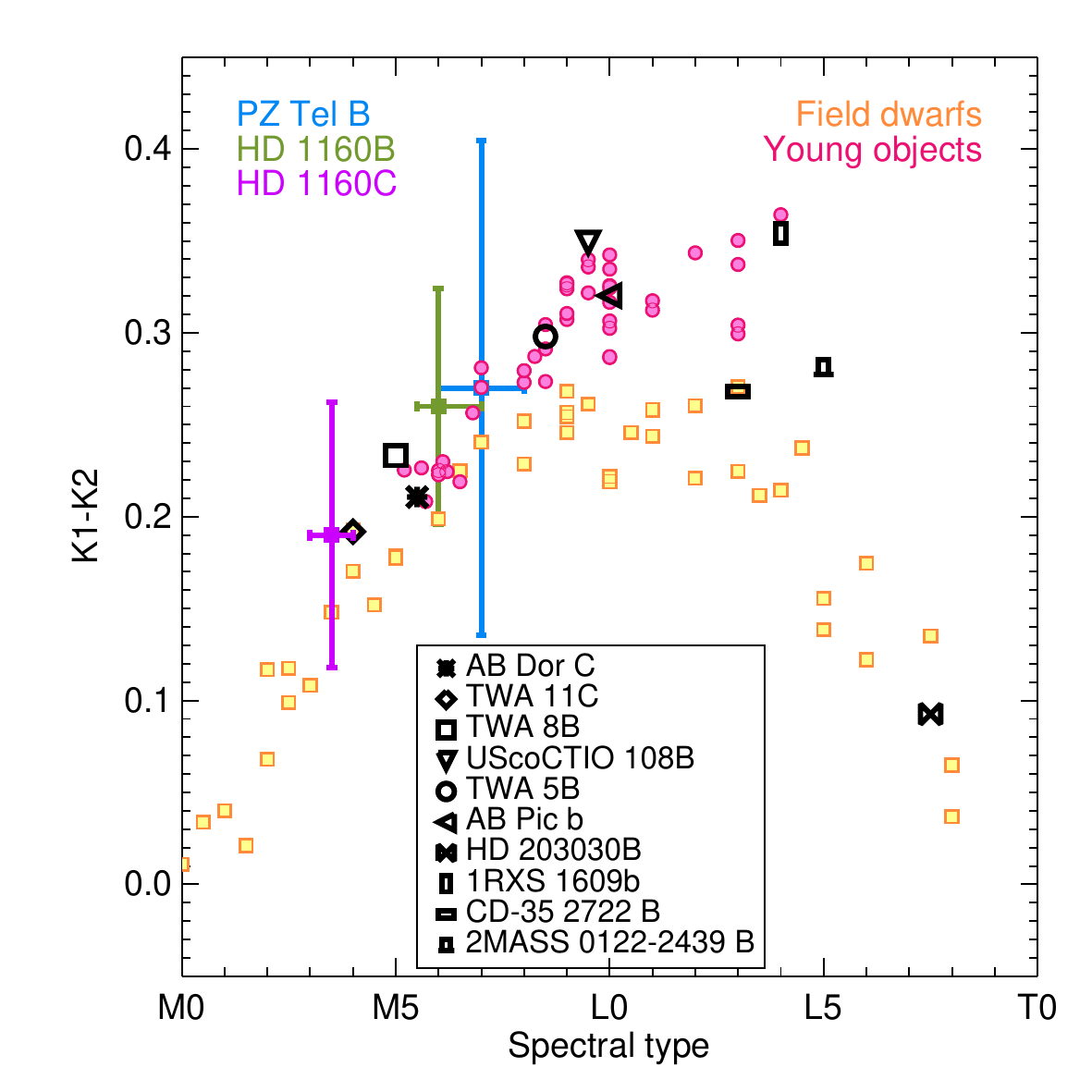}
\caption{Comparison of the $K1-K2$ colors of PZ~Tel~B and of HD~1160~B and C to those of field dwarfs (yellow squares), young brown dwarfs at the M/L transition (pink circles), and companions to nearby young stars (symbols).}
\label{Fig:K1K2}
\end{figure}

We report in Fig.~\ref{Fig:K1K2} the $K1-K2$ colors of HD~1160~B and C, and PZ~Tel~B and then compare them to those of standard field dwarfs synthesized from IRTF and NIRSPEC spectra \citep{2005ApJ...623.1115C, 2009ApJS..185..289R, 2003ApJ...596..561M}. {For HD~1160~C, we relied on the spectral type estimate derived by \citet{Nielsen2012} from its IRTF/SpeX NIR spectra.} We added the synthetic colors of young companions computed from their spectra \citep{Bonnefoy2014b, 2013ApJ...772...79A, 2008ApJ...689L.153L, 2011ApJ...729..139W, 2007ApJ...665..736C, 2013ApJ...774...55B, 2006ApJ...651.1166M}, low-gravity objects \citep{2013ApJ...772...79A} and Upper Sco dwarfs at the M/L transition \citep{2014MNRAS.442.1586D, 2008MNRAS.383.1385L}. For the latter, we followed the reclassification of the object proposed by \cite{Bonnefoy2014b}. The remaining young and/or low-gravity objects were all classified at NIR wavelengths. The location of PZ~Tel~B and HD~1160~B in these diagrams is consistent with their spectral type derivation based on the NIR spectra. The slight offset with respect to the sequence of field dwarfs may be due to the reduced collision-induced absorption of H$_{2}$, so that the spectrum in the $K$ band is more curved. We conclude that the photometry of PZ~Tel~B and HD~1160~B is consistent with the spectral type derived from the NIR spectra. The $K1-K2$ color can be used with $YJH$-band low-resolution spectra to identify young L-type companions, provided that a high quality photometry (errors $\leq$ 0.05~mag) can be extracted from {each IRDIS channel}.

\subsection{Synthetic spectra}
\label{subsubsec:synthspec}

We first compared the SED of PZ~Tel~B and HD~1160~B and C to those of synthetic spectra from the 2013 release of the BT-Settl\footnote{``BT'' refers to the source of water vapor line list 
used in these models \citep{Barber2006} and ``Settl'' to models accounting for dust
formation via a parameter-free cloud model \citep{Rossow1978}.} models \citep{2012RSPTA.370.2765A}. We also considered the restrained grid of synthetic spectra corresponding to the 2014 release of these models (Allard et al., in prep.), as well as the grid of DRIFT-PHOENIX models \citep{2008MNRAS.391.1854H}. The range of parameters for each model grid is summarized in Table~\ref{Tab:atmomodchar}. The BT-Settl14 models consider two $\alpha$-element enrichments. The DRIFT-PHOENIX and BT-Settl13 models {allow us to explore} the effect of the metallicity on the emergent spectra.

\begin{figure*}[t]
\centering
\includegraphics[width=12.3cm]{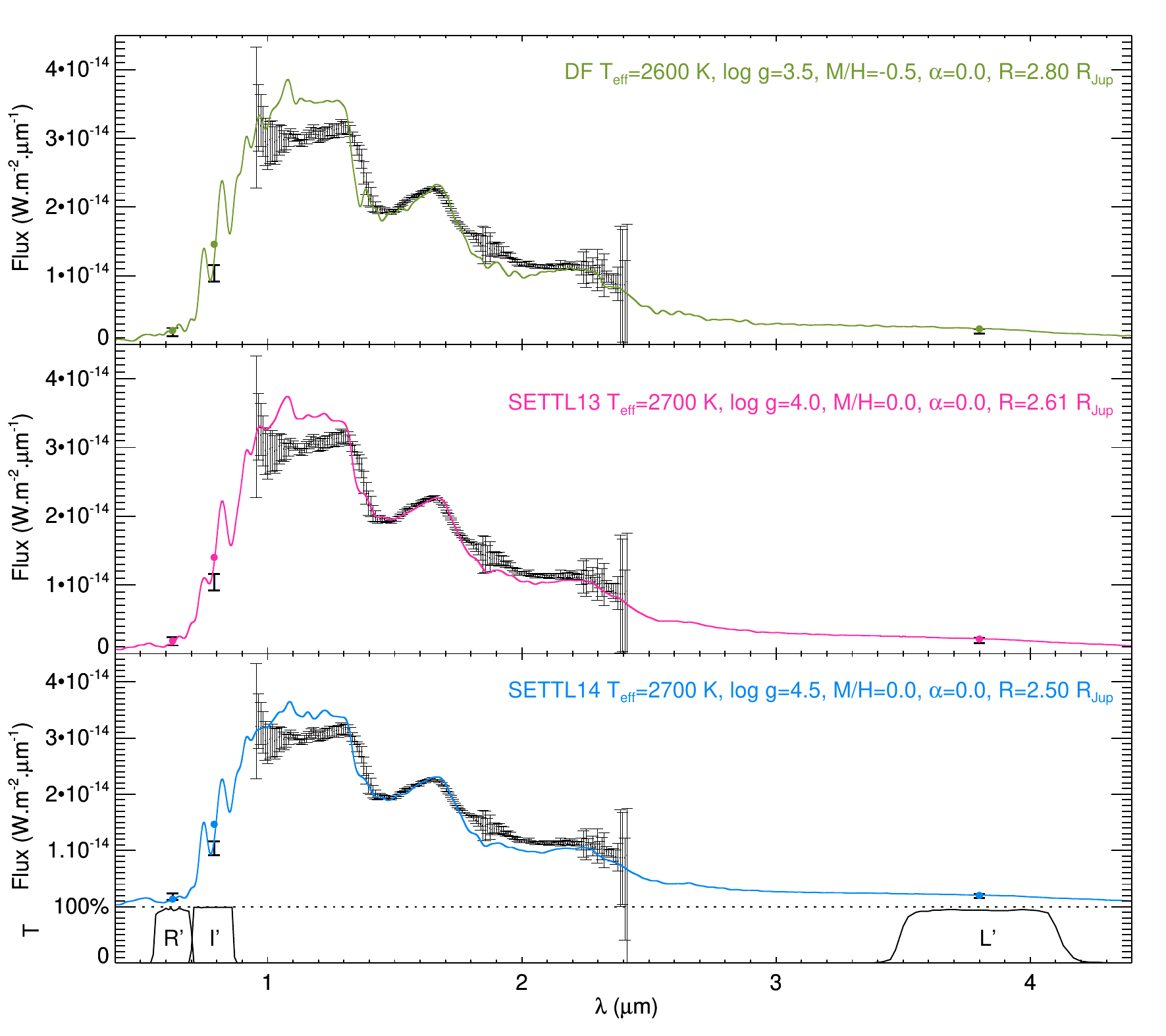}
\caption{{First version of the PZ~Tel~B SED (black) compared to best-fitting synthetic spectra for three atmospheric models (see text).}}
\label{Fig:SEDmodPZTelB}
\end{figure*}

\begin{figure*}[!t]
\centering
\includegraphics[width=12.3cm]{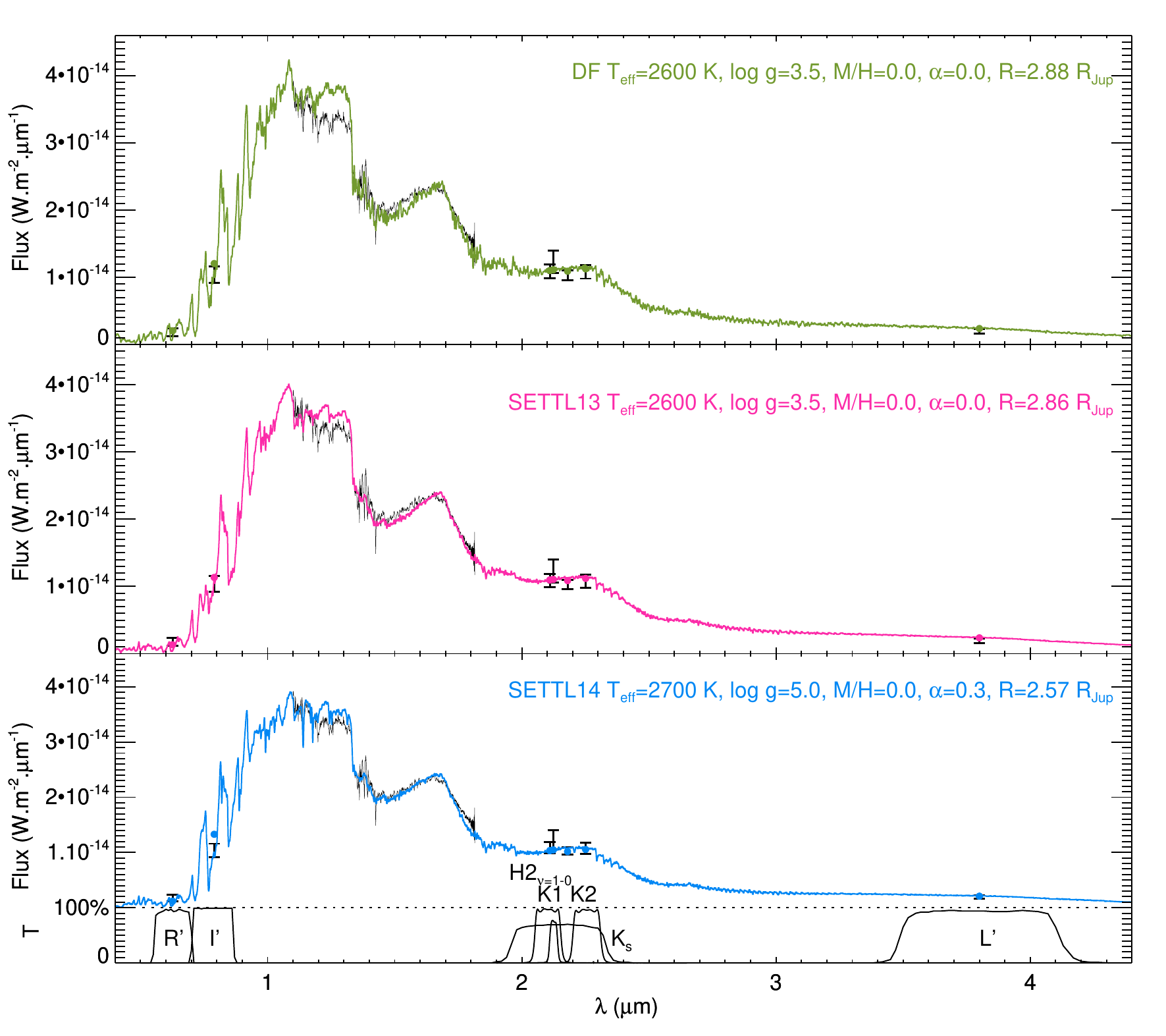}
\caption{{Same as Fig.~\ref{Fig:SEDmodPZTelB} but for the second version of the SED of PZ~Tel~B.}}
\label{Fig:SEDmodPZTelB2}
\end{figure*}

\begin{figure*}[t]
\centering
\includegraphics[width=12.3cm]{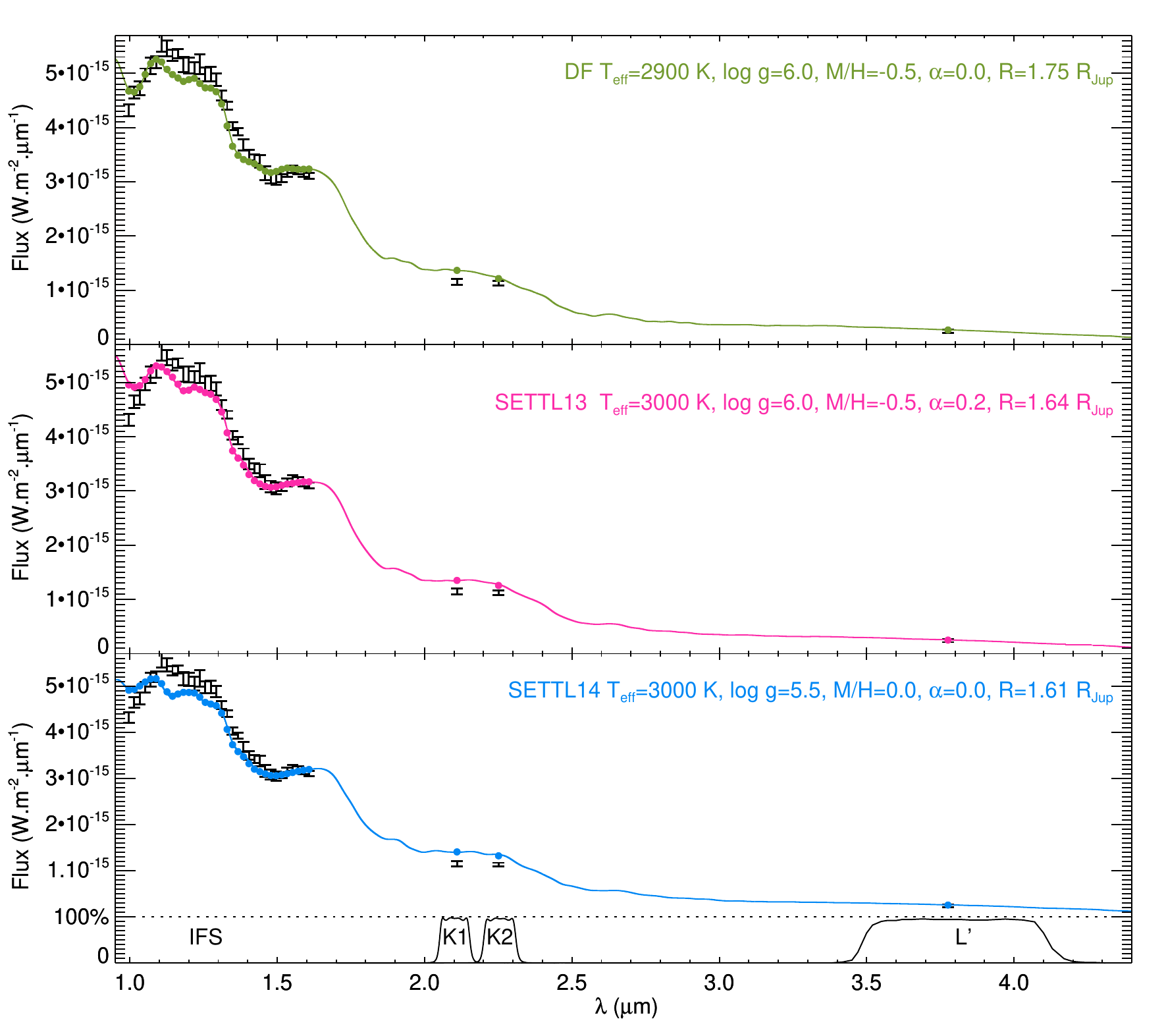}
\caption{{Comparison of the SED of HD~1160~B (black) to best-fitting synthetic spectra for three atmospheric models (see text).}}
\label{Fig:SEDmodHD1160B}
\end{figure*}

\textit{PZ~Tel~B.} We compared the $I^{\prime}$ and $R^{\prime}$ (ZIMPOL), $L^{\prime}$-band photometry, and LRS spectrum of PZ~Tel~B jointly to the three grids of synthetic spectra (hereafter version 1). We also considered an alternative version of the SED (version 2) made of the MRS spectrum longward of 1.1~$\muup$m (not affected by the flux excess) the $I^{\prime}$ and $R^{\prime}$ photometry, and the photometry longward of 2~$\muup$m. We used the G' goodness-of-fit indicators to identify the best-fitting template. The fitting procedure is described in \cite{2008ApJ...678.1372C} and Vigan et al. (2015, in press). The normalization factor used to fit the predicted surface flux onto the measured fluxes of the companion can be expressed as $(R/d)^{2}$ with $R$ and $d$ the source radius and the distance. Assuming the Hipparcos distance \citep[51.5~pc,][]{VanLeeuwen2007}, we were therefore able to derive a radius estimate for the companion for each fitting solution. The atmospheric parameters, G' values, radii, and Monte-Carlo $f_{MC}$ test of the fit robustness \citep[see][and Vigan et al. 2015, in press]{2008ApJ...678.1372C} for the companion and each model grid are reported in Table~\ref{Tab:sedresults}. The best-fitting solutions are shown in Figs.~\ref{Fig:SEDmodPZTelB} and \ref{Fig:SEDmodPZTelB2}. The three most probable (highest $f_{MC}$) solutions, as well as the solution found with a $\chi^{2}$ test, were compared visually and discarded if necessary. We derived $T_{\rm{eff}}$\,=\,2700\,$\pm$100~K from the models. They also indicate that the companion has a reduced surface gravity ($\leq$4.5~dex). The comparison to the models shows that the LRS spectrum lacks fluxes in the $J$ band. The MRS spectrum is represented better by the models at these wavelengths. The imperfect fit of the optical photometry may be related to non-reproducibilities of the models in the red-optical domain \citep{Bonnefoy2014b}. The log($g$) agrees with the value found by \cite{Schmidt2014}. The higher $T_{\rm{eff}}$ found here may be linked to the reduced water-band absorptions found in our spectra (SINFONI, SPHERE) with respect to the spectrum extracted by \cite{Schmidt2014}, to the extended wavelength coverage, and/or to our different fitting method (reddening considered to be negligible here). The existence of best-fitting solutions for solar and subsolar metallicities (Table~\ref{Tab:sedresults}) indicates that this parameter could not be constrained.

\begin{table*}[t]
\centering
\caption{\label{Tab:sedresults} Fitting solutions with the highest $f_{MC}$ values for HD~1160~B and PZ~Tel~B SED and the three sets of atmospheric models using the G' goodness-of-fit indicator.}
\begin{tabular}{lllllllll}
  \hline \hline
Object &   Model        & $T_{\rm{eff}}$ & log($g$) & [M/H] & $[\alpha]$  & $R$ & G' & $f_{MC}$ \\
                        &                                       & (K)   & (dex) & (dex) & (dex) & ($R_{\rm{J}}$) & & \\
  \hline
HD 1160 B &   BT-Settl14    & 3000 & 5.5 & -- & 0.0 & 1.61 & 6.21 & 0.98 \\
 &   BT-Settl13    & 3000 & 6.0 & $-$0.5 & 0.2 & 1.64 & 3.75 & 0.22 \\
 &   DRIFT-PHOENIX    &  2900 & 6.0 & $-$0.5 & -- & 1.75 & 4.18 & 0.24 \\\hline
PZ Tel B -- v1 & BT-Settl14    & 2700 & 4.5 & -- & 0.0 & 2.50 & 10.30 & 0.30 \\ %MAJ - 2eme solution
                 &   BT-Settl13    &  2700 & 4.0 & 0.0 & 0.0 &  2.61 & 7.61 & 0.30 \\ %MAJ - 1st solution
 & DRIFT-PHOENIX    &  2600 & 3.5 & $-$0.5 & -- & 2.80 & 12.70 & 0.19 \\ %MAJ - 2eme solution
  \hline
PZ Tel B -- v2 & BT-Settl14    & 2700 & 5.0 & -- & 0.3 & 2.57 & 1.84 & 0.81 \\ %MAJ - 1eme solution
         &   BT-Settl13    &  2600 & 3.5 & 0.0 & 0.0 & 2.86 & 1.88 & 0.16 \\ %MAJ - 2eme solution
 & DRIFT-PHOENIX    &  2600 & 3.5 & 0.0 & -- & 2.88 & 3.68 & 0.20 \\ %MAJ - 3eme solution
  \hline
\end{tabular}
\end{table*}

\textit{HD~1160~B.} {We used the same procedure to compare the $K1$, $K2$ (SPHERE), $L^{\prime}$ (NaCo) band photometry of HD~1160~B, and the IFS spectrum of the companion to the models. The synthetic spectra were first degraded to the IFS resolution\footnote{The flux values corresponding to the first and last two channels of IFS were removed from the fit since they appeared to be affected by systematic uncertainties not well accounted for in our error bars.} before the average flux into each IFS channel and filter passband was computed. We did not intend to repeat the analysis for HD~1160~C since its temperature was already accurately determined semi-empirically from its NIR spectrum by \cite{Nielsen2012} and spectral type to $T_{\rm{eff}}$ conversion scales.} {We found a $T_{\rm{eff}}$\,=\,3000\,$\pm$\,100~K for HD~1160~B (Fig.~\ref{Fig:SEDmodHD1160B} and Table~\ref{Tab:sedresults}). This agrees well with the $T_{\rm{eff}}$\,=\,2990$^{+70}_{-110}$~K that can be derived from spectral type and the conversion scale of \cite{2003ApJ...593.1093L}. The surface gravity could not be constrained with the presently available data. At low spectral resolutions like for the IFS data used in the fit ($R$\,$\sim$\,30), a good indicator for the surface gravity of substellar objects is the spectral shape in the $\sim$1.5--1.8~$\muup$m range \citep[see, e.g.,][]{Barman2011a, Chilcote2015}, but this spectral range is outside the IFS coverage. Follow-up observations with the LSS mode of IRDIS will allow us to cover the full range of the $H$ band and to constrain the surface gravity of HD~1160~B (Fig.~\ref{Fig:SEDmodHD1160B}). The BT-Settl13 and DRIFT-PHOENIX grids point toward a subsolar metallicity for the companion. Additional medium-resolution spectra of the companion would be needed to confirm this point.}

\subsection{Revised masses}
\label{subsubsec:masses}

We used the new estimated temperatures of PZ~Tel~B and HD~1160~B to revise previous mass estimates.

\textit{PZ~Tel~B.} {The $T_{\rm{eff}}$ of PZ~Tel~B corresponds to a mass of $M$\,=\,45$^{+9}_{-7} M_{\rm{J}}$ using \cite{2015arXiv150304107B} ``hot-start'' evolutionary tracks. These tracks are based on the interior models of \cite{1998A&A...337..403B} and use the BT-Settl14 models as boundary limits. We caution that the tracks do not explore the impact of the initial conditions on the mass predictions due to the lack of ``cold-start'' and ``warm-start'' evolutionary models \citep[e.g.,][]{2007ApJ...655..541M, 2012ApJ...745..174S} in the brown-dwarf mass range. We assumed a mean system age of 24\,$\pm$\,2~Myr \citep{Jenkins2012}, which agrees with the most recent estimates for the age of the $\beta$~Pictoris moving group \citep{Binks2014, 2014ApJ...792...37M, 2014MNRAS.445.2169M}. We used the mean contrast in the $K_s$ band (5.26\,$\pm$\,0.06~mag) from \cite{Mugrauer2012}, the $K$-band magnitude of the star reported in Sect.~\ref{subsec:REMobs}, and a $BC_{K}$\,=\,3.05\,$\pm$\,0.10~mag from \cite{2010ApJ...722..311L} to derive log($L/L_{\odot}$)\,=\,$-$2.51\,$\pm$\,0.10~dex. This corresponds to a mass of $M$\,=\,59$^{+13}_{-8} M_{\rm{J}}$ according to the evolutionary tracks of \citet{2015arXiv150304107B}. These mass predictions are consistent one to each other within the error bars. They fall above those reported in \cite{Biller2010} and \cite{Schmidt2014}, because of the older system age adopted here. We reach similar conclusions comparing the NaCo $L^{\prime}$ band to the \cite{1998A&A...337..403B} tracks.}

\begin{figure}
\centering
\includegraphics[trim = 5mm 2mm 2mm 5mm,clip,width=.41\textwidth]{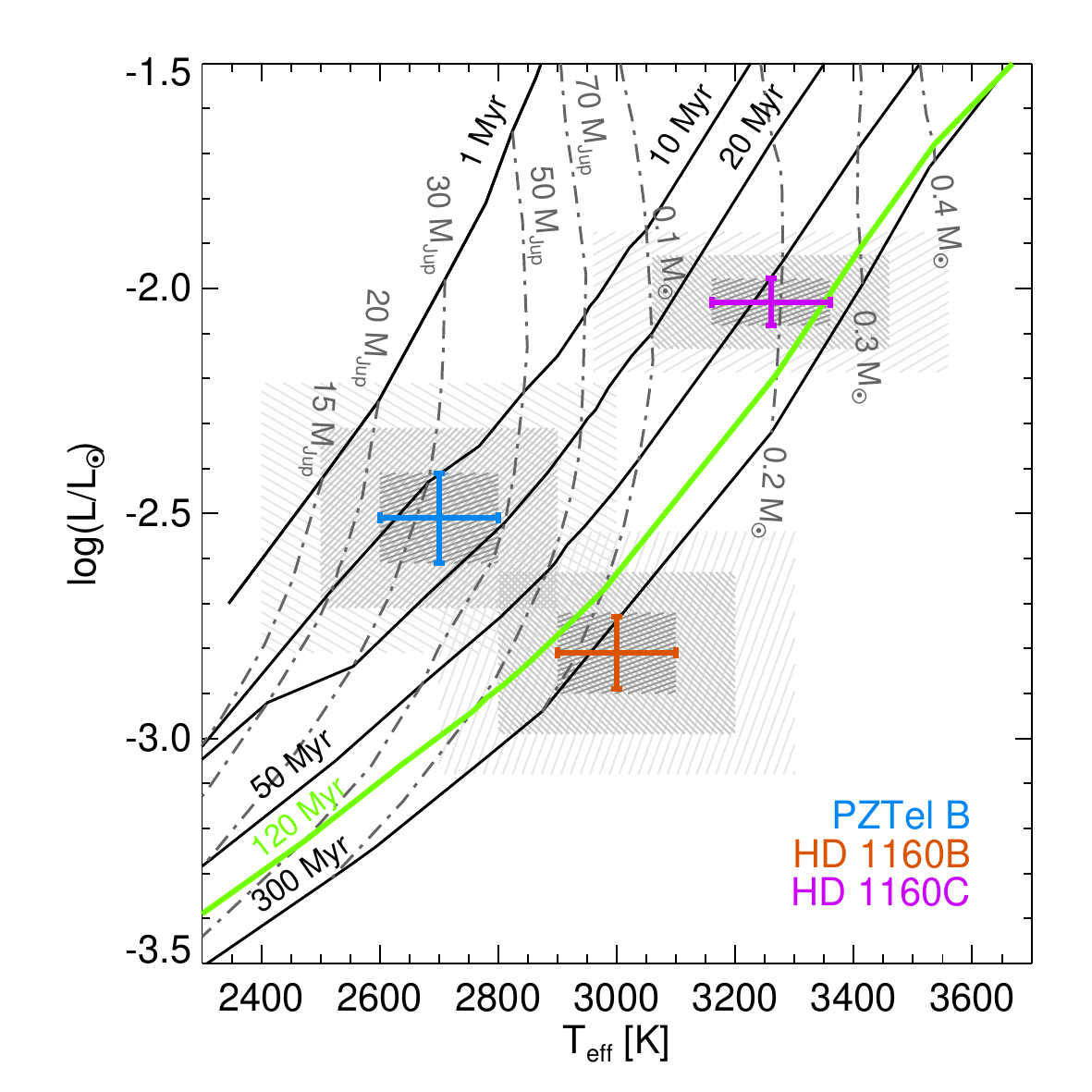} 
\caption{Location of the companions with respect to the predictions of the \cite{2015arXiv150304107B} models. The concentric hatched areas show the 1, 2, and 3$\sigma$ values for the estimated parameters.}
\label{Fig:HRDiag}
\end{figure}

\textit{HD~1160~C.} {We used the $T_{\rm{eff}}$ vs. spectral type conversion scale of young objects from \cite{Pecaut2013} and the spectral type determined by \cite{Nielsen2012} to re-estimate a $T_{\rm{eff}}$\,=\,3260\,$\pm$\,100~K for HD~1160~C. The conversion scale valid for old objects \citep{Pecaut2013} yields close results ($T_{\rm{eff}}$\,=\,3300\,$\pm$\,100~K). \cite{Nielsen2012} estimate that HD~1160~C has a $BC_{K}$\,=\,2.72\,$\pm$\,0.06~mag using the estimated spectral type of the companion and the relations tabulated in \cite{2004AJ....128.1733G}. We preferred to use the relations of \cite{Pecaut2013} instead of those of \cite{2004AJ....128.1733G} because the latter are not valid for objects later than M6. The NICI $J$-band photometry of the system appears to be at odds with the $J$-band photometry of HD~1160~B derived from SPHERE IFS data (see below). Nevertheless, the NICI $K_s$-band photometry of both companions agree with the SPHERE IRDIS photometry in the $K12$ filter pair. Consequently, we used the NICI $K_s$-band photometry of HD~1160~C, as well as the bolometric corrections reported in \cite{Pecaut2013} to find log($L/L_{\odot}$)\,=\,$-$2.05\,$\pm$\,0.06~dex.} 

\textit{HD~1160~B.} \cite{Nielsen2012} show that HD~1160~B has redder $J-K_{s}$ colors than the sequence of field dwarfs. Using the extracted IFS spectrum of the companion and the NICI filter passbands, we instead found $\Delta J$\,=\,7.43\,$\pm$\,0.03~mag and $M_{J}$\,=\,9.53\,$\pm$\,0.11~mag. We assessed that both the $K1$ and $the K2$ photometry of the companion match $M_{K_{s}}$\,=\,8.88\,$\pm$\,0.12~mag using the NICI and SPHERE filter passbands and spectra of Upper Sco M5.5--M7 dwarfs \citep{2014MNRAS.442.1586D}. The $K$-band photometry is consistent within error bars with the value derived in \cite{Nielsen2012}. The revised color $J-K$\,=\,0.50\,$\pm$\,0.07~mag is bluer than expected ($J-K$\,=\,0.90\,$\pm$\,0.07~mag) for a M6.0$^{+1.0}_{-0.5}$ dwarf. This may be due to the methods used to estimate the new photometry. However, such blue colors correspond to those of mid-M subdwarfs and could thus be due to subsolar metallicity, as already discussed in Sect.~\ref{subsubsec:synthspec}. We relied on the SPHERE $K_{s}$-band photometry in the following. We estimated a $BC_{K}$\,=\,3.00$^{+0.05}_{-0.01}$~mag for HD~1160~B using the corrections from \cite{2010ApJ...722..311L} and the spectral type estimate from Sect.~\ref{subsubsec:empstud}. We derived log($L/L_{\odot}$)\,=\,$-$2.81$\pm$\,0.10~dex.  

We compare HD~1160~B and C $T_{\rm{eff}}$ and luminosities to \cite{2015arXiv150304107B} evolutionary tracks in Fig.~\ref{Fig:HRDiag}. Both companions fall marginally (within 2$\sigma$) onto the 100--120~Myr isochrone. HD~1160~B lies at the substellar boundary. In comparison, PZ~Tel~B still falls onto the 20 Myr isochone, in agreement with the host star age. We reach similar conclusions using the \cite{1998A&A...337..403B} models and the $L^{\prime}$-band magnitudes\footnote{The predictions of \cite{2015arXiv150304107B} models for the NaCo filter set were not available at the time of the submission of this article.} of the objects (Sect.~\ref{subsec: NaCoim}). 

This questions previous age estimates of HD~1160~A\footnote{We note that for the statistical study of the NICI survey of young B and A stars, \citet{Nielsen2013} derive an older age of 92~Myr, with 68\% confidence between 36 and 178~Myr, based on a Bayesian analysis.}. The age of the star is difficult to estimate because of the small number of reliable age indicators for this target. \cite{Nielsen2012} argue that HD~1160~A cannot be older than 300~Myr because of the expected main-sequence lifetime of an A0 star \citep{2000A&A...358..593S}. {The kinematics of HD~1160 are still compatible (2.8$\sigma$ in $U$, $<$1$\sigma$ in $V$, and 2.3$\sigma$ in $W$) with the Octans-Near association \citep{2013ApJ...778....5Z}, for which \cite{2015MNRAS.447.1267M} estimate a 30--40 Myr lithium age.} \cite{Nielsen2012} show that the location of HD~1160~A and C in HR diagrams is consistent with both the young associations and the Pleiades, but indicates a younger age than the Pleiades. Nevertheless, this analysis relied on the $J$-band photometry of HD~1160~C, which may be biased. The revised photometry of the system and our analysis of the spectro-photometry instead suggest that the system has a subsolar metallicity. In that case, HD~1160~A falls onto the 10--300~Myr isochrones in HR diagrams \citep{Nielsen2012}. {Using the constraints from the kinematics and the isochrones, we considered an age of $100_{-70}^{+200}$~Myr for the system in the following.}  

For the mass of HD~1160~C, we estimate  $M_{C}$\,=\,205$_{-51}^{+72}$ and $244^{+113}_{-70}$~$M_{\rm{J}}$ from its $T_{\rm{eff}}$ and bolometric luminosity, respectively, assuming the \cite{2015arXiv150304107B} evolutionary models. For HD~1160~B, estimates are $M_{B}$\,=\,107$_{-38}^{+59}$ and 79$_{-40}^{+65}$~$M_{\rm{J}}$, respectively. {The masses for HD~1160~B are revised to higher values with respect to the estimate of 33$^{+12}_{-9}$ derived by \citet{Nielsen2012}, because of a combination of a brighter J-band magnitude estimate (by 1.22~mag) and of an age range estimate covering older values (up to 300~Myr against 100 Myr). The larger error bars compared to the estimate in \citet{Nielsen2012} stem from the larger uncertainties in our estimate of the system age. When considering the youngest possible ages, the masses fall in the substellar regime.} They marginally agree with those of \cite{Nielsen2012}. Nonetheless, we cannot exclude that HD~1160~B is a low-mass star for the oldest possible ages. The radius and $T_{\rm{eff}}$ of HD~1160~B estimated from the synthetic spectra (Sect.~\ref{subsubsec:synthspec}) are notably well reproduced for ages $\geq $100~Myr. We note that the new mass estimates of both companions agree with the predictions of a formation scenario by disk instability in a disk of mass half of the stellar mass \citep[see Fig.~C.1 in][]{Bonnefoy2014a}.

\begin{table}[t]
\centering
\caption{\label{tab:masses} {Masses for PZ~Tel~B and HD~1160~BC derived in this work compared to literature estimates.}}
\begin{tabular}{lccc}
\hline \hline
Parameter & PZ~Tel~B & HD~1160~B & HD~1160~C \\
        & ($M_{\rm{J}}$)  & ($M_{\rm{J}}$) & ($M_{\rm{J}}$) \\
\hline
log($L/L_{\odot}$) & 59$^{+13}_{-8}$ & 79$_{-40}^{+65}$ & $244^{+113}_{-70}$ \\  
$T_{\rm{eff}}$ & 45$^{+9}_{-7}$ & 107$_{-38}^{+59}$  & 205$_{-51}^{+72}$ \\  
\hline
Literature & 36\,$\pm$\,6\tablefootmark{a} & 33$^{+12}_{-9}$\tablefootmark{b} & 220$^{+30}_{-40}$\tablefootmark{b} \\
\hline
\end{tabular}
\tablefoot{\tablefoottext{a}{\citet{Biller2010}.} \tablefoottext{b}{\citet{Nielsen2012}.}}
\end{table}

Given the new mass estimates in Table~\ref{tab:masses}, PZ~Tel~B and potentially HD~1160~B can be added to the scarce population of companions with mass ratios with respect to their host star below 0.1 \citep{Raghavan2010}. Recently, such low-mass companions have been reported around intermediate-mass stars in the Sco-Cen stellar association \citep{Hinkley2015b} and the F-type star HD~984 \citep{Meshkat2015}. The observed statistical properties (e.g., mass ratio, eccentricity) of these objects can be compared to the predictions of formation scenarios \citep[e.g.,][]{Bate2009, Raghavan2010}. For both PZ~Tel and HD~1160 systems, a formation mechanism based on the collapse and the fragmentation of a dense molecular cloud seems plausible. Gravitational instabilities in a disk may also explain the formation of the companions in both systems, provided subsequent dynamical interactions excited the measured eccentricity for PZ~Tel~B and early fragmentation in the disk occurred for the HD~1160 system to meet the requirement of a high mass for the disk \citep[see above and Fig.~C.1 in][]{Bonnefoy2014a}.

\section{Constraints on the orbit of PZ~Tel~B}
\label{sec:orbitpztelb}

\begin{figure}[t]
\centering
\includegraphics[trim = 9mm 6mm 20mm 19mm,clip,width=.41\textwidth]{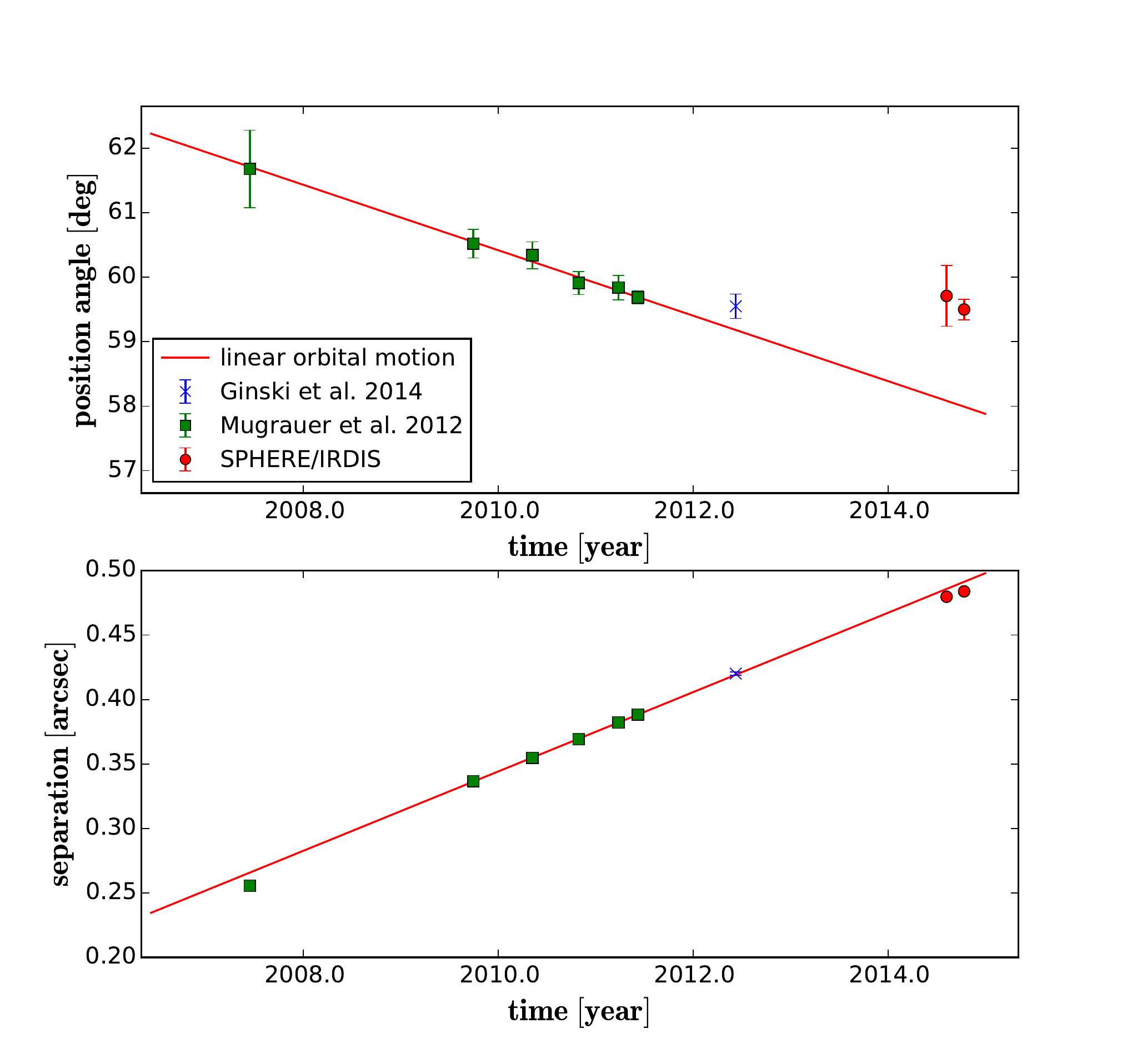}
\caption[]{Relative astrometric measurements of the PZ~Tel system. The linear motion fit (red solid line) was computed by taking only the data points between 2009 and 2011 into account.}
\label{fig: rel-astrometry-pztel}
\end{figure}

In \cite{Ginski2014}, it was noted that possible deceleration of the companion is detectable in the 2012 VLT/NaCo data. This was done by comparing its relative astrometric position in 2012 to the position obtained by linear fitting of the data taken between 2009 and 2011. The deviation from simple linear motion was on the 2$\sigma$ level in position angle. In Fig.~\ref{fig: rel-astrometry-pztel}, we show the previous relative astrometric measurements of the system from \cite{Mugrauer2012} and \cite{Ginski2014}, along with the same linear fit and our new SPHERE/IRDIS measurements. The IRDIS measurements are clearly deviant from the linear motion fit in position angle (9.4$\sigma$). In addition, we also observe a highly significant\footnote{Error bars in Fig.~\ref{fig: rel-astrometry-pztel} are much smaller than the displayed markers.} deviation from simple linear motion in separation (25.5$\sigma$). It is in principle possible that a small systematic offset exists between the NaCo and IRDIS data since the reference fields used for astrometric calibration are different. However, given that the current trend in PA was already detected in the NaCo data and that the significance of the detected deceleration is very high, we conclude that the companion is decelerating, which is consistent with it moving toward apastron.\footnote{It should be noted that previously deceleration of the companion was detected by \cite{Mugrauer2010} between the earliest 2007 observation epoch and the later 2009 epoch. This is also visible in Fig.~\ref{fig: rel-astrometry-pztel} as a deviation of this epoch from the linear motion fit.}

We then used the two astrometric measurements taken with SPHERE/IRDIS to further constrain the orbital elements of the system taking the previous astrometric measurements with VLT/NaCo presented in \cite{Mugrauer2012} and \cite{Ginski2014} into account. For this purpose, we considered a least-squares Monte Carlo (LSMC) fitting technique, which was presented in detail in \cite{Ginski2013}. The fitting algorithm created $5 \times 10^6$ random sets of orbit parameters, which were drawn from uniform distributions, without considering the astrometric data. These parameter sets were then used as starting points for a least-squares minimization. To constrain the parameter space, we fixed the system mass to 1.2~$M_\odot$ as done in \cite{Ginski2014}. We also set an upper limit to the semi-major axis assuming that the system is stable against disruptions in the galactic disk. To estimate this value, we utilized the relation $a_{\mathrm{max}}\mathrm{[au]}\,=\,1000\,M_{\mathrm{tot}}/M_{\odot}$ as given in \cite{Close2003} (where $M_{\mathrm{tot}}$ is the total system mass in solar masses, and $a_{\mathrm{max}}$ the maximum stable semi-major axis). This put an upper limit on the semi-major axis of 23.3$''$ ($\sim$1200\,au). In addition, we limited our orbit fitting to longitudes of the ascending node $\Omega$ between 0 and 180$^{\circ}$, since no precise radial velocity measurements of the companion are available.

\begin{figure*}[!t]
\centering
\includegraphics[trim = 12mm 3mm 21mm 13mm,clip,height=0.28\textheight]{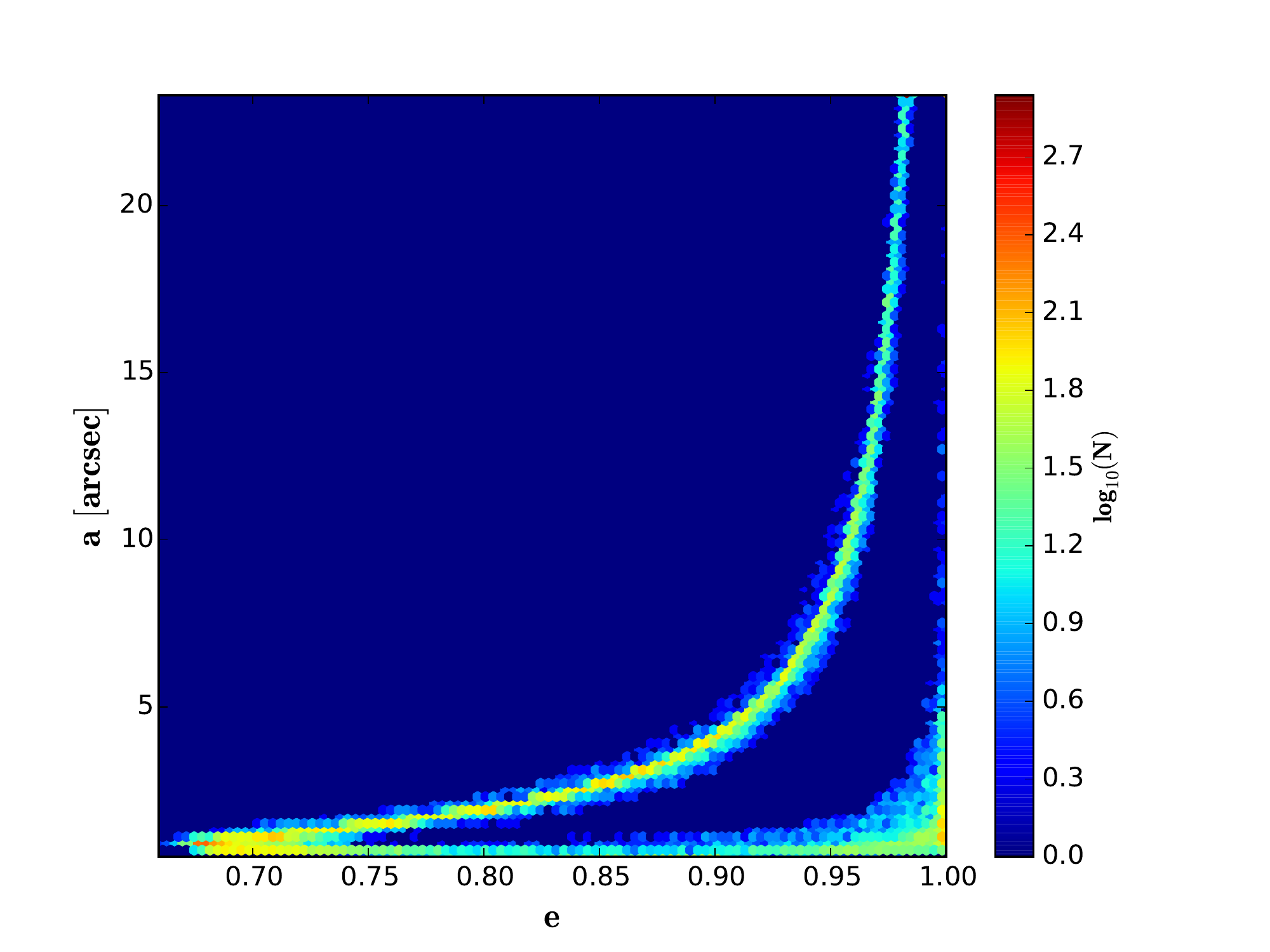}
\includegraphics[trim = 9mm 3mm 21mm 13mm,clip,height=0.28\textheight]{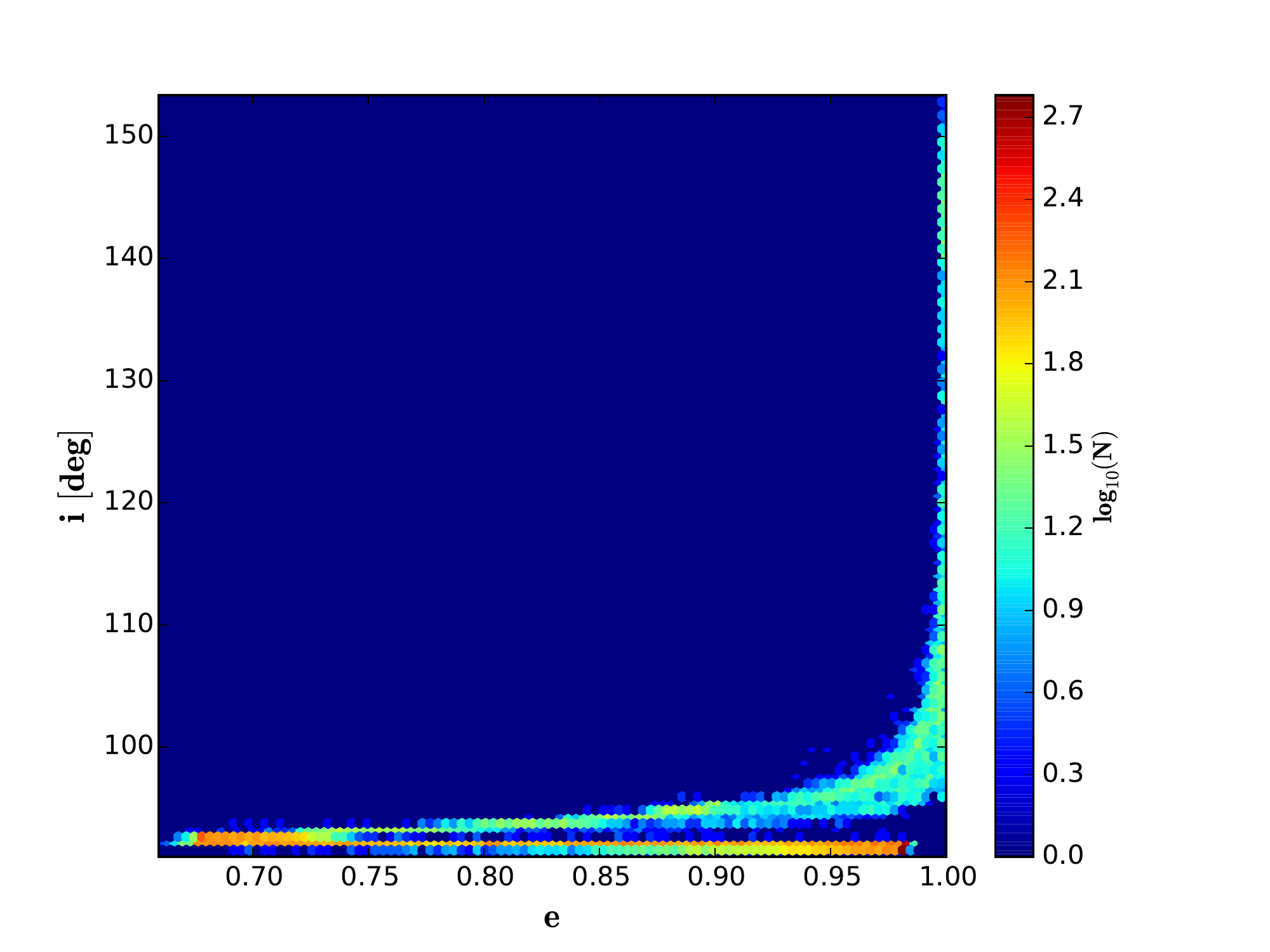}
\includegraphics[trim = 9mm 3mm 24mm 13mm,clip,height=0.28\textheight]{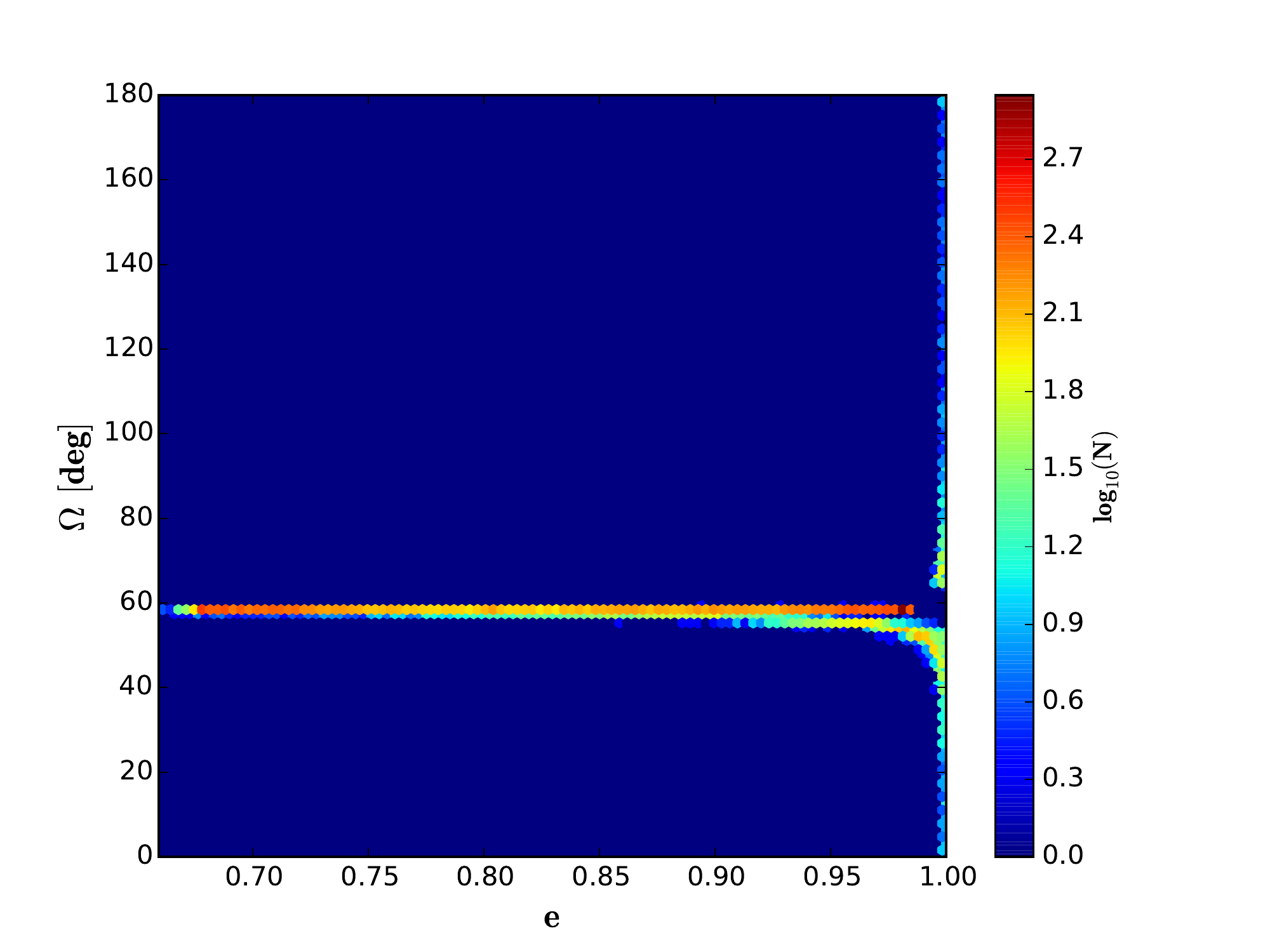}
\includegraphics[trim = 10mm 3mm 21mm 13mm,clip,height=0.28\textheight]{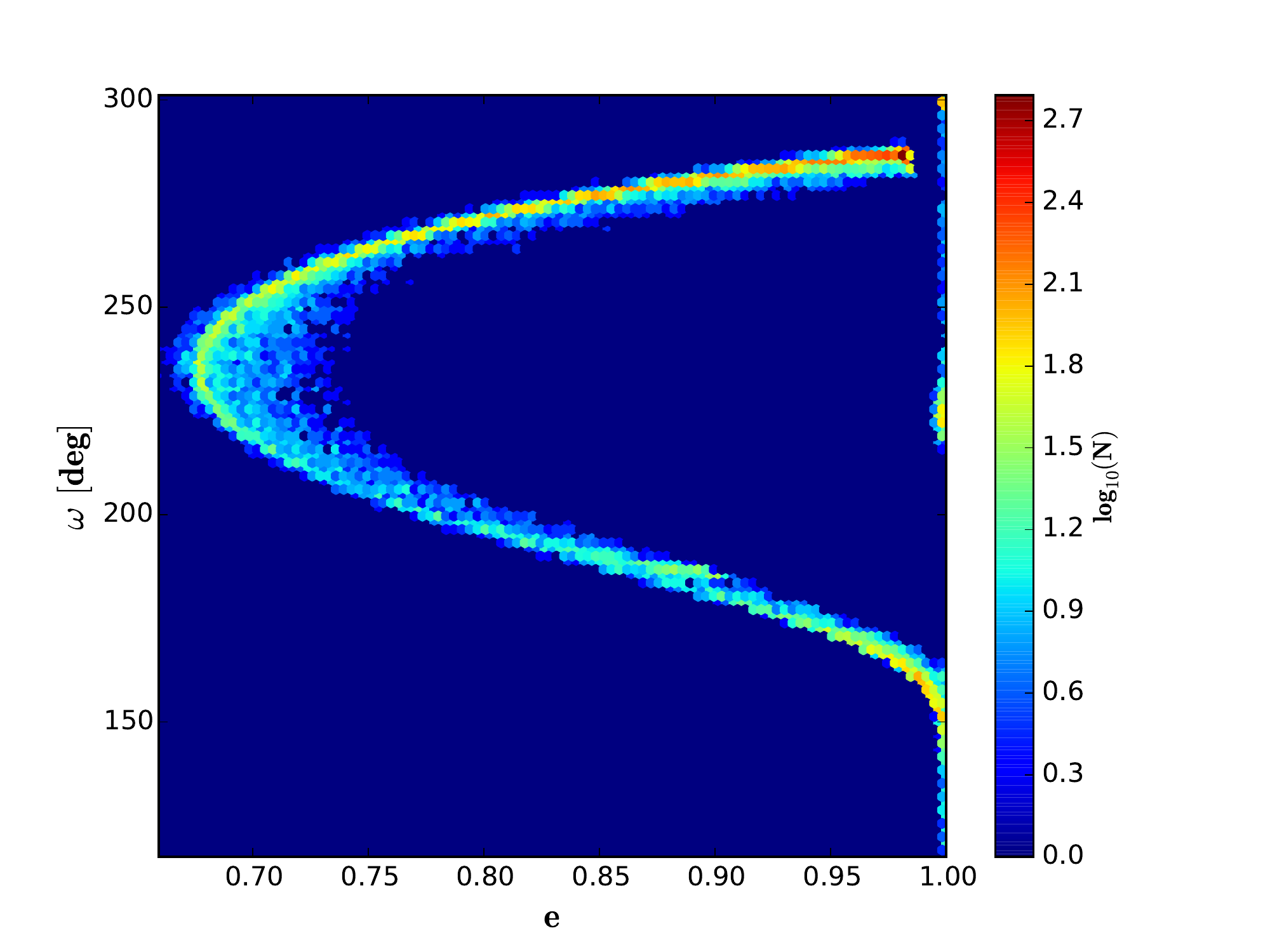}
\includegraphics[trim = 0mm 3mm 21mm 13mm,clip,height=0.28\textheight]{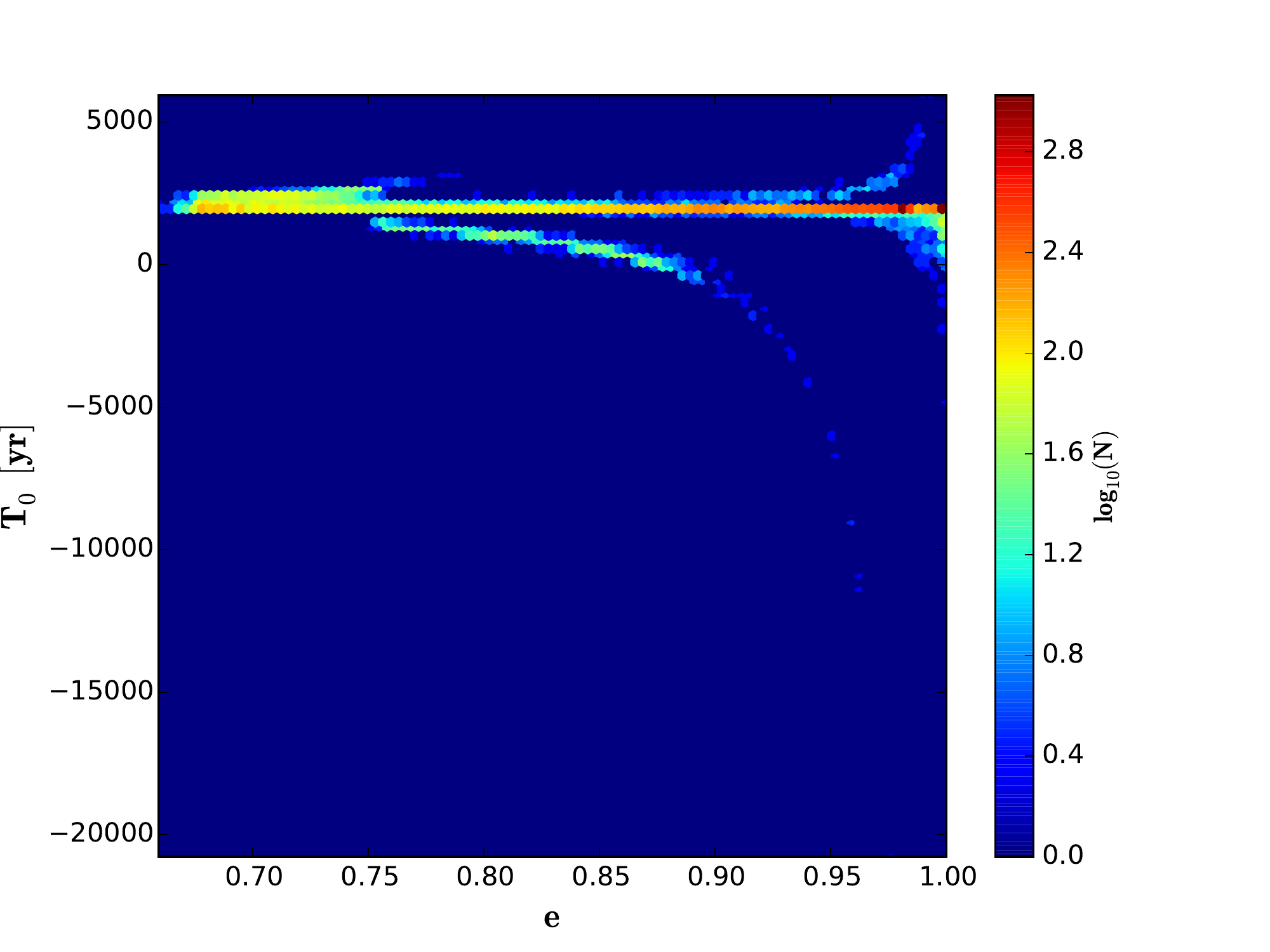}
\caption[]{{Orbital elements as function of eccentricity for all solutions with $\chi^2_{\mathrm{red}} \, \leq \, 2$ out of 5~000~000 runs of our LSMC fit of the PZ~Tel~B astrometric data. \textit{From left to right and from top to bottom} are represented the semi-major axis, inclination, longitude of the ascending node, argument of periastron, and time at periastron passage. Logarithmic density of solutions is indicated by color. }} \label{fig:orbit-corr}
\end{figure*}

\begin{figure*}
\centering
\includegraphics[trim = 14mm 2mm 23mm 13mm,clip,width=0.41\textwidth]{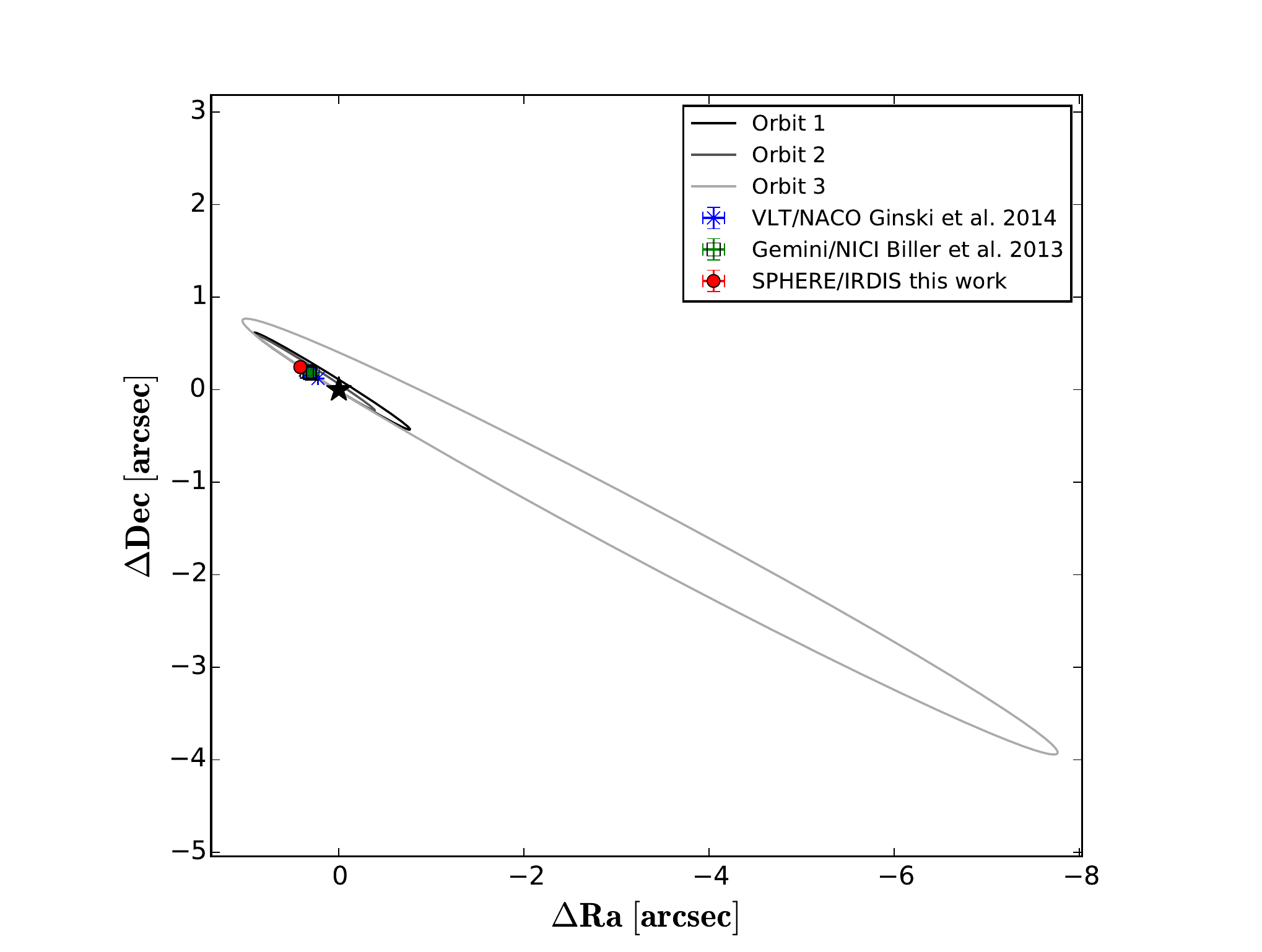}
\includegraphics[trim = 14mm 2mm 23mm 13mm,clip,width=0.41\textwidth]{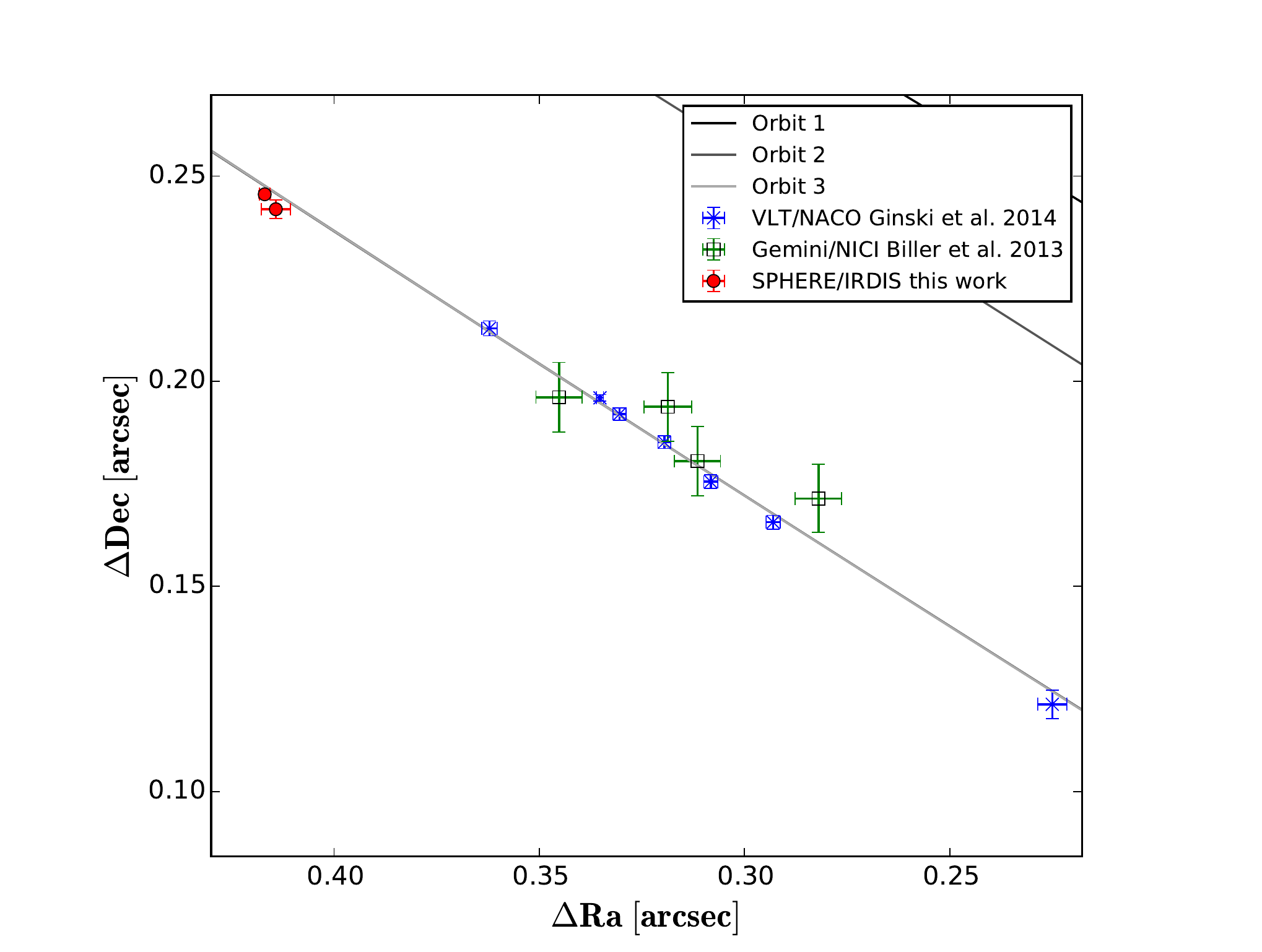}
\caption[]{Top 3 best-fitting orbits out of 5~000~000 runs of our LSMC fit of the PZ~Tel~B astrometric data. Solid lines represent the apparent orbits. The \textit{right panel} is zoomed in on the data points. The corresponding orbit elements are listed in Table~\ref{tab: orbit-elements}. We show the data points taken with VLT/NaCo as given in \cite{Ginski2014} (blue crosses), as well as the data points taken with Gemini/NICI given in \cite{Biller2013} (empty green boxes), together with our SPHERE/IRDIS measurements (red circles). For the orbit fitting, only the VLT/NaCo and SPHERE/IRDIS data points were considered.} 
\label{orbits}
\end{figure*}

The results of the fitting procedure are displayed in Fig.~\ref{fig:orbit-corr}. Shown are all orbit solutions with $\chi^2_{red} \, \leq \, 2$. We also show the three best-fitting orbits in Fig.~\ref{orbits}, along with their corresponding parameters in Table~\ref{tab: orbit-elements}. Compared to the previous analysis of the system in \cite{Ginski2014}, the possible parameter space for orbital fitting with the astrometric measurements can be constrained somewhat further. We find a lower limit for the semi-major axis of 0.52$''$ (previously 0.38$''$) and a corresponding lower limit for the eccentricity of 0.66 (previously 0.62). Orbit periods range between 126.6 and 37938.3~yr. The results of our new analysis show two distinguishable branches for orbits considering semi-major axis and eccentricity, which are visible in the top lefthand panel of Fig.~\ref{fig:orbit-corr}. One branch shows a strongly increasing semi-major axis with increasing eccentricity. The other branch exhibits small semi-major axes ($\leq$1.1$''$) even for eccentricities up to $\sim$0.9 and then also rises quickly toward large semi-major axes. In the first case, the orbit inclination is well constrained between $\sim$91$^{\circ}$ and 94$^{\circ}$, while in the latter the inclination rises with increasing eccentricities to values of up to 100$^{\circ}$ until eccentricities of up to 0.95 and then even higher inclinations of up to 153.4$^{\circ}$. With further monitoring in the next few years, it might well be possible to exclude or confirm especially short period orbits in this second branch.

It is also useful to look into possible orbit constraints for the less extreme eccentricity cases with $e$\,$\leq$\,0.9. If indeed the orbit eccentricity were not higher than this value, the orbit would already be well constrained. In particular, the inclination would be limited to values between 91.0$^{\circ}$ and 96.1$^{\circ}$, and the longitude of the ascending node would lie between 55.1$^{\circ}$ and 59.1$^{\circ}$ (with a possible change in direction of +180$^{\circ}$ depending on radial velocity). In this scenario and given that the host star is seen almost from its equator (Sect.~\ref{sec:literaturepztela}), the PZ~Tel system is close to a configuration with spin-orbit alignment. In addition, semi-major axes would lie between 0.52$''$ and 4.72$''$, i.e. orbital periods between 126.9 and 3455.9~yr.

More than two-thirds of all orbit solutions (72.6\,\%) have a time of the periastron passage between 1994 and 2005. This is very consistent with the non-detection of the companion in VLT/NaCo images of epoch 2003.55 as reported in \cite{Masciadri2005}. In fact, all our recovered orbits exhibit a separation smaller than 0.16$''$ for this observation epoch, which is the minimum separation for which a companion as massive as PZ~Tel~B would have been detected in the corresponding study.

\begin{table}[t]
 \centering
  \caption{Parameters and $\chi^2_{\mathrm{red}}$ of the best-fitting orbits for PZ~Tel~B shown in Fig.~\ref{orbits}.}
  \begin{tabular}{@{}lccc@{}}
  \hline   
        \hline
        Orbital solution                        & 1                                             &  2                                              & 3                             \\
        \hline
        $a$\,[$''$]                     & 1.504                         &       0.951                                   & 13.409                  \\
        $e$                                                                     & 0.755                           & 0.686                                 & 0.969                   \\
        $P$\,[yr]                                               & 622.2                         &       313.1                                   &       16563.3         \\      
        $i$\,[$^{\circ}$]                                       & 92.05                         &       92.34                                   &       91.83                   \\
        $\Omega$\,[$^{\circ}$]          & 58.43                         &       58.25                                   &       58.58                   \\
        $\omega$\,[$^{\circ}$]          & 264.38                        &       241.99                          & 286.54                  \\
        $T_0$\,[yr]                             & 2000.0                        & 1997.5                          & 2003.8                \\
        $\chi^2_{\mathrm{red}}$         &       1.220                           &       1.220                                   &       1.223                   \\
 \hline\end{tabular}

\label{tab: orbit-elements}
\tablefoot{The notations for the orbit elements refer to the semi-major axis, eccentricity, orbital period, inclination, longitude of the ascending node, argument of periastron, and time of periastron passage.}
\end{table}

\begin{figure}[t]
\centering
\includegraphics[trim = 6mm 2mm 19mm 13mm,clip,width=0.41\textwidth]{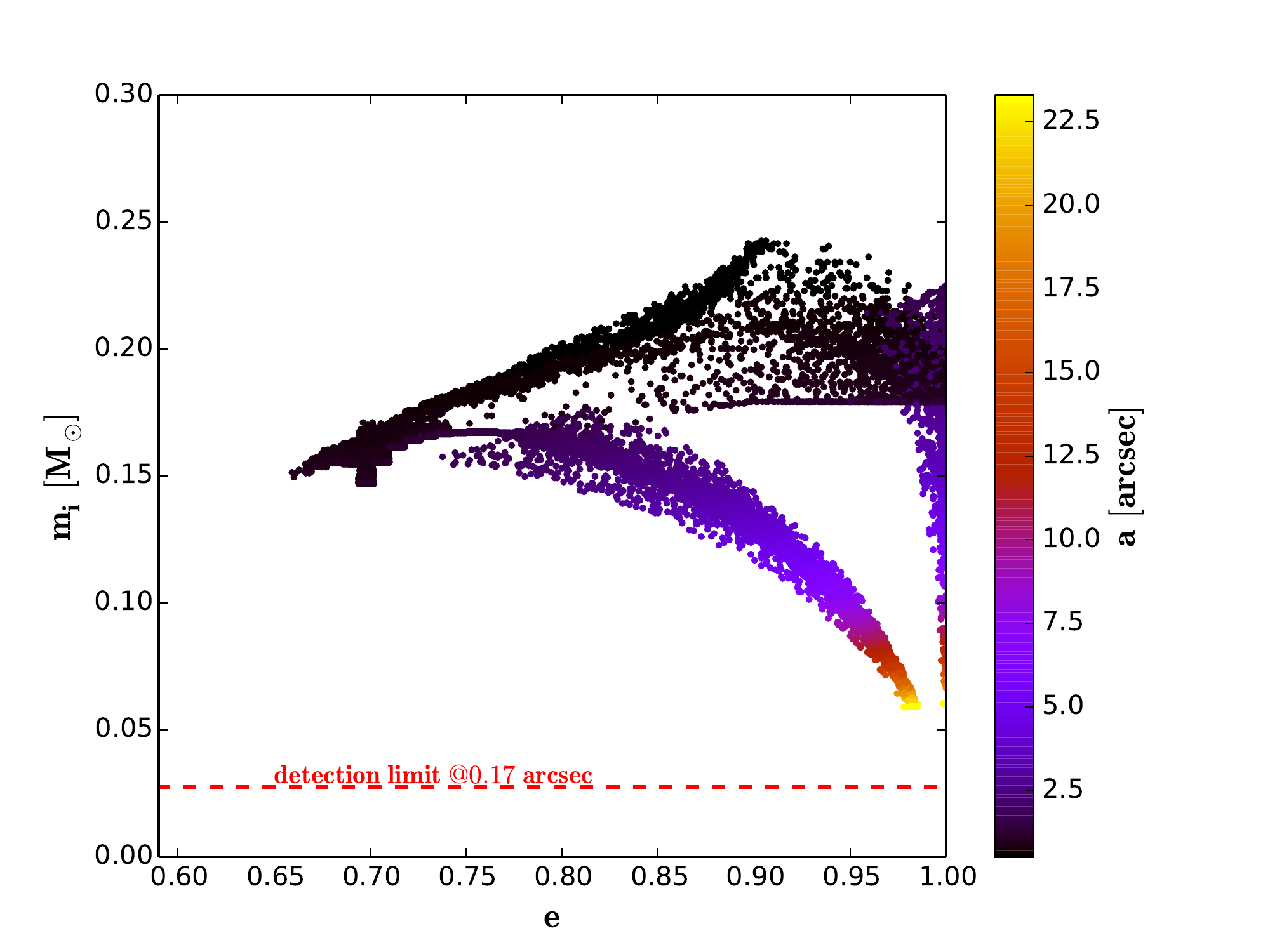}
\caption[]{Minimum mass of an unseen inner companion that would cause a false positive eccentricity signal in the relative astrometry of PZ~Tel~A and B by astrometric displacement of PZ~Tel~A due to their common orbit around their center of mass. The minimum mass is shown for all the recovered orbits with $\chi^2_{\mathrm{red}} \, \leq \, 2$ (see text). We also indicate the minimum mass of a putative inner companion detected in the IRDIS $K2$-band images at the angular separation at which it would need to reside.}
\label{fig:pearce-pztel}
\end{figure}

The best-fitting orbits that we recovered are shown in Fig.~\ref{orbits} with the corresponding orbit elements listed in Table~\ref{tab: orbit-elements}. Compared to previous studies, it is notable that with the SPHERE/IRDIS data points, the best-fitting orbits are less eccentric. Specifically, the previously best-fitting orbit for this system exhibited an eccentricity greater than 0.99, while in our new analysis the best-fitting orbit only exhibits an eccentricity of 0.755. This orbit configuration is much more likely to produce a long-term stable orbit than the high eccentricity case. In addition to the VLT measurements, we show astrometric data points obtained with Gemini/NICI by \cite{Biller2013} overplotted on our best-fitting orbits. They show a larger intrinsic scatter and also larger uncertainties compared to the considered VLT/NaCo and SPHERE/IRDIS measurements. However, they also agree with the best-fitting orbits within $\sim$1.1$\sigma$.

In a recent study, \cite{Pearce2014} discuss the problems that additional undetected inner companions could cause for the orbit determination of known direct imaging companions. In particular, they point out that an inner companion on a much shorter period than the outer direct imaging companion would introduce an astrometric ``wobble'' to its host star, owing to their orbital motion around their common center of mass. This astrometric signal would then influence the relative astrometric measurements between the host star and the direct imaging companion. This could lead to apparent orbits derived from this relative astrometry that appear eccentric, while in fact the true orbit of the outer companion is circular or nearly circular.

To validate the high eccentricities found for the PZ~Tel system, we used the formalism introduced in \cite{Pearce2014} and computed the minimum mass and distance of such a putative inner companion for each of our recovered orbit solutions.\footnote{A similar computation for an exemplary orbit has already been done in \cite{Pearce2014}; however, the results depend on the semi-major axis and eccentricity of the apparent orbits, thus it is sensible to repeat the calculation for all orbit solutions. The minimum mass is also a function of the maximum epoch difference of all astrometric data.} We then compared these masses with the detection limits derived from our deep IRDIS $K2$-band images (Sect.~\ref{sec:detlims}). The results are shown in Fig.~\ref{fig:pearce-pztel}. In principle, inner companions with masses between $\sim$0.06 and $\sim$0.25~$M_\odot$ could cause a false positive eccentricity signal in case of the PZ~Tel system. However, all of these putative companions would have been detected in our IRDIS images. Thus, we can conclude that the observed high eccentricities of PZ~Tel~B are genuine.

   \begin{figure*}[t]
   \centering
   \includegraphics[trim = 15mm 5mm 4mm 5mm, clip,width=.41\textwidth]{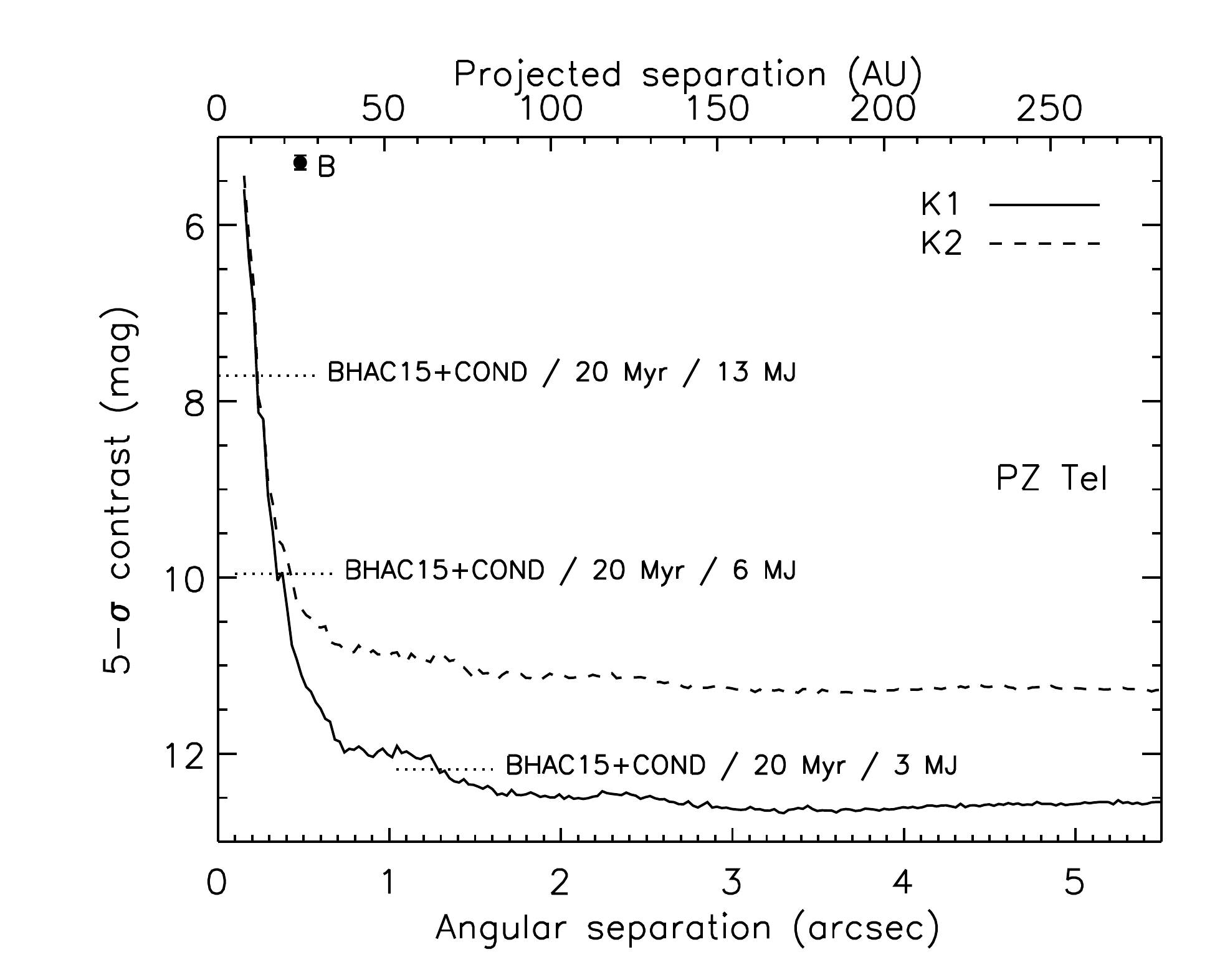}
   \includegraphics[trim = 15mm 5mm 4mm 5mm, clip,width=.41\textwidth]{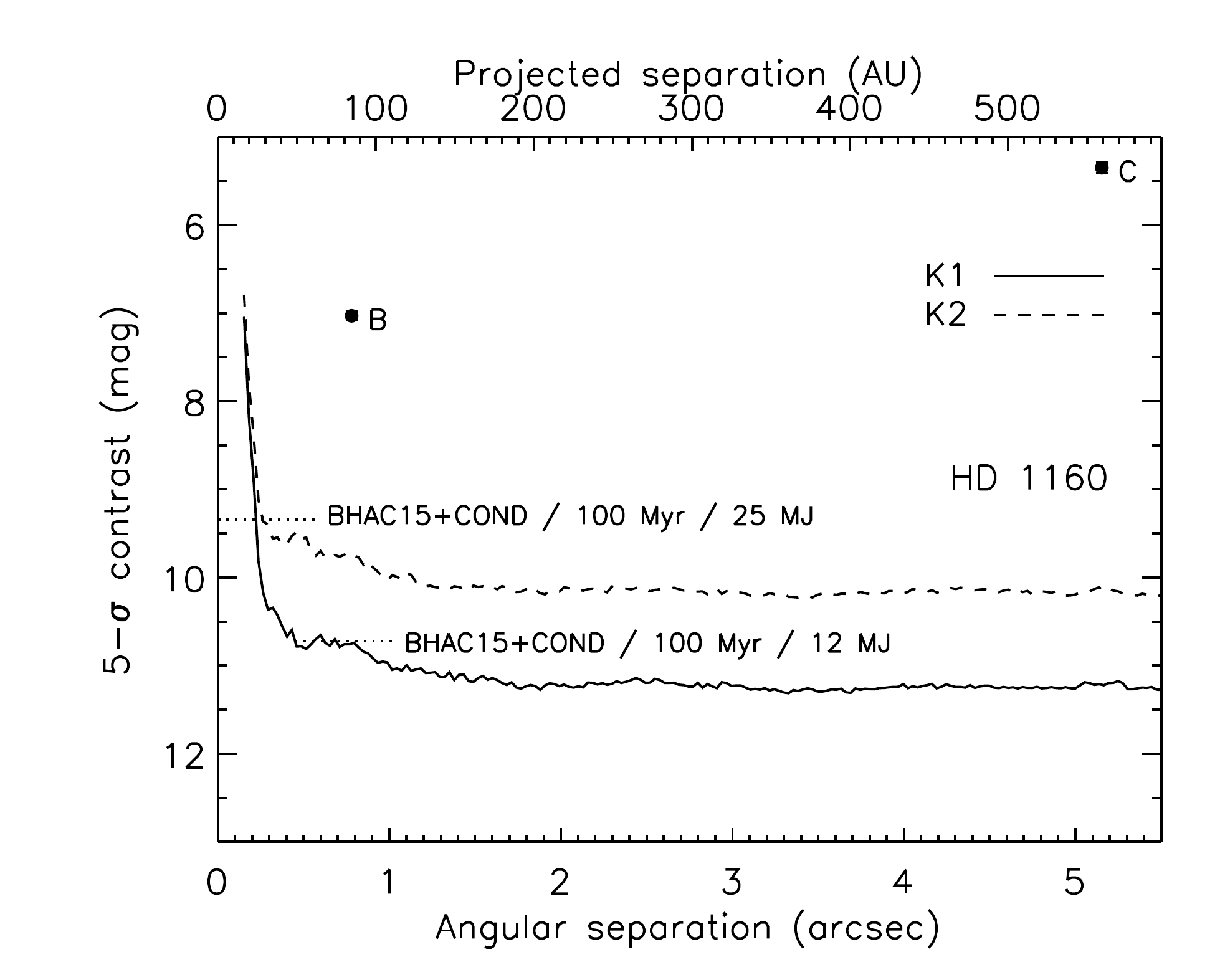}
      \caption{SPHERE/IRDIS detection limits for the images in the $K12$ filter pair analyzed with TLOCI using ADI only for the systems PZ~Tel (\textit{left}) and HD~1160 (\textit{right}). The mass limits are estimated from the atmospheric and evolutionary ``hot-start'' model BHAC15+COND \citep{2015arXiv150304107B}. The relative magnitude of the companions and the mass predictions are derived for the $K1$ band.}
         \label{fig:detectionlimits}
   \end{figure*}
   
      \begin{figure*}[t]
   \centering
   \includegraphics[trim = 15mm 5mm 4mm 5mm, clip,width=.41\textwidth]{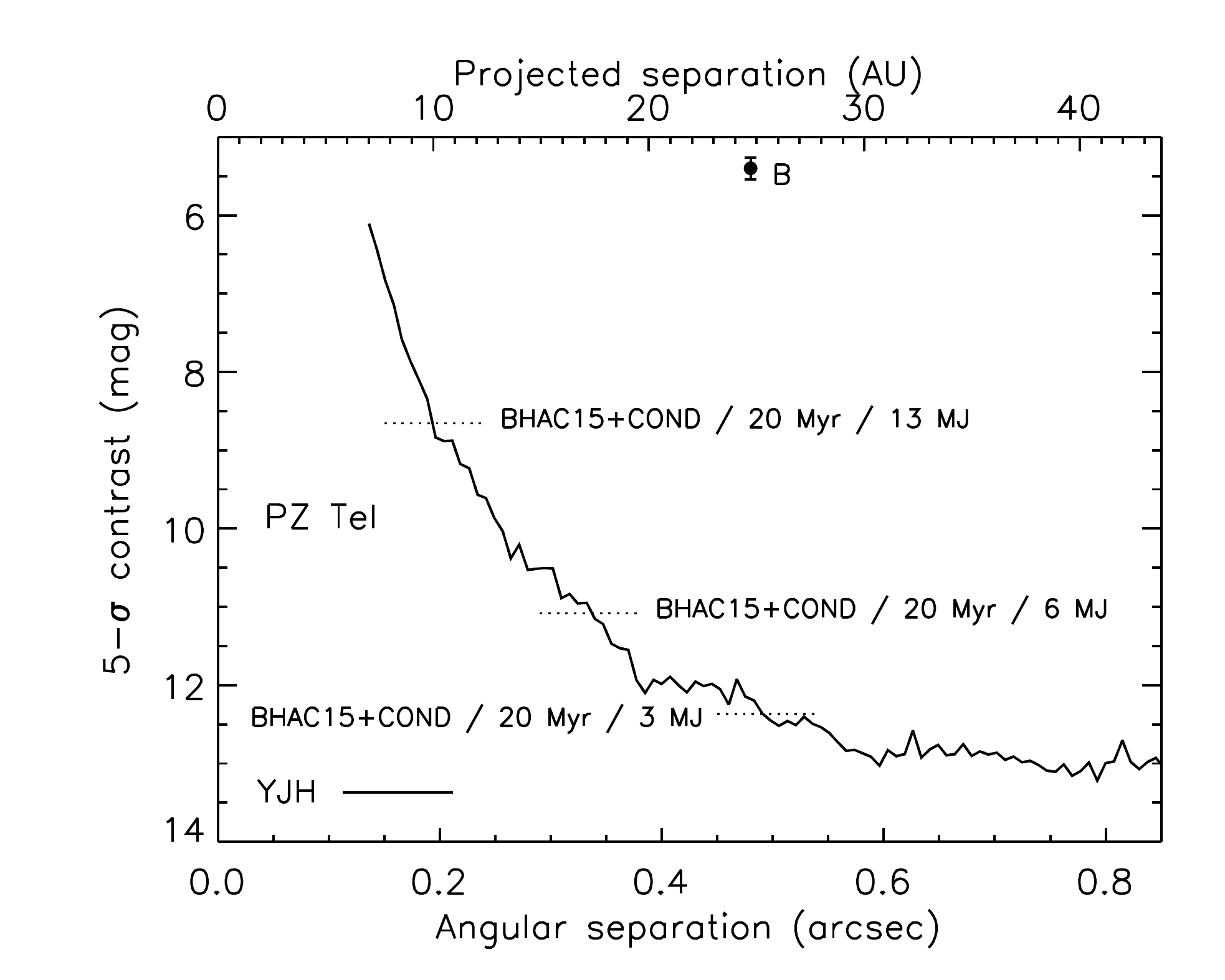}
   \includegraphics[trim = 15mm 5mm 4mm 5mm, clip,width=.41\textwidth]{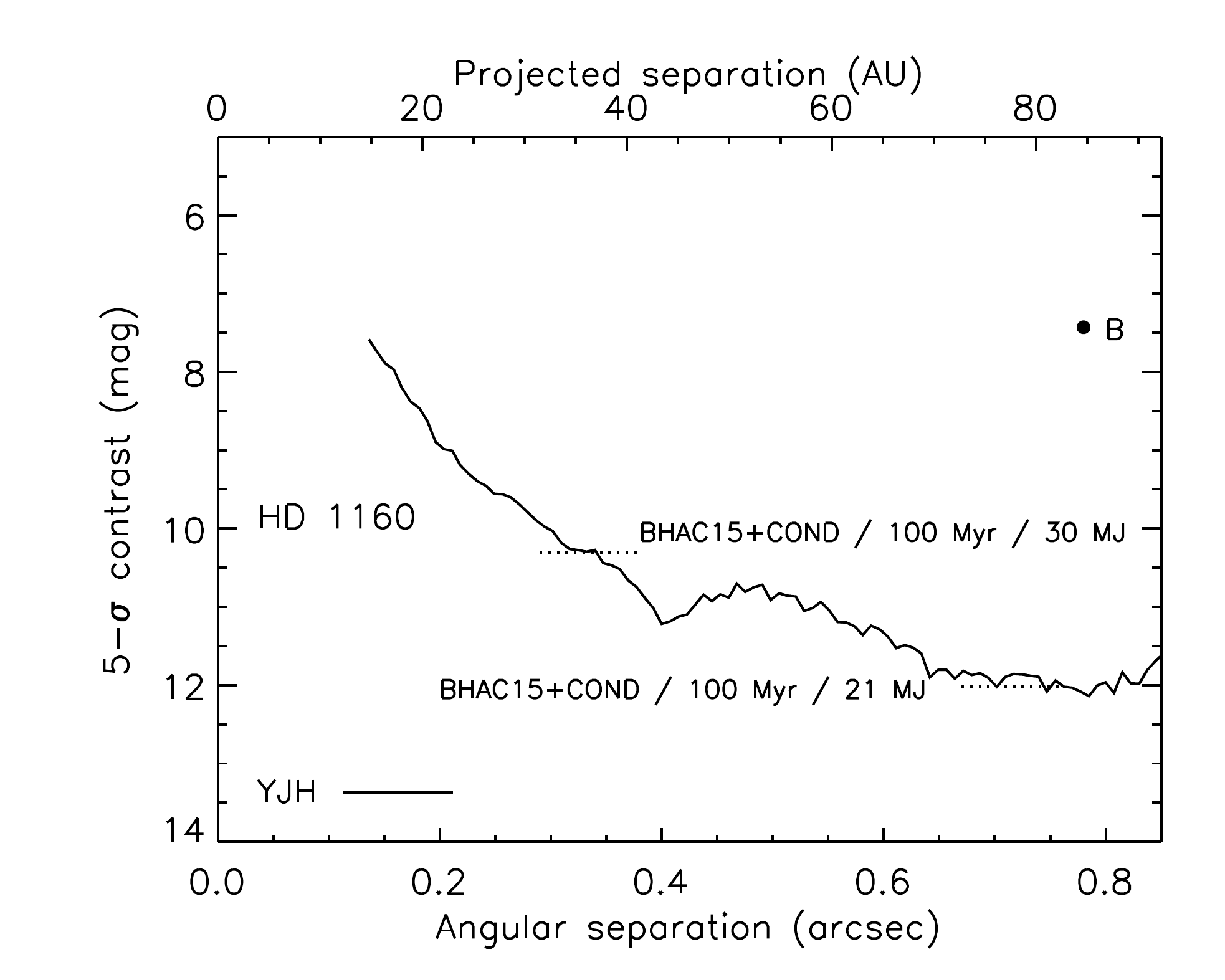}
      \caption{{SPHERE/IFS detection limits for the $YJH$ spectral data cubes analyzed with PCA using ADI only for the systems PZ~Tel (\textit{left}) and HD~1160 (\textit{right}). The resulting spectral data cube was collapsed before the estimation of the detection limits. The mass limits are estimated from the atmospheric and evolutionary ``hot-start'' model BHAC15+COND \citep{2015arXiv150304107B}. The plotted relative magnitude of the companions is in the $J$ band and the mass derived from model predictions in this band.}}
         \label{fig:ifsdetectionlimits}
   \end{figure*}

\section{Constraints on putative additional companions}
\label{sec:detlims}

We discuss the constraints on putative additional companions in both PZ~Tel and HD~1160 systems using the sensitivity of the IRDIS images in the $K12$ filter pair and in the IFS $YJH$ data sets. We present in Fig.~\ref{fig:detectionlimits} the detection limits for both IRDIS $K1$ and $K2$ images processed with TLOCI using ADI alone (Sect.~\ref{sec:irdisreduction}). Figure~\ref{fig:ifsdetectionlimits} shows the detection limits measured in the IFS data analyzed with ADI using a PCA algorithm (Sect.~\ref{sec:ifsdataanalysis}). {The PCA algorithm was} applied to each spectral channel. The detection limits were measured in the collapsed PCA data cube. For deriving the mass limits, we used the predictions of the atmosphere and evolutionary ``hot-start'' model BHAC15+COND \citep{2015arXiv150304107B} derived in the IRDIS filters and the IFS bands.

{Before analyzing the detection limits, we emphasize that the contrast performance shown in Figs.~\ref{fig:detectionlimits} and \ref{fig:ifsdetectionlimits} is not representative of the SPHERE requirements ($\sim$12.5--15~mag at 0.5--1$''$). We recall that the primary goals of the commissioning runs were to test and validate the instrument modes on sky. The IRDIFS\_EXT observations used here are snapshots and cover small field rotations (Table~\ref{tab:obs}). The short integration times and the relative faintness of PZ~Tel and HD~1160 in the NIR bands ($\sim$6--7~mag) are the main limiting factors to the contrast performance at large angular separations ($\gtrsim$1$''$), where the dominant source of noise is thermal background. The thermal background increases with wavelength, hence the poorer contrast performance observed for IRDIS in the $K2$ band with respect to the $K1$ band. The small field rotations limit the performance at small angular separations ($\lesssim$1$''$), because of the algorithm throughput that counterbalances the gain in contrast offered by the PSF subtraction.}

For the PZ~Tel system, the field rotation for the data is small (9$^{\circ}$). Despite this, the IRDIS detection limits in the $K1$ filter allow brown dwarf companions ($M$\,$>$\,13~$M_{\rm{J}}$) to be excluded outside 0.25$''$ and massive ($>$6~$M_{\rm{J}}$) giant planet companions {outside $\sim$0.4$''$}. The IFS data are sensitive to brown dwarf companions outside 0.2$''$ and giant planets more massive than 3~$M_{\rm{J}}$ {beyond $\sim$0.5$''$}. Detection limits on additional companions were presented in \citet{Mugrauer2010, Biller2013, Ginski2014} based on VLT/NaCo $K_s$-band images or Gemini/NICI broad-band data in the $H$ band\footnote{\label{note:asdidetlimits}We note that \citet{Biller2010, Biller2013} derived detection limits combining ADI and dual-band differential imaging in narrow-band images in the $H$ band. Nevertheless, the detection limits do not account for the throughput of the dual-band differential imaging, because the latter depends on the assumed spectral properties for the companions.}. We achieve contrasts with IRDIS in the $K1$ narrow-band filter deeper than the contrasts reported in \citet{Ginski2014} {with NaCo} in the $K_s$ broad-band filter at separations beyond $\sim$0.3$''$.

For the HD~1160 system, the data were obtained in better conditions (field rotation $\sim$18.5$^{\circ}$). Nonetheless, the star is fainter in the NIR with respect to PZ~Tel. {Thus, the contrasts that we achieve with IRDIS are higher in the speckle-limited regime and lower in the background-limited regime.} The sensitivity to low-mass companions is worse with respect to PZ~Tel because of the older age and farther distance of HD~1160. We can reject brown dwarf companions outside 0.25$''$ based on the IRDIS detection limits. The sensitivity of the IFS data allows us to exclude {massive ($>$25~$M_{\rm{J}}$) brown dwarf companions beyond $\sim$0.5$''$}. The contrasts that we measured for the IRDIS $K1$ narrow-band data are deeper than the contrasts in broad $H$ band presented in \citet{Nielsen2013}\footnote{We do not consider the combined ADI and dual-band imaging detection limits also reported in this paper (see note~\ref{note:asdidetlimits}).} at close-in separations ($\lesssim$0.75$''$) and are more sensitive to putative additional low-mass companions.

\section{Conclusions}
We have presented first light results of the new planet finder VLT/SPHERE for the young systems PZ~Tel and HD~1160. These data were analyzed with complementary literature and unpublished data from existing facilities (VLT/SINFONI, VLT/NaCo, Gemini/NICI, Keck/NIRC2) to provide new insight into both systems.

We first studied the short-term and long-term photometric variability of the host star PZ~Tel~A based on REM and literature data. The results for the rotational period agree with previous studies. We also estimated updated photometry for the star at optical and NIR wavelengths. We used the NIR updated photometry of the host star with optical and NIR spectrophotometry measured for the companion PZ~Tel~B to analyze the SED and physical characteristics of the latter. We confirmed a M7$\pm$1 spectral type and a low surface gravity log($g$)\,$<$\,4.5~dex, but found a higher effective temperature $T_{\rm{eff}}$\,=\,2700\,$\pm$\,100~K. The metallicity could not be constrained. Assuming the stellar age of 24\,$\pm$\,2~Myr suggested by \citet{Jenkins2012}, {we proposed a revised mass of PZ~Tel~B of} 45$^{+9}_{-7}$~$M_{\rm{J}}$ based on its effective temperature and 59$^{+13}_{-8}$~$M_{\rm{J}}$ based on its bolometric luminosity. We used the new SPHERE/IRDIS astrometry to update the constraints on the orbital parameters of the companion. We confirmed its deceleration and the high eccentricity of its orbit $e$\,$>$\,0.66. {The deep SPHERE/IRDIS detection limits suggest} that this high eccentricity is genuine and not induced by an unseen companion very close to the star. The time of periastron passage for more than 70\% of the possible orbits is consistent with the non-detection of the companion in 2003 VLT/NaCo data \citep{Masciadri2005}. For eccentricities below 0.9, we constrain the inclination to nearly edge-on values (within $\sim$6$^{\circ}$) and the longitude of the ascending node to values in the range $\sim$55--59$^{\circ}$. {Further astrometric monitoring in the coming years will provide} upper limits on the orbital eccentricity. Based on ``hot-start'' evolutionary models, we were able to exclude brown dwarf companions ($M$\,$>$\,13~$M_{\rm{J}}$) outside 0.2$''$ and {massive ($>$3~$M_{\rm{J}}$)} giant planet companions beyond $\sim$0.5$''$.

For the HD~1160 system, we derived the first NIR spectral classification ($R$\,$\sim$\,30) of HD~1160~B and found a spectral type M6.0$^{+1.0}_{-0.5}$. This spectral type is earlier than the estimate by \citet{Nielsen2012} based on the NICI photometry (L0$\pm$2). The $J-K_s$ color derived from the SPHERE data agrees with the colors of field dwarfs, although it is slightly bluer with respect to dwarfs of the same spectral type. This could be a hint that there is subsolar metallicity for the companion. Medium-resolution spectroscopy is needed to confirm this point. We constrained the effective temperature of HD~1160~B to a value of 3000\,$\pm$\,100~K but not its surface gravity. For HD~1160~C, we estimated $T_{\rm{eff}}$\,=\,3260\,$\pm$\,100~K, which agrees with the results of \citet{Nielsen2012}. The effective temperature and bolometric luminosity of HD~1160~B and C marginally agree (within 2$\sigma$) with an age of 100--120~Myr, which corresponds to the upper limit of the range derived by \citet{Nielsen2012} (50$^{+50}_{-40}$~Myr). Assuming an age for the host star of 100$^{+200}_{-70}$~Myr, we assessed a mass for HD~1160~B of 107$^{+59}_{-38}$~$M_{\rm{J}}$ (based on effective temperature) and 79$^{+65}_{-40}$~$M_{\rm{J}}$ (based on bolometric luminosity). For the lowest values in the age range, our mass estimates are in the substellar regime and marginally agree with the results of \citet{Nielsen2012}. Nevertheless, the uncertainties on the system age do not allow us to exclude that HD~1160~B is a low-mass star. The corresponding mass estimates for HD~1160~C are 205$^{+72}_{-51}$~$M_{\rm{J}}$ and 244$^{+113}_{-70}$~$M_{\rm{J}}$, respectively. These estimates agree with the values reported in \citet{Nielsen2012}, but have larger error bars due to the larger error bars adopted for the stellar age. {Further astrometric follow-up of the system may reveal} orbital motion for both companions thanks to the improved astrometric accuracy of the SPHERE measurements ($<$3~mas in separation and $\leq$0.2$^{\circ}$ for the parallactic angle) with respect to previous data \citep{Nielsen2012}. The SPHERE detection limits {allow  excluding} brown dwarf companions outside 0.25$''$, assuming ``hot-start'' evolutionary models.

\begin{acknowledgements}
      We thank the anynomous referee for a constructive report on the manuscript and Daniel Rouan for helpful suggestions. The authors are grateful to the Consortium and the ESO Paranal Staff for making SPHERE a reality. The authors warmly thank Andrea Bellini and Jay Anderson for kindly providing the catalog positions of the stars in the 47~Tuc field before their publication. We are also grateful to ESO for releasing the commissioning data for publication. A.-L.M., S.M., D.M., R.G., S.D., R.U.C., and A.Z. acknowledge support from the ``Progetti Premiali'' funding scheme of the Italian Ministry of Education, University, and Research. {A.-L.M. thanks the MPIA for support during the last stages of this work.} A.V., M.B., G.C., and D.M. acknowledge financial support from the French National Research Agency (ANR) through the GUEPARD project grant ANR10-BLANC0504-01, the CNRS-D2P PICS grant, and the Programmes Nationaux de Plan\'etologie et de Physique Stellaire (PNP \& PNPS). D.E. acknowledges the financial support of the Swiss National Science Fundation (SNSF) through the National Centre for Competence in Research ``PlanetS''. {A.Z. acknowledges support from the Millennium Science Initiative (Chilean Ministry of Economy), through grant ``Nucleus RC130007''.} This research has benefitted from the SpeX Prism Spectral Libraries, maintained by Adam Burgasser at http://pono.ucsd.edu/$\sim$adam/browndwarfs/spexprism. This research made use of the SIMBAD data base, as well as the VizieR catalog access tool, both operated at the CDS, Strasbourg, France. The original description of the VizieR service was published in Ochsenbein et al. (2000, A\&AS 143, 23).\\
      
      SPHERE is an instrument designed and built by a consortium consisting of IPAG (Grenoble, France), MPIA (Heidelberg, Germany), LAM (Marseille, France), LESIA (Paris, France), Laboratoire Lagrange (Nice, France), INAF -- Osservatorio di Padova (Italy), Observatoire de Gen\`eve (Switzerland), ETH Zurich (Switzerland), NOVA (Netherlands), ONERA (France), and ASTRON (Netherlands), in collaboration with ESO. SPHERE was funded by ESO, with additional contributions from the CNRS (France), MPIA (Germany), INAF (Italy),
FINES (Switzerland), and NOVA (Netherlands). SPHERE also received funding from the European Commission Sixth and Seventh Framework Programs as part of the Optical Infrared Coordination Network for Astronomy (OPTICON) under grant number RII3-Ct-2004-001566 for FP6 (2004--2008), grant number 226604 for FP7 (2009--2012), and grant number 312430 for FP7 (2013--2016).\end{acknowledgements}

\bibliographystyle{aa}
\bibliography{biblio}

\end{document}